\documentclass[aps,toolkits,twocolumn]{revtex4}

\usepackage{graphicx,bm}
\usepackage{amsmath,amssymb}
\usepackage{mathrsfs,verbatim,subfigure}
\usepackage{graphics,ulem}

\def\eg{{\it e.g.}}
\def\cf{{\it cf.}}
\def\ie{{\it i.e.}}

\newcommand{\beq}{\begin{equation}}
\newcommand{\eeq}{\end{equation}}
\newcommand{\bea}{\begin{eqnarray}}
\newcommand{\eea}{\end{eqnarray}}

\newcommand{\avg}[1]{\left< #1 \right>} 

\begin{document}

\title{Massive Yang-Mills for Vector and Axial-Vector
Spectral Functions at Finite Temperature}

\author{Paul M. Hohler}
\email{pmhohler@comp.tamu.edu}
\affiliation{Cyclotron Institute and Department of Physics\,\&\,Astronomy,
Texas A{\&}M University, College Station, Texas 77843-3366, USA }

\author{Ralf Rapp}
\email{rapp@comp.tamu.edu}
\affiliation{Cyclotron Institute and Department of Physics\,\&\,Astronomy,
Texas A{\&}M University, College Station, Texas 77843-3366, USA }

\date{\today}

\begin{abstract}
The hadronic mechanism which leads to chiral symmetry restoration is
explored in the context of the $\rho\pi a_1$ system using Massive Yang-Mills,
a hadronic effective theory which governs their microscopic interactions.
In this approach, vector and axial-vector mesons are implemented as gauge
bosons of a local chiral gauge group. We have previously shown that this
model can describe the experimentally measured vector and axial-vector
spectral functions in vacuum. Here, we carry the analysis to finite
temperatures by evaluating medium effects in a pion gas and
calculating thermal spectral functions. We find that the spectral peaks
in both channels broaden along with a noticeable downward mass shift
in the $a_1$ spectral peak and negligible movement of the $\rho$ peak.
The approach toward spectral function degeneracy is accompanied by a
reduction of chiral order parameters, \ie, the pion decay constant and
scalar condensate. Our findings suggest a mechanism where the chiral
mass splitting induced in vacuum is burned off.
We explore this mechanism and identify future
investigations which can further test it.
\end{abstract}

\maketitle

\section{Introduction}

Chiral symmetry is a basic symmetry of quantum chromodynamics (QCD),
the theory of the strong nuclear force. This symmetry is spontaneously
broken in the vacuum via the formation of quark condensates. At higher
temperatures, the symmetry is restored as the condensates melt passing
through a pseudo-critical region~\cite{Borsanyi:2010bp,Bazavov:2011nk}
around $T_{\rm pc}\simeq$~160\,MeV. This behavior is imprinted on the
hadronic spectra when heating up the vacuum. Despite these
well-established properties of QCD,
the hadronic mechanism leading to chiral restoration is not fully
understood to date.

The only in-medium hadronic spectral function which is experimentally
accessible is that of the light vector mesons, dominated by the $\rho$,
which can be probed via low-mass dilepton spectra in heavy-ion
collisions~\cite{Porter:1997rc,Adamova:2006nu,Arnaldi:2008fw,Agakishiev:2011vf,Geurts:2012rv}.
Theory calculations reveal that the experimental data are
compatible with a $\rho$ melting scenario whereby the spectral
peak broadens without a significant shift in its mass, see, \eg,
Refs.~\cite{Rapp:2009yu,Rapp:2013nxa}.
It has been conjectured that this broadening is an indication of chiral
symmetry restoration based on QCD and chiral sum rules~\cite{Hohler:2013eba}.
However, for a definitive assertion of restoration, the in-medium spectral
function of the chiral partner of the $\rho$, namely the light
axial-vector meson $a_1$, needs to be calculated to test spectral
degeneracy. This spectral function is difficult to measure
experimentally. Thus one is led to study the $\rho \pi a_1$ system
theoretically to establish chiral restoration.

Recent sum rule analyses indicate that the experimental dilepton
measurements are consistent with an approach towards restoration and
spectral degeneracy between the vector ($V$) and axial-vector ($AV$)
channels~\cite{Hohler:2013eba}. This degeneracy is imprinted on the
$a_1$ spectral function by a mass shift toward the $\rho$ mass,
accompanied by spectral broadening. This analysis suggests a possible
mechanism for restoration in the hadronic spectrum. However, other
analyses suggest that the sum rules can be satisfied in medium without
a mass shift or a need to approach spectral degeneracy~\cite{Ayala:2014rka}.
Therefore, it is in order to explore the mechanism from a microscopic
perspective.

A theoretically appealing construction is to implement the $\rho$ and $a_1$
as the gauge bosons of a local chiral gauge group~\cite{Yang:1954ek}.
This combines chiral effective theories, which have had considerable
success in describing pion driven low-energy properties of
QCD~\cite{Gasser:1983yg}, with the general field theoretical concept
of gauge symmetries.
This has been realized in two formalisms, Massive
Yang-Mills (MYM)~\cite{Gomm:1984at} and Hidden Local
Symmetry (HLS)~\cite{Bando:1987br,Harada:2003jx},
which have been shown to be on-shell equivalent.

Early applications of MYM successfully describe the tree level masses and
decays of the $\rho$ and $a_1$ mesons~\cite{Gomm:1984at,Song:1993ae,Ko:1994en}.
These studies were extended to finite
temperatures~\cite{Song:1993af,Song:1993ipa,Pisarski:1995xu} revealing
a reduction of the $a_1$ mass along with an increase in the $\rho$ mass.
However, these calculations do not survive the scrutiny of a loop calculation
needed to determine the vacuum spectral functions which have been
experimentally measured with high precision through $\tau$ leptons decaying
into multi-pion states~\cite{Barate:1998uf,Ackerstaff:1998yj}.
An in-medium mass reduction in the $a_1$ spectral function is also compatible
with the HLS framework~\cite{Harada:2008hj}, though a satisfactory description
of vacuum spectral functions has not yet been demonstrated either.

The complications in describing the vacuum spectral functions with MYM and HLS
triggered investigations which abandon the local implementation of the gauge
group in favor of a global one~\cite{Urban:2001ru,Wagner:2008gz,Parganlija:2010fz}.
These models have also been extended to include temperature effects; using a loop
expansion, a broadening of the $\rho$ spectral peak without a mass shift
was found in Ref.~\cite{Urban:2001uv}, whereas Ref.~\cite{Struber:2007bm},
using a 2-particle irreducible truncation scheme for the in-medium masses,
found a slightly increasing $\rho$-mass to degenerate with a dropping 
$a_1$-mass above the critical temperature.

In a recent work, we have re-evaluated MYM as an effective theory to
describe the light-meson chiral properties, and were able to overcome
the initial difficulties associated with the vacuum spectral
functions~\cite{Hohler:2013ena}.
This was made possible by two critical ingredients: a fully dressed
$\rho$ propagator in the calculation of the $a_1$ self-energy, and the
inclusion of a chirally invariant continuum. In the present paper,
motivated by this development and the theoretical appeal of MYM, we
implement this approach in a finite-temperature pion gas, examining the
medium modifications of the spectral functions to deduce information
on the hadronic mechanism of chiral restoration.

Our paper is organized as follows. In Sec.~\ref{sec:MYM}, we briefly
introduce MYM. In Sec.~\ref{sec:calc}, we recall our strategy of
implementing a broad $\rho$ spectral function into the vacuum MYM
framework, and extend our earlier work to the linear $\sigma$ model
and with further applications to vacuum observables.
In Sec.~\ref{sec:rhoFT}, the in-medium $\rho$ and $a_1$ spectral functions
are calculated including polarization and 3-momentum dependencies.
In Sec.~\ref{sec:chirest}, we illustrate and quantify
the progression towards chiral symmetry restoration by computing
the pion decay constant and the scalar condensate.
In Sec.~\ref{sec:disc}, we conduct a more general discussion on
the mechanism of chiral restoration in light of our and related works
on the subject, and identify promising directions of future work.
We summarize and conclude in Sec.~\ref{sec:concl}.

\section{The Massive Yang-Mills Lagrangian}
\label{sec:MYM}
In the MYM approach~\cite{Gomm:1984at,Song:1993ae}, the chiral pion
Lagrangian is locally gauged under the symmetry
group $SU(2)_L \times SU(2)_R$.
In this paper, we will explore both the linear and
non-linear realization of this symmetry.
The explicit form of the Lagrangian, including the
higher derivative ``non-minimal" terms, is given by
\begin{equation}\label{eq:lag}
\begin{split}
\mathcal{L}_{\rm MYM}& = \frac{\tilde{f}_\pi^2}{8} ( {\rm Tr}[D_\mu U^\dag D^{\mu} U]
+ \tilde{m}_\pi^2{\rm Tr}[U+U^\dag-2] )\\
& - \frac{1}{2} {\rm Tr}[F_L^2 + F_R^2]  + m_0^2 {\rm Tr}[A_L^2+A_R^2] \\
& - i \xi \ {\rm Tr}[D_\mu U D_\nu U^\dag F_L + D_\mu U^\dag D_\nu U F_R]\\
& +\gamma \ {\rm Tr}[F_L U F_R U^\dag] \ .
\end{split}
\end{equation}
Implicit in this Lagrangian is a derivative expansion in the chiral
pion fields but not in the gauge fields.

In the non-linear realization, the pion fields are generated
by expanding the ansatz for the field
$U$=$\exp\left [{2\pi i}/{F_\pi} \right ]$
with $\pi = \pi^a \tau_a/\sqrt{2}$, where $\tau_a$ are the Pauli
matrices -- the generators of the symmetry group.
The gauge coupling, $g$, figures in the Lagrangian through the
covariant derivative,
$D_\mu U$=$\partial_\mu U - i g (A_{L \mu} U - U A_{R \mu})$
where $A_{L/R}^{\mu} = \left(A_{L/R}^{\mu}\right)^a \tau_a/\sqrt{2}$.
Vector and axial-vector combinations of the gauge fields are
defined as
\begin{equation}
V_\mu = A_{L \mu} + A_{R \mu}\, , \quad A_\mu = A_{L \mu} - A_{R \mu} \ .
\end{equation}
The last term in the Lagrangian induces a contribution to the gauge
fields' kinetic terms which can be absorbed by a field redefinition.
Thus the physical $\rho$ and $a_1$ fields, $\rho_\mu$ and $a_\mu$,
are defined as
\begin{eqnarray}
\rho_\mu = \kappa_V^{-1} V_\mu && a_\mu = \kappa_A^{-1} A_\mu \\
{\rm with} \quad \kappa_V = (1-\gamma)^{-1/2} &&
\kappa_A = (1+\gamma)^{-1/2} \ .
\nonumber
\end{eqnarray}
Mass terms, $m_0$ and $\tilde{m}_\pi$, for the physical fields are
added which explicitly break the gauge and chiral symmetries.
The unphysical $a_\mu\partial^\mu \pi$ coupling is removed by the
shift $a_\mu\rightarrow a_\mu + \alpha Z_\pi \partial_\mu \pi$.
The physical pion mass and decay constant can then be defined
in terms of the Lagrangian couplings and the pion field
renormalization, $Z_\pi$, as $M_\pi$=$Z_\pi \tilde{m}_\pi$=139.6\,MeV
and $F_\pi$=$\tilde{f_\pi}/ Z_\pi$=131\,MeV with
\begin{equation}
Z_\pi = \left(1-\frac{g_{\rho\pi\pi}^2 F_\pi^2}{2 M_\rho^2}\right)^{-1/2}.
\end{equation}

The coupling to external electro-weak (EW) fields is accomplished by
gauging the fields
under the same symmetry group as the mesons, with the coupling
constant fixed by the charge of the pion.
This includes the addition of kinetic terms for the EW fields,
new chirally and gauge invariant interactions between
the EW and meson fields of a form similar to the
Kroll, Lee, and Zumino coupling~\cite{Kroll:1967it},
{\it i.e.}, $F_L^{\rm EW} F_L$, as well as interactions
of the form of the non-minimal terms. All couplings
associated with the external fields are dictated by gauge
symmetry and related to the parameters in Eq.~(\ref{eq:lag}).
In the end, the Lagrangian contains
four parameters: the gauge coupling, $g$, the bare gauge boson
mass, $m_0$, and two non-minimal couplings, $\gamma$ and $\xi$. These can
be traded for the phenomenologically more pertinent parameters of the single-
and triple-derivative $\rho\pi\pi$ couplings, $g_{\rho\pi\pi}$ and
$g_{\rho\pi\pi}^{(3)}$, and the bare masses of the $\rho$ and $a_1$,
$M_\rho$ and $M_{a_1}$. Expressing them in terms of the Lagrangian
parameters leads to
\begin{eqnarray}
g_{\rho\pi\pi}=\frac{1}{\sqrt{2}} g \kappa_V; &&
g_{\rho\pi\pi}^{(3)} = \frac{4 \kappa_V \xi}{\sqrt{2} F_\pi^2 Z_\pi^4}
- \frac{F_\pi^2 g^3}{8 \sqrt{2} m_0^4 \kappa_V} \nonumber \\
M_\rho^2 = \kappa_V^2 m_0^2; && M_{a_1}^2 =
\kappa_A^2 (m_0^2+\frac{1}{2} g_{\rho\pi\pi}^2 F_\pi^2 Z_\pi^2) \  ,
\end{eqnarray}
which highlights the chiral splitting between the bare $\rho$ and
$a_1$ masses through the spontaneous symmetry breaking represented
by the pion decay constant.
The interaction vertices of the physical fields are then obtained
by expanding the Lagrangian; they are spelled out in
Appendix~\ref{sec:appVert}.

In the linear realization of the pion Lagrangian, an explicit $\sigma$
field is kept which may be viewed as resumming a set of $\pi\pi$
$s$-wave interactions of the non-linear formalism.
This is achieved by evaluating the Lagrangian in Eq.~(\ref{eq:lag})
with the linear ansatz
$U=\sqrt{2}/F_\pi(\sigma \mathbb{I}/Z_\pi+\sqrt{2}i \pi)$,
where $\mathbb{I}$ is the unit matrix in isospin space. Two additional
Lagrangian terms must be added to describe the scalar potential
\begin{equation}
\mathcal{L} = -\frac{\tilde{f_\pi^2}}{8}\left(\mu^2 {\rm Tr}
\left[U^\dag U\right] + \frac{\lambda }{4} \tilde{f_\pi^2}\,{\rm Tr}
\left[\left(U^\dag U\right)^2\right]\right) \, .
\end{equation}
The potential parameters $\mu$ and $\lambda$ are fixed by the vacuum
pion and sigma masses, inducing a spontaneous breaking of the gauge
symmetry resulting in a vacuum expectation value (VEV) for the $\sigma$
field, $\sigma_0$, related to the chiral condensate. The physical $\sigma$
field is then redefined relative to the vacuum,
$\sigma \rightarrow \sigma +\sigma_0$.
The VEV of the $\sigma$ field is related to the pion decay constant
as $\sigma_0 = F_\pi Z_\pi/\sqrt{2}$.

While the physical $\rho$ and $a_1$ fields are defined in
the same manner as in the non-linear case, the field
renormalizations and their masses (as well as the pion
renormalization, $Z_\pi$) can be expressed in terms of
the VEV $\sigma_0$ as
\begin{eqnarray}
\label{eq:linpara}
\kappa_{V,A} &=& (1\mp\frac{2\gamma\sigma_0^2}{F_\pi^2 Z_\pi^2})^{-1/2}
\nonumber\\
Z_\pi^2 &=& 1+\frac{g_{\rho\pi\pi}^2 \sigma_0^2}{M_\rho^2}
\nonumber\\
g_{\rho\pi\pi} &=& g \kappa_V / \sqrt{2}
\\
g_{\rho\pi\pi}^{(3)} &=& \frac{4 \kappa_V \xi}{\sqrt{2} F_\pi^2 Z_\pi^4}
- \frac{\sigma_0^2 g^3}{4 \sqrt{2} m_0^4 Z_\pi^2 \kappa_V}
\nonumber\\
M_\rho^2 &=& \kappa_V^2 m_0^2
\nonumber\\
M_{a_1}^2 &=& \kappa_A^2 \left(m_0^2+g_{\rho\pi\pi}^2 \sigma_0^2\right) \ .
\nonumber
\end{eqnarray}
Evaluating these expressions with the vacuum value for $\sigma_0$ results
in the same expressions as in the non-linear case. The relevant vertices
of the linear realization are also given in Appendix~\ref{sec:appVert}.

\section{Vacuum spectral functions in the vector and axial-vector channels}
\label{sec:calc}
We are interested in the $V$ and $AV$ spectral functions as probed by an external
EW current. The correlators of these currents are defined as
\begin{equation}
\Pi_{V,A}^{\mu\nu}(q^2) = - i \int d^4 x \ e^{i x q}
\avg{T \vec{J}_{V,A}^{\mu}(x) \vec{J}_{V,A}^{\nu}(0) } \ .
\end{equation}
In each channel, there are two independent polarization functions,
for the 4D-transverse and 4D-longitudinal components.
The correlators can be expressed in terms of
these functions as
\begin{equation}
\Pi_{V,A}^{\mu\nu}(q^2) = \Pi_{V,A}^T(q^2) \left(-g^{\mu\nu} +
\frac{q^\mu q^\nu}{q^2}\right)
+ \Pi_{V,A}^L(q^2) \frac{q^\mu q^\nu}{q^2} \, .
\end{equation}
Because of gauge invariance, $\Pi_V^L = 0$, while
$\Pi_A^L$  plays an important role in verifying that the
chiral properties inherent in the Lagrangian are maintained
once loops are included.

\begin{figure*}[!t]
\includegraphics[width=.95\textwidth]{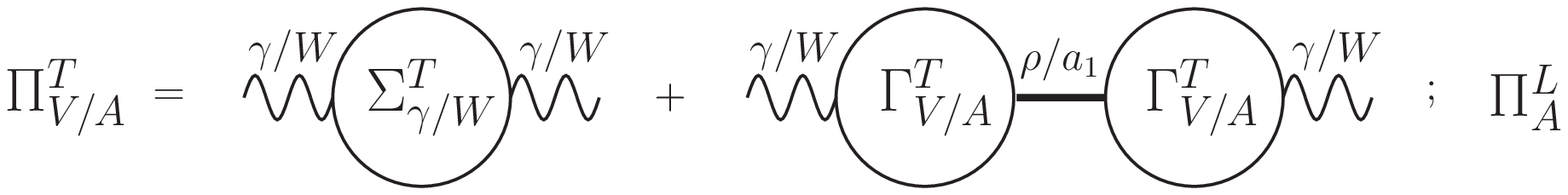}
\caption{Schematic depiction of the contributions to the EW
current-current correlators. The bold lines in the center indicate fully
resummed propagators.}
\label{fig:correlator}
\end{figure*}

The current-current correlators with external EW fields
can be constructed from the sum of a direct term
and one which transitions through
a resonant meson. For example, for the $V$ channel,
the correlator is comprised of the photon self-energy,
$\Sigma_{\gamma}^T$, plus a term which converts a photon
into a $\rho$ meson, $\Gamma_V^T$, propagates the $\rho$,
and finally converts the $\rho$ back into a photon. This,
and the processes for the $AV$ channels, are depicted
diagrammatically in Fig.~\ref{fig:correlator}.
For the transverse mode, the correlators are
expressed as
\begin{equation}
\Pi_{V/A}^T (p^2) = \Sigma_{\gamma/W}^T (p^2) + \left(\Gamma_{V/A}^T(p^2)\right)^2
\!\! D_{V/A}^T(p^2) ,
\end{equation}
with the propagators given by
\begin{equation}
D_{V,A}^T\left(p^2\right) =
\left(p^2 - M_{\rho, a_1}^2 - \Sigma_{\rho, a_1}^T\left(p^2\right)\right)^{-1} .
\end{equation}
The transition functions, $\Gamma_{V/A}^T$, describe the conversion
of a photon/W boson into a $\rho$/$a_1$ meson and are comprised
of a contact contribution and a self-energy given by
\begin{eqnarray}
\Gamma_V^T(p^2) &=& \frac{p^2}{g_{\rho\pi\pi}} + \Sigma_{\gamma \rho} (p^2), \\
\Gamma_A^T(p^2) &=& \frac{p^2}{g_{\rho\pi\pi}} - \frac{g_{\rho\pi\pi} F_\pi^2
M_{a_1} Z_\pi}{2 M_\rho} + \Sigma_{W a_1}^T(p^2) .
\end{eqnarray}

The $V$ and $AV$ transverse spectral functions are defined
in terms of the 4D-transverse current-current correlator
associated with the external EW fields,
$\rho_{V/A}^T \equiv -{\rm Im}\Pi_{V/A}^T/\pi$.
Due to the gauging procedure,
the EW self-energies can be expressed in terms of the mesonic
self-energies as
\begin{equation}
\begin{split}
\Sigma_\gamma^T = \frac{1}{g_{\rho\pi\pi}^2} \Sigma_\rho^T (p^2), &
\quad  \Sigma_{\gamma \rho} = -\frac{1}{g_{\rho\pi\pi}}
\Sigma_\rho^T (p^2) , \\
\Sigma_W^T = \frac{M_\rho^2 Z_\pi^2}{g_{\rho\pi\pi}^2 M_{a_1}^2}
\Sigma_{a_1}^T (p^2) , &
\quad \Sigma_{W a_1} = - \frac{M_\rho Z_\pi}{g_{\rho\pi\pi}
M_{a_1}} \Sigma_{a_1}^T (p^2) \, .
\end{split}
\end{equation}
Using these relations,
the spectral functions emerge in a form suggestive
of (axial-) vector meson dominance, (A)VMD,
\begin{equation}
\label{eq:vmdcor}
\rho_{V,A}^T (p^2) = - \frac{C_{V,A}}{\pi} {\rm Im} D^T_{V,A}(p^2)
\end{equation}
with the (A)VMD-like couplings given by
\begin{equation}
C_V = \frac{M_\rho^4}{g_{\rho\pi\pi}^2}\ , \quad
C_A = \frac{M_{a_1}^2 M_\rho^2}{g_{\rho\pi\pi}^2 Z_\pi^2} \ .
\end{equation}
Note that (A)VMD is not explicitly implemented, but this structure is a consequence
of both the mesons and external fields being employed as gauge bosons of the same
gauge group.

\begin{figure}[!t]
  \centering
	\includegraphics[width=.45\textwidth]{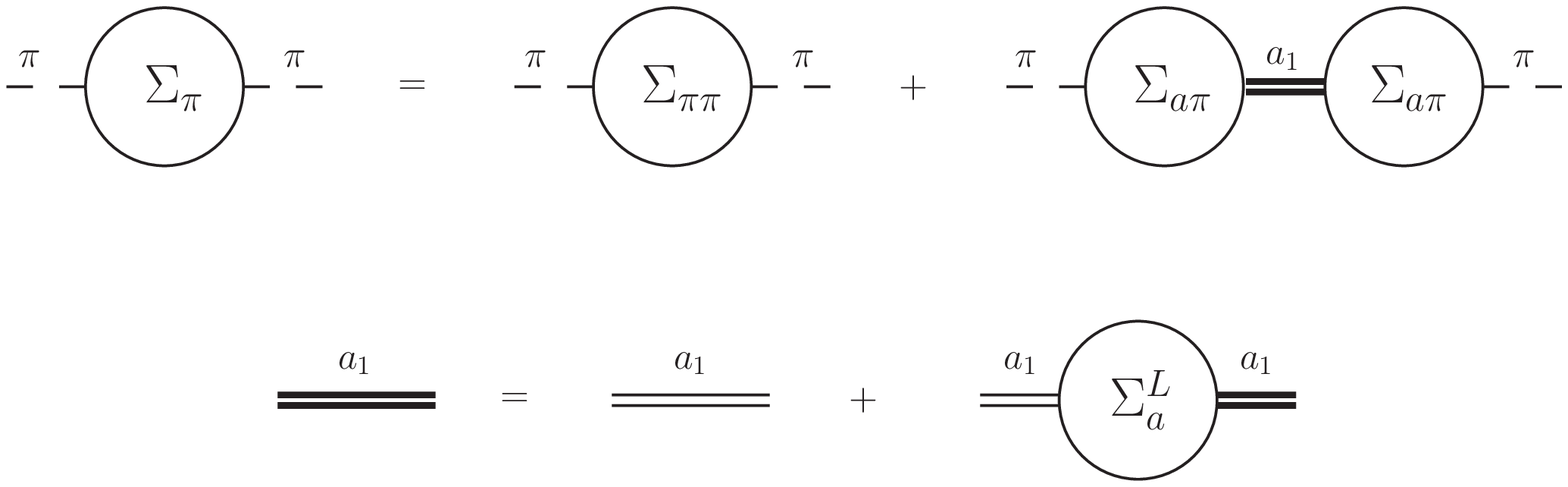}
\caption{Schematic depiction of the contributions
to the pion and $a_1$ self-energies, $\Sigma_\pi$, $\Sigma_{a_1}^L$.
The bold $a_1$ line represents
a resummed $a_1$ propagator in the longitudinal channel.}
\label{fig:pionprop}
\end{figure}
The 4D-longitudinal $AV$ polarization function can be constructed
in a manner similar to the transverse mode but with intermediate
$a_1$ and $\pi$ propagators (\cf~Fig.~\ref{fig:correlator}),
\begin{equation}
\begin{split}
\Pi_A^L (p^2) &= \Sigma_{W}^L (p^2) + \left(\Gamma_A^L (p^2)\right)^2
D_A^L (p^2) \\&+ \frac{p^2 F_\pi^2}{2} F^2 (p^2) D_\pi(p^2) \, .
\end{split}
\end{equation}
The propagators read
\begin{eqnarray}
D_A^L (p^2) &=& [M_{a_1}^2 - \Sigma_{a_1}^L(p^2)]^{-1} , \\
D_\pi (p^2) &=& [p^2 - M_\pi^2 - \Sigma_{\pi}(p^2)]^{-1} \, ,
\label{eq:piprop}
\end{eqnarray}
while the pion self-energy, $\Sigma_\pi$, contains a direct
contribution and one with an intermediate $a_1$ propagator,
\begin{equation}
\Sigma_{\pi}(p^2) \equiv
\Sigma_{\pi\pi}(p^2) + p^2 \Sigma_{a\pi}^2(p^2) D_A^L(p^2) \ ,
\end{equation}
see Fig.~\ref{fig:pionprop}.
The transition functions, $\Gamma_A^L$ and $F$, where the latter
connects the $W$ boson to the pion, are given by:
\begin{equation}
\begin{split}
\Gamma_A^L (p^2) &= \frac{g_{\rho\pi\pi} F_\pi^2 Z_\pi M_{a_1}}
{2 M_\rho} + \Sigma_{W a_1}^L(p^2) ,\\
F (p^2) &= 1-\frac{\sqrt{2}}{F_\pi} \Sigma_{W \pi}(p^2)\\
&+\frac{\sqrt{2}}{F_\pi} D_A^L(p^2) \Gamma_A^L(p^2)
\Sigma_{a \pi}(p^2).
\end{split}
\end{equation}

A spectral function can also be defined for the 4D-longitudinal
$AV$ polarization function:
$\rho_A^L = -{\rm Im}\Pi_A^L/\pi$.
As with the transverse mode, because of the
the gauge symmetry, the EW self-energies
can be expressed in terms of the meson self-energies,
\begin{equation}
\begin{split}
\Sigma_W^L (p^2) &= \frac{M_\rho^2 Z_\pi^2}{g_{\rho\pi\pi}^2
M_{a_1}^2} \Sigma_{a_1}^L (p^2), \\
\Sigma_{W a_1}^L (p^2) &= - \frac{M_\rho Z_\pi}{g_{\rho\pi\pi}
M_{a_1}} \Sigma_{a_1}^L (p^2), \\
\Sigma_{W \pi} (p^2) &= \frac{M_\rho Z_\pi}{g_{\rho\pi\pi} M_{a_1}}
\Sigma_{a \pi} (p^2) \  .
\end{split}
\end{equation}
This allows
the spectral function
to be written in a compact form in terms of the
longitudinal $a_1$ propagator and the $\pi$ propagator,
\begin{equation}
\label{eq:rhoaL}
\rho_A^L (p^2) = -\frac{C_A}{\pi} {\rm Im}D_A^L (p^2)
- \frac{p^2 F_\pi^2}{2\pi} {\rm Im} \left[D_\pi(p^2) F^2(p^2)\right] \ .
\end{equation}
The coupling constant $C_A$ is the same as above, and $F$
is expressed in condensed form as
\begin{equation} \label{eq:Feq}
F(p^2) = 1-\frac{\sqrt{2}}{F_\pi} C_A^{1/2} \Sigma_{a\pi}(p^2) D_A^L(p^2) \, .
\end{equation}
In all expressions above, $\Sigma_{a\pi}$ denotes the self-energy
associated with the transition between the $a_1$ and $\pi$ states.

\begin{figure}[!t]
  \centering
	\includegraphics[width=.45\textwidth]{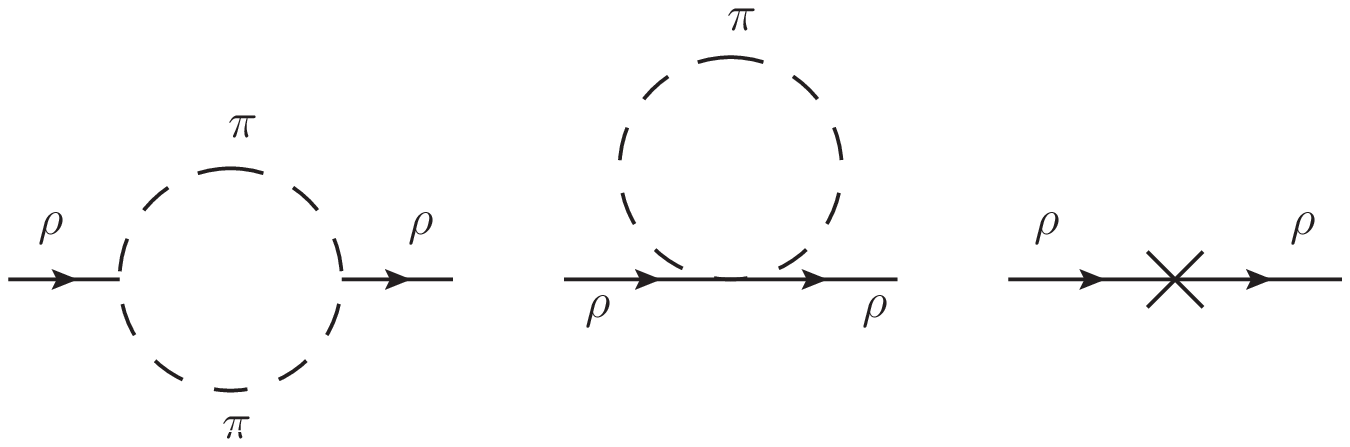}
	\includegraphics[width=.45\textwidth]{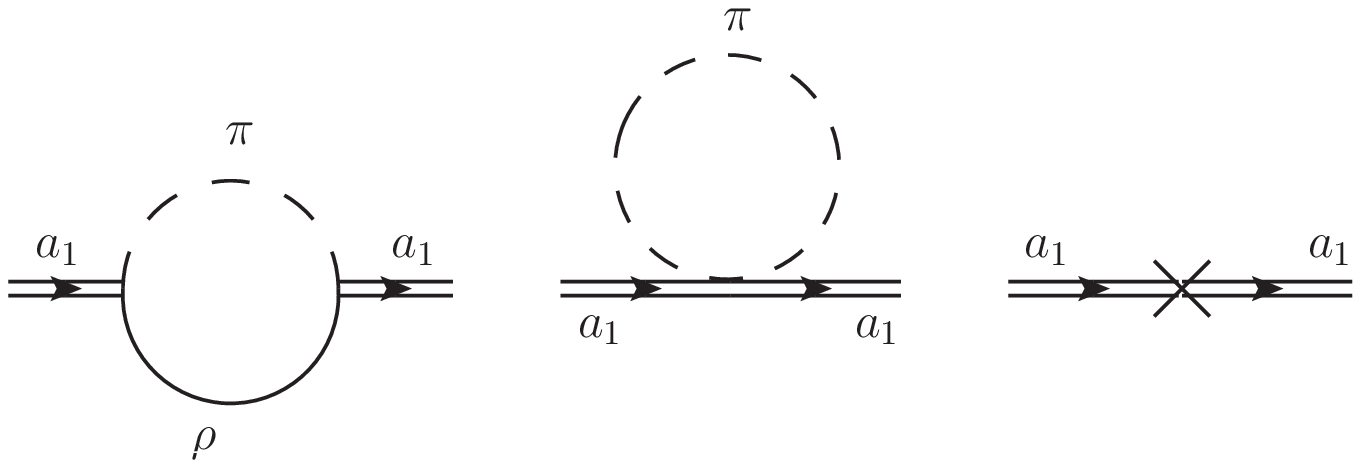}
\caption{Feynman diagrams for the $\rho$ (upper row) and $a_1$
(lower row) self-energies; crosses denote counter-terms.}
\label{fig:dia1}
\end{figure}

\begin{figure}[!t]
  \centering
  \includegraphics[width=.45\textwidth]{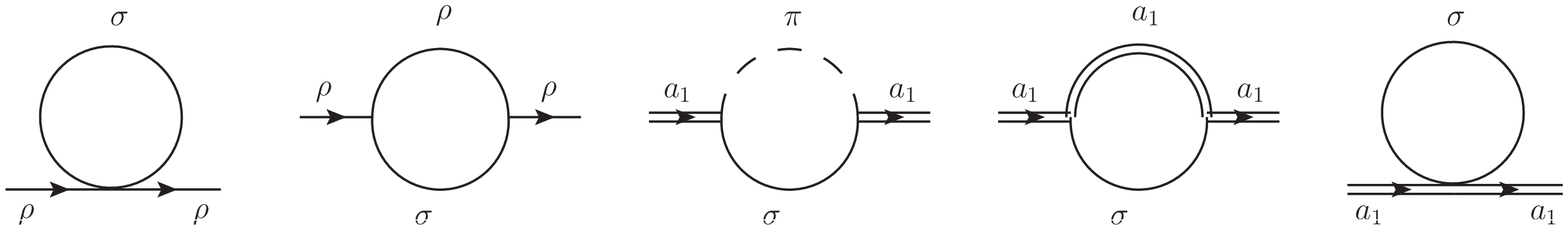}
 \caption{Additional diagrams contributing to the
 $V$ and $AV$ self-energies due to the inclusion of an
 explicit $\sigma$ field.}
\label{fig:axialLin}
\end{figure}

To calculate the $\rho$ and $a_1$ self-energies,
we perform a loop expansion in addition to the derivative expansion
associated with the Lagrangian, thereby  including the
diagrams depicted in Fig.~\ref{fig:dia1}.
In the linear realization, the explicit $\sigma$ field precipitates
the additional diagrams depicted in Fig.~\ref{fig:axialLin}
for the $V$ and $AV$ channels. In both realizations, the $\pi a_1$
loop contribution to the $\rho$ self-energy is not included,
as we have verified that it's contribution to the vacuum spectral
functions is negligible~\cite{Hohler:2013ena}.
This then requires a modification of the $\rho\rho\pi\pi$ vertex
coupling in the $\pi$-tadpole diagram in order to preserve gauge
invariance.
The self-energies, $\Sigma_{\pi\pi}$ and $\Sigma_{a\pi}$, are determined
from the same diagrams as used in the $AV$ channel with the appropriate
external legs.

The loop integrals are rendered finite by dimensional regularization with
the inclusion of suitable counter-terms. The latter are based on terms
already present in the Lagrangian, thus preserving the chiral symmetry.
Because of the momentum dependence of the divergences, higher derivative
Lagrangian terms are also needed for the counter-terms. For
the non-linear realization the counter-term Lagrangian needed
to regulate the $V$ and $AV$ self-energies is given
by:
\begin{equation}
\begin{split}
\mathcal{L}_{\rm CT} &= \frac{\tilde{f}_\pi^2}{8}
\left( \delta Z_\pi^{(2)} {\rm Tr}[D_\mu U^\dag D^{\mu} U] \right.\\
&\left.+ \frac{1}{4} \delta Z_\pi^{(4)} {\rm Tr}\left[\left(D_\mu D_\nu U^\dag
+D_\nu D_\mu U^\dag\right) \right.\right.\\
& \left. \quad \quad \quad \times \left(D_\mu D_\nu U + D_\nu D_\mu U\right)\right]\\
&+ \delta m_\pi^2{\rm Tr}[U+U^\dag-2]\Big)
 - \frac{1}{2} \delta Z_A^{(2)} {\rm Tr}[F_L^2 + F_R^2] \\
& + \delta Z_A^{(4)} {\rm Tr} \left[D^\mu F_{\mu \nu}^L
D_\lambda F_L^{\lambda \nu} + D^\mu F_{\mu \nu}^R
D_\lambda F_R^{\lambda \nu}\right]\\
& + \delta Z_A^{(6)} {\rm Tr} \left[D_\mu D^\lambda
F_{\lambda\nu}^L D^\mu D_\sigma F_L^{\sigma \nu} \right.\\
& \quad\quad\quad\quad +
D_\mu D^\lambda
F_{\lambda\nu}^L D^\nu D_\sigma F_L^{\sigma \mu}\\
& \quad\quad\quad\quad + D_\mu D^\lambda
F_{\lambda\nu}^R D^\mu D_\sigma F_R^{\sigma \nu} \\
& \left.\quad\quad\quad\quad +D_\mu D^\lambda
F_{\lambda\nu}^R D^\nu D_\sigma F_R^{\sigma \mu}\right]\\
& +\delta\gamma^{(2)} \ {\rm Tr}[F_L U F_R U^\dag]\\
&+ \delta\gamma^{(4)} {\rm Tr} \left[D^\mu F_{\mu\nu}^L
U D^\lambda F_{\lambda \nu}^R U^\dag\right]\\
& + \delta\gamma^{(6)} {\rm Tr} \left[ D^\mu D^\lambda
F_{\lambda\nu}^L U D_\mu D_\sigma F_R^{\sigma\nu} U^\dag\right.\\
& \left. \quad\quad\quad\quad + D^\mu D^\lambda F_{\lambda \nu}^L
U D^\nu D^\sigma F_{\sigma\mu}^R U^\dag \right] \ ,
\end{split}
\end{equation}
where
\begin{equation}
\begin{split}
D_\mu F_{\lambda \nu} &= \partial_\mu F_{\lambda\nu} - i g
\left[ A_\mu, F_{\lambda\nu}\right] ,\\
D_\mu D_\lambda F_{\nu\sigma} &= \partial_\mu \left(D_\lambda F_{\nu\sigma}
\right)
- i g \left[A_\mu, D_\lambda F_{\nu\sigma}\right] \ .
\end{split}
\end{equation}
This can be expanded in terms of the physical fields
as described for the MYM Lagrangian. The terms relevant for the
self-energies are detailed in Appendix \ref{sec:counter}.
In the linear realization, we must add the counter-terms akin to the
scalar potential
\begin{equation}
\mathcal{L}_{CT} = -\frac{\tilde{f}_\pi^2}{8}
\left(\delta \mu^2 {\rm Tr}\left[U^\dag U\right]
+\delta \lambda \frac{\tilde{f}_\pi^2}{4} {\rm Tr}
\left[\left(U^\dag U\right)^2\right]\right) ,
\end{equation}
whose parameters are chosen so that the vacuum loop
contribution from the lollipop diagrams in Fig.~\ref{fig:sigmaTadb}
is zero. This is equivalent to preventing
the vacuum masses to be renormalized by these diagrams.

Each counter-term coupling has an infinite contribution
needed to cancel off the divergences from the loops and a
finite contribution, {\it e.g.} $\delta Z_\pi^{(2)} = \delta
Z_\pi^{(2) \infty} + \bar{\delta Z}_\pi^{(2)}$ The finite contributions of
the terms $\delta m_\pi$, $\delta Z_\pi^{(2)}$, and $\delta Z_\pi^{(4)}$
are determined by requiring that $M_\pi$ and $F_\pi$ are renormalized
to their physical values. This is achieved by implementing
the conditions $\Sigma_{\pi\pi}(M_\pi^2) = 0$,
$\Sigma_{\pi\pi}'(M_\pi^2) = 0$, and
$\Sigma_{a_1 \pi}(M_\pi^2) = 0$.
The remaining six counter-term couplings are used to fit the vacuum
spectral functions to experiment.

As discussed in Ref.~\cite{Hohler:2013ena}, we find that an agreement
with the data can be achieved when the finite width of the $\rho$
is included in the calculation of the $a_1$ self-energy. This is
done by using the fully resummed $\rho$ propagator in the $a_1$
self-energy's $\pi\rho$ loop.  However, this alone compromises
the chiral properties at loop level which can be checked by employing
a generalized form of PCAC~\cite{Urban:2001ru,Weinberg:1996kr},
\begin{equation}
\label{eq:fpiT}
\int \frac{\rho_A^L}{s} ds = \frac{F_\pi^2}{2} \ .
\end{equation}
We have verified that this equation is exactly satisfied
for a zero-width $\rho$ in the chiral limit ($M_\pi=0$),
facilitated by the following relations between the self-energies
$\Sigma_{a_1}^L$, $\Sigma_{a\pi}$, and $\Sigma_{\pi\pi}$,
\begin{equation}
\label{eq:pcacrel}
\Sigma_{a_1}^L = \frac{F_\pi^2}{2}\frac{M_{a_1}^4}{C_A}
\frac{\Sigma_{\pi\pi}}{p^2}\,, \quad\quad
\Sigma_{a\pi} = \frac{F_\pi}{\sqrt{2}}\frac{M_{a_1}^2}{C_A^{1/2}}
\frac{\Sigma_{\pi\pi}}{p^2}\, .
\end{equation}
For finite pion mass, deviations to these relations
at order  ${\cal O}(M_\pi^2)$ develop due to pion tadpole
contributions.

\begin{figure}[!tb]
\centering
\includegraphics[width=.45\textwidth]{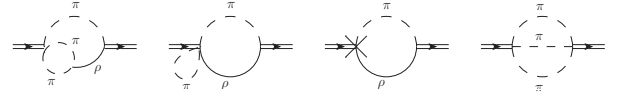}
\caption{Vertex correction diagrams needed to recover PCAC
in the presence of a pion-dressed $\rho$ propagator.
The external legs represent $a_1$ or $\pi$ to generate the
self-energies $\Sigma_{ij}$.
The first three diagrams show the corrections to the left-hand vertex
while the complete set includes the corrections to the right-hand
vertex and to both simultaneously.}
\label{fig:vertexdia}
\end{figure}
Using a broad $\rho$ in the loops of the $AV$ channel creates
violations of Eq.~(\ref{eq:pcacrel}) even in the chiral limit,
upon integrating over the invariant-mass distribution of the $\rho$.
It turns out that PCAC can be restored by including the vertex
correction diagrams to the axial-vector self-energy shown in
Fig.~\ref{fig:vertexdia}. These are constructed such that
a sub-diagram which is part of the vertex correction has the
same structure as the diagrams in the $\rho$ self-energy.
Thus note that in the far right diagram, the 2 pions are in a
relative $P$-wave to mimic a $\rho$ meson.
In order to recover the necessary relations between
the self-energies, the couplings between the external
$a_1$/$\pi$ and the $3\pi$ (first and last diagram in
Fig.~\ref{fig:vertexdia}) and $\rho3\pi$ states 
(second diagram in Fig.~\ref{fig:vertexdia}),
as well as the counter-terms 
(third diagram in Fig.~\ref{fig:vertexdia}), 
must be judiciously
chosen, different from their Lagrangian values.
This is expected since the full $\rho$ propagator
is only a {\it partial} resummation.
The deviation of the couplings from their Lagrangian
values appropriately captures this resummation.
More details on how this is implemented can be found in
Appendix~\ref{sec:VCdisc}.

With the use of the resummed $\rho$ propagator and accompanying vertex
corrections, a new divergence is introduced into the $a_1$ self-energy which
needs to be regulated. The problem arises from the fact that the
sub-loops which are similar to the $\rho$ self-energy grow as $M^6$
for large invariant $\rho$ masses due to higher derivatives in the
couplings. This growth, especially in the vertex corrections, is not
compensated by the propagators resulting in divergent diagrams. To regulate
these $a_1$ self-energy contributions in the spirit of effective field theory,
we introduce a hard cut-off on the invariant $\rho$ mass when integrating
over the $\rho$ spectral function by means of a dispersion relation, as
detailed in Appendix~C. The cut-off characterizes the scale
where the theory breaks down, while the resulting vertex corrections still
satisfy the relations of Eq.~(\ref{eq:pcacrel}), thus preserving the chiral
symmetry. We set the cut-off to $\Lambda_{\rm cut} =1.5$\,GeV, based on
the mass range where the $AV$ width starts to grow rapidly (larger values
still allow fits of similar quality to the $\tau$-decay data, but become
increasingly sensitive to the higher derivative terms).

Before turning to the fit of the vacuum $V$ and $AV$ spectral functions
using the $\rho$ and $a_1$ propagators from MYM, two further ingredients
beyond MYM are needed for a description of the data toward higher energies.
The first is a contribution from the high-energy continuum, which saturates
at a level corresponding to the perturbative quark-antiquark limit.
Following our previous work~\cite{Hohler:2012xd,Hohler:2013ena}, we
adopt a chirally invariant continuum, \ie, identical for the two channels,
with a functional form~\cite{Shuryak:1993kg}
\begin{equation}
\rho_{\rm cont} (s) = \frac{1}{8\pi^2}\left(1+\frac{\alpha_s}{\pi}\right)
\frac{s}{1+\exp[(E_0-\sqrt{s})/\delta]} \ ,
\end{equation}
which gradually ramps up to the  expected high-energy perturbative limit.
The second contribution is from the excited states, $\rho'$ and $a_1'$
which we include through a phenomenological Breit-Wigner resonance ansatz.
These states chiefly represent 4- and 5-pion states coupling through resonant
states, not captured by the MYM framework nor by the continuum. This is quite
evident from the data in the $V$ channel, but more subtle in the $AV$
channel. Thus, the $\rho'$ can be straightforwardly fitted to the data, while
$a_1'$ can be inferred from a quantitative analysis of Weinberg sum rules (WSRs),
in the spirit of Ref.~\cite{Hohler:2012xd}. Clearly, the continuum and the
resonant states cannot be uniquely separated.

\begin{figure}[!tb]
  \centering
        \includegraphics[width=.45\textwidth]{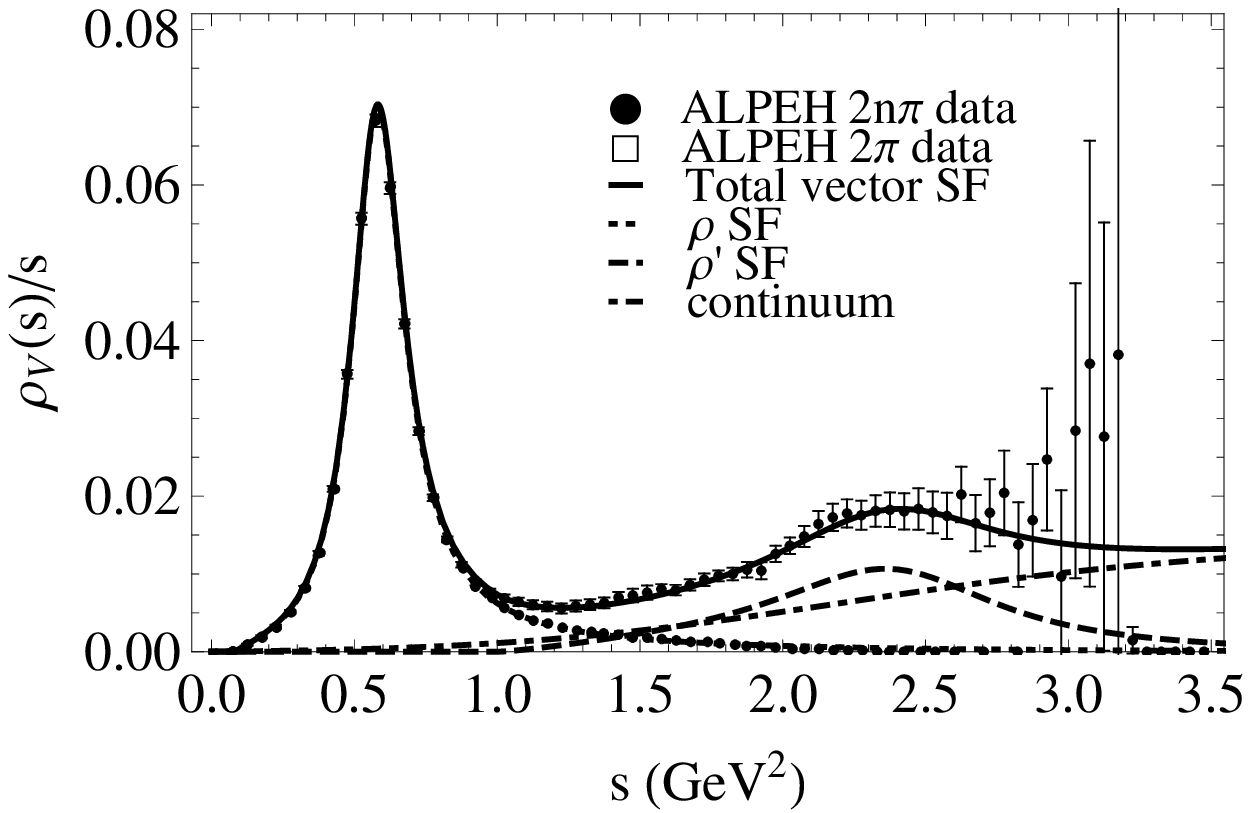}

\vspace{0.3cm}

        \includegraphics[width=.45\textwidth]{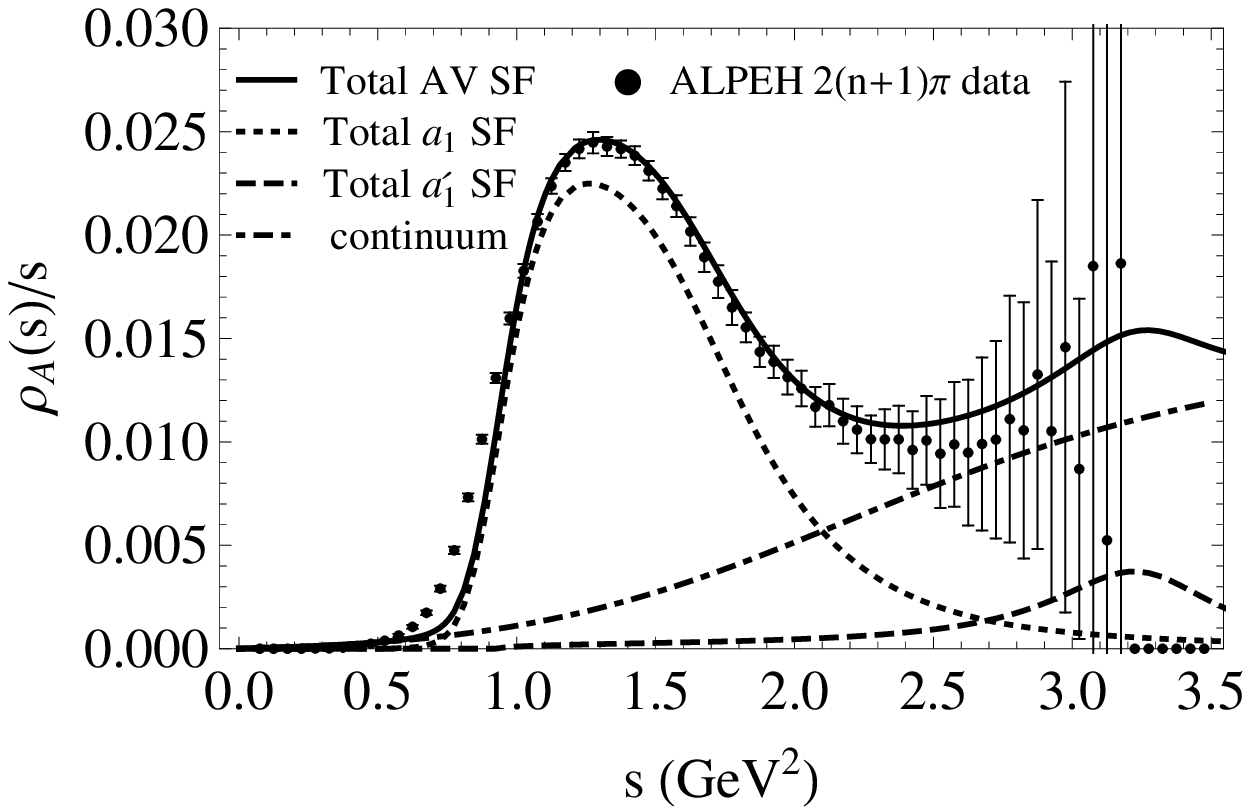}
\caption{Spectral functions in the vector (upper panel) and axial-vector
(lower panel) channels as calculated from our MYM approach in non-linear
realization (dashed lines), supplemented with a chirally invariant continuum
(dotted lines) and first excited states (dash-dotted lines)~\cite{Hohler:2012xd},
compared to hadronic $\tau$-decay data~\cite{Barate:1998uf}.}
  \label{fig:sf}
\end{figure}
\begin{figure}[t!]
  \centering
    \includegraphics[width=.45\textwidth]{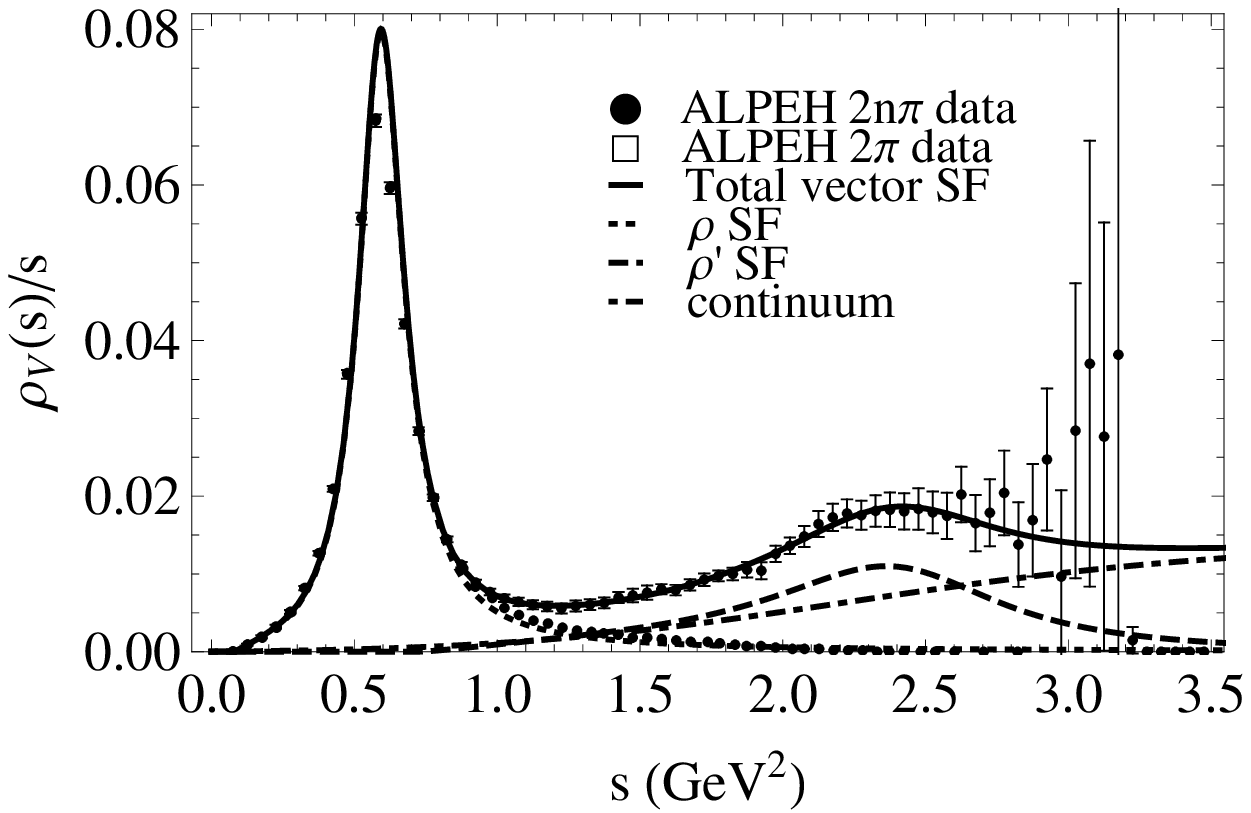}
	\includegraphics[width=.45\textwidth]{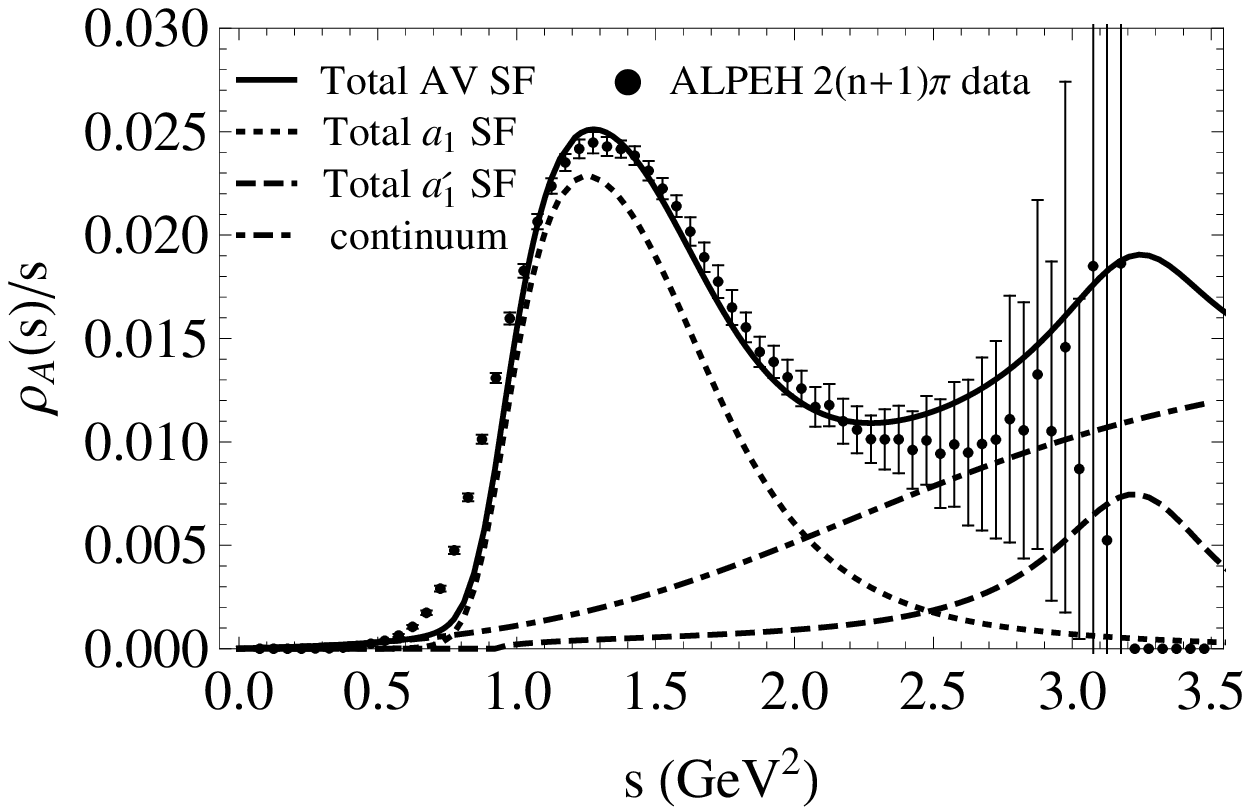}
\caption{Spectral functions in the vector (upper panel) and axial-vector
(lower panel) channels as calculated from our MYM approach in linear
realization (dashed lines), supplemented with a chirally invariant continuum
(dotted lines) and first excited states (dash-dotted lines)~\cite{Hohler:2012xd},
compared to hadronic $\tau$-decay data~\cite{Barate:1998uf}.}
\label{fig:sflin}
\end{figure}

Our final fit to the experimental spectral distributions as measured via hadronic
$\tau$ decays  by the ALEPH collaboration~\cite{Barate:1998uf} is displayed in
Figs.~\ref{fig:sf} and \ref{fig:sflin} for the non-linear and linear
realization, respectively.
The values of the Lagrangian parameters used in the spectral function fits
are listed  in Table~\ref{tab:para} while the values for the
finite contribution to the counter-terms can be found in Table~\ref{tab:counter}.
We note that $M_\rho^2 g_{\rho\pi\pi}^{(3)}$
is small compared to $g_{\rho\pi\pi}$ as required for a meaningful derivative
expansion of the effective theory.
The same can be said for the counter-terms when noting that
the relevant energy scale of $\delta Z_\pi^{(4)}$ is $M_\pi$
rather than $M_\rho$ as with the other couplings.
For the linear realization, the mass of the $\sigma$ field
is needed, which was determined to be $M_\sigma = 0.80$ GeV.
In both realizations the cut-off scale is set
to $\Lambda_{\rm cut} = 1.5$\,GeV and
the continuum parameters ($\alpha_s$=0.5, $E_0$=1.55\,GeV and $\delta$=0.22\,GeV)
have been slightly varied compared to Ref.~\cite{Hohler:2012xd} to optimize
the fit.
\begin{table*}[t!]
\begin{tabular}{c|c|c|c|c|c|c|c|c|c}
& $g_{\rho\pi\pi}$ & $g_{\rho\pi\pi}^{(3)}$ &$M_\rho$&$M_{a_1}$&
$M_\sigma$ & $\Lambda_{\rm cut}$ & $\alpha_s$&$E_0$&$\delta$\\
\hline
Non-Linear& 6.01 & 0.02 & 0.86 & 1.20 & -- & 1.5 &
0.5 & 1.55 & 0.22 \\
Linear & 6.36 & -0.57  & 0.92  & 1.26  & 0.8  &
1.5  & 0.5 & 1.55  & 0.22 \\
\end{tabular}
\caption{Parameters used to obtain the vacuum fits to the experimental spectral
functions in the vector and axial-vector channels in
Figs.~\ref{fig:sf} and \ref{fig:sflin}. The masses, cutoff and continuum parameters
($E_0,\delta$) are in units of GeV, $g_{\rho\pi\pi}^{(3)}$ in GeV$^{-2}$.  }
\label{tab:para}
\end{table*}
\begin{table*}[t!]
\begin{tabular}{c|c|c|c|c|c|c|c|c|c}
& $\delta Z_A^{(2)}$ & $\delta Z_A^{(4)}$ &$\delta Z_A^{(6)}$
& $\delta \gamma^{(2)}$ & $\delta \gamma^{(4)}$ & $\delta \gamma^{(6)}$
&$\delta Z_\pi^{(2)}$ & $\delta Z_\pi^{(4)}$ & $\delta m_\pi$\\
\hline
Non-Linear& 1.91 & 0.54 & -0.06 & 1.76 & -0.31 & -0.08 &
1.94 & 14.70 & 0.06 \\
Linear & 1.14 & 0.72  & -0.10  & 0.92  & -1.13  &
0.19  & 1.14 & 10.06  & 0.05 \\
\end{tabular}
\caption{Counter-term parameters used to obtain the vacuum fits to the experimental spectral
functions in the vector and axial-vector channels in
Figs.~\ref{fig:sf} and \ref{fig:sflin}. $\delta Z_A^{(4)}$, $\delta \gamma^{(4)}$,
and $\delta Z_\pi^{(4)}$ are in units of GeV$^{-2}$; $\delta Z_A^{(6)}$ and
$\delta \gamma^{(6)}$ are in units of GeV$^{-4}$, and $\delta m_\pi$ is in units
of GeV.}
\label{tab:counter}
\end{table*}

\begin{figure}[t!]
  \centering
    \includegraphics[width=.45\textwidth]{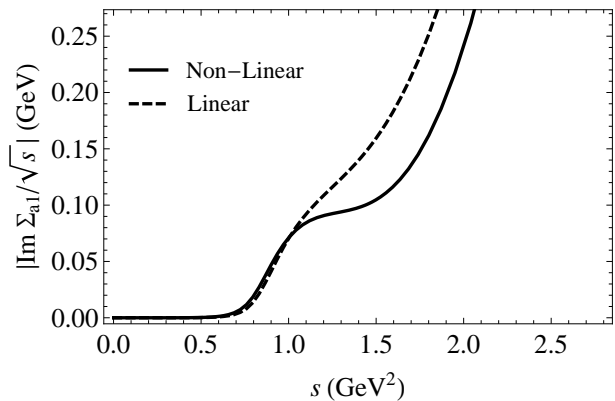}
\caption{Imaginary part of the axial-vector self-energy for
the non-linear (solid curve) and linear (dot-dashed curve)
realization of MYM.}
\label{fig:ImSigmaA}
\end{figure}

\begin{figure}[t!]
  \centering
	\includegraphics[width=.45\textwidth]{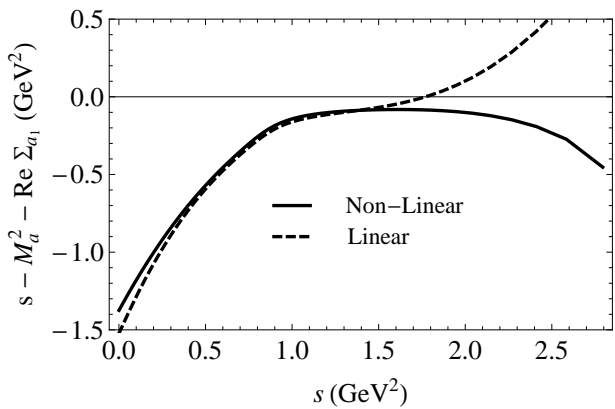}
\caption{Real part of the inverse axial-vector propagator in vacuum
for the non-linear (solid line) and linear (dot-dashed linear)
realizations of chiral symmetry within MYM.}
\label{fig:ReSigmaA}
\end{figure}

Our fit within the loop-augmented MYM to the vacuum $AV$ spectral function
presented here implies several specific features. One of them is illustrated
in Fig.~\ref{fig:ImSigmaA} where the ``width" of the $a_1$,
$\Gamma_a=-{\rm Im} \Sigma_a(s)/\sqrt{s}$, is plotted as a function of
energy. In the vicinity of the resonance peak, $s\simeq1-1.5$\,GeV$^2$,
it amounts to about 100-150\,MeV, quite a bit smaller than the ``apparent"
width of the peak of $\sim$300-400\,MeV.
In order to generate sufficient spectral strength across the peak
region, the real part of the $a_1$ self-energy needs to be tuned such
that the real part of the inverse of the $a_1$ propagator,
$s - M_{a_1}^2 - {\rm Re} \Sigma_{a_1}$, stays in the vicinity of
zero over an extended range in energy, \cf~Fig.~\ref{fig:ReSigmaA}.
This results in the spectral function being effectively given by
$1/{\rm Im}\Sigma_{a_1}^T$, generating the necessary spectral strength
through a sufficiently small ${\rm Im}\Sigma_{a_1}$ across the peak region.
The remaining $\sim$10\% peak strength is provided by the low-energy tails
of the continuum and/or $a_1'$ resonance, implying them to have non-vanishing
3-pion components (experimentally the peak region is known to be saturated
by 3-pion states).
As we already indicated in our previous work~\cite{Hohler:2013ena},
this construction is dictated by the lack of strength in the
$AV$ coupling, $C_A$, which is fixed by the gauge principle, \ie,
it cannot be adjusted independently from the $V$ coupling. Within
the basic MYM setup, such a construction appears to be unavoidable;
it is present in both non-linear and linear realiziations of chiral
symmetry, although somewhat less pronounced in the
latter \footnote{Interestingly,
very recent observations of a new $a_1(1420)$ state by the COMPASS
collaboration~\cite{Adolph:2015pws} may alleviate this issue.
This new state is not included in our current model, but it is
expected to provide additional strength in the $3\pi$ spectral
function in the region beyond the nominal $a_1$ mass. Thereby,
the spectral shape of the $a_1(1260)$ would not have to be as broad,
thus limiting the extent of a tuning. Further studies of the quantitative
role of the new $a_1(1420)$ in the vacuum $AV$ spectral function are
needed but beyond the scope of the present work.}.
Earlier works which considered a global gauge symmetry do not
have this feature as their couplings controlling the spectral
height decouple in the $V$ and $AV$ channels.
We will come back to this issue below, \ie, in how far it affects the
robustness of the predictions for the finite-temperature spectral functions.
As mentioned above, the strong growth of the width beyond $s=2$\,GeV$^2$,
which is also common to both realizations, indicates the break-down
of the effective theory and is the basis for our choice of the sharp
cut-off when integrating over pertinent spectral functions.

To further test the vacuum model, we calculate additional
observables, including the radiative $a_1$ decay width, the $D/S$
ratio of the $a_1\to\rho\pi$ decay, the iso-vector $\pi\pi$ phase
shift, and the pion electromagnetic (EM) formfactor.
The radiative $a_1$ decay
is found to be $\Gamma_{a_1 \rightarrow \pi\gamma}$=244\,keV
for the non-linear realization and 42\,keV for the linear case;
especially the latter is quite a bit below the available datum extracted
from proton-nucleus collisions, 640$\pm$246\,keV~\cite{Zielinski:1984au}.
These are tree-level results, as the hadronic loops in the current
calculation, which potentially contribute to the radiative decay,
vanish at the photon point. Additional
hadronic loop corrections, which are not included in our scheme, might improve
the situation, in particular pion exchange diagrams as discussed in
Ref.~\cite{Roca:2006am}.
For the $D/S$ ratio of the $a_1\rho\pi$ decays, loop corrections figure through
the broad $\rho$ and associated vertex correction.
When evaluated at $\sqrt{s} = 1.23$\,GeV, we find -0.101 for the non-linear
realization \footnote{The value for the non-linear realization is
slightly different from our previous study~\cite{Hohler:2013ena}
because there the $\pi a_1$ loop in the $\rho$ self-energy was included
and here it is not.} and -0.052 for the linear realization. The Particle
Data Group reports an average value of $0.062\pm0.02$ (three out of the
four listed measurements are around -0.12, while the fourth one is
at $-0.043\pm0.01$)~\cite{Agashe:2014kda}.

\begin{figure}[t!]
  \centering
	\includegraphics[width=.45\textwidth]{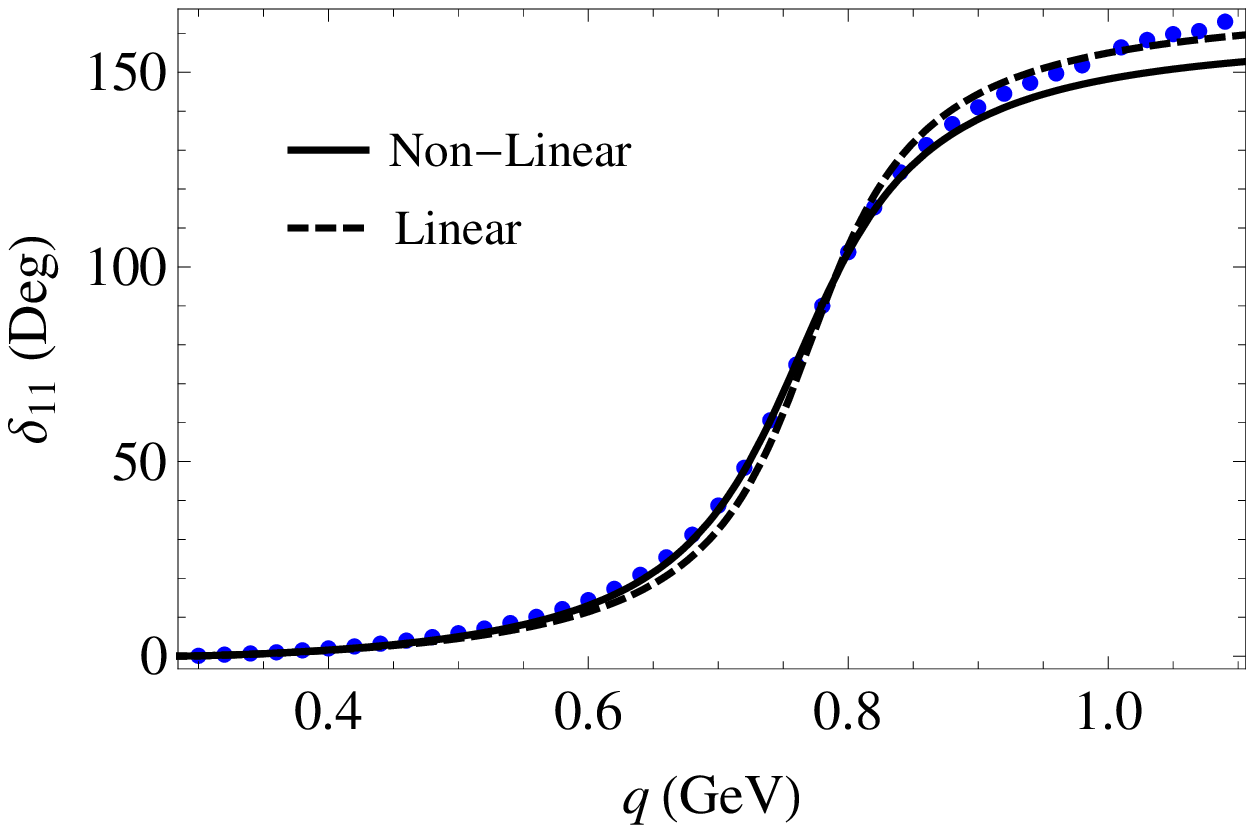}
    \includegraphics[width=.45\textwidth]{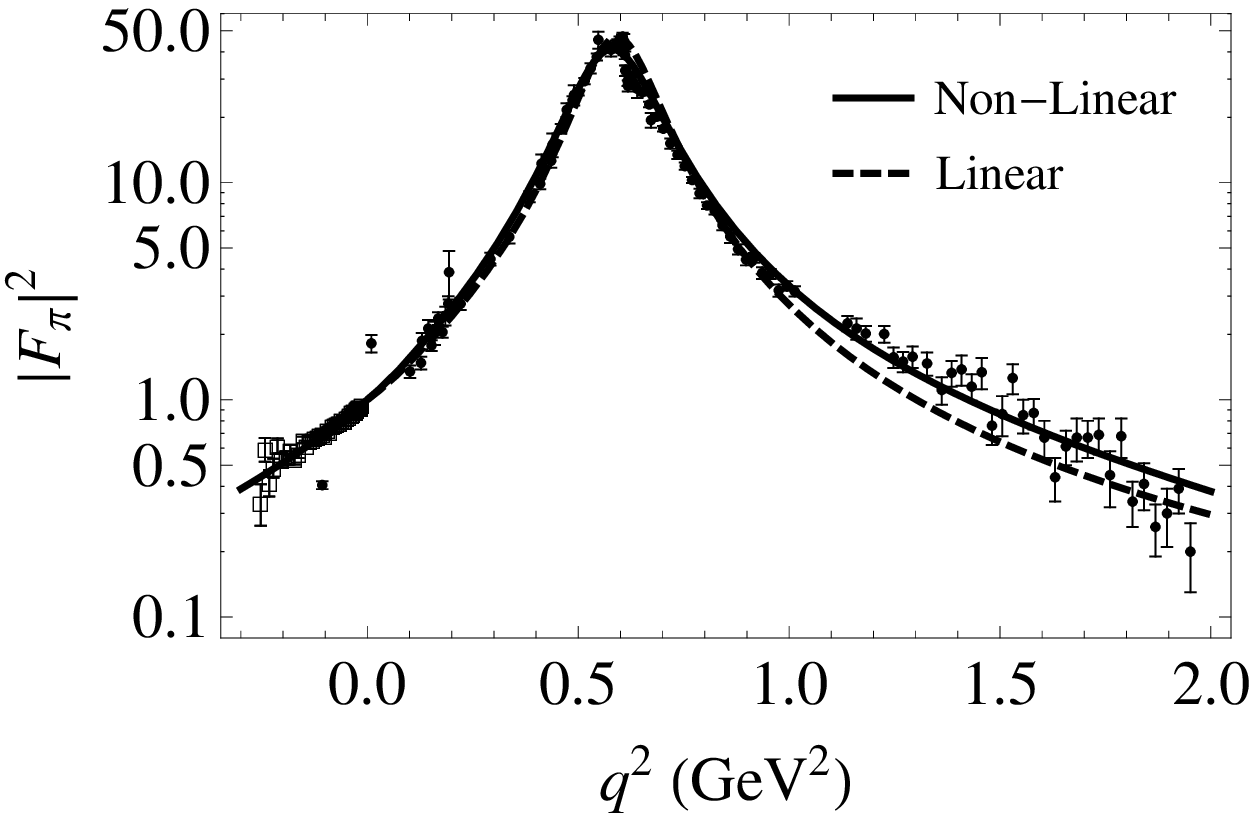}
\caption{Upper panel: $\pi\pi$ scattering phase shift in the iso-vector channel
compared with experimental measurements (points)~\cite{Froggatt:1977hu}.
Lower panel: Pion EM formfactor compared to experimental measurements
(points)~\cite{Amendolia:1983di,Amendolia:1984nz,Barkov:1985ac}. In each panel,
results of the non-linear (solid curve) and linear (dashed curve) realization
are shown.}
\label{fig:pheno}
\end{figure}

In extension of our previous work~\cite{Hohler:2013ena}, we here
calculate the $\pi\pi$ phase shift and the pion EM formfactor,
$|F_\pi|^2$, from our vector correlator and compare them to
data~\cite{Froggatt:1977hu,Amendolia:1983di,Amendolia:1984nz,Barkov:1985ac},
in Fig.~\ref{fig:pheno}.
Both quantities agree reasonably well with experiment in both
chiral realizations.
In particular, the formfactor below the 2-pion threshold, extending
into the space-like region, is not directly constrained by the vacuum
spectral functions. The overall agreement with these data suggests
that the vacuum model has reasonably well incorporated the relevant
physics.

\section{In-medium spectral functions}
\label{sec:rhoFT}
At finite temperature, the breaking of Lorentz symmetry splits
the 4D-transverse polarization functions at finite 3-momentum
into 3D-transverse, $\Pi^\perp$, and longitudinal, $\Pi^\parallel$,
components. The polarization functions in $V$ and $AV$ channels can
thus be decomposed as
\begin{equation}
\Pi^{\mu\nu}_{V,A}\left(p_0,|\vec{p}|\right) = \Pi^{\perp}_{V,A} P_T^{\mu\nu} +
\Pi^{\parallel}_{V,A} P_L^{\mu\nu} + \Pi^{L}_{V,A} \frac{p^\mu p^\nu}{p^2} \ ,
\end{equation}
with the standard 3D-transverse ($P_T^{\mu\nu}$) and -longitudinal
($P_L^{\mu\nu}$) projection operators
\begin{eqnarray}
P_T^{\mu\nu} &=& \delta^{ij} - \frac{p^i p^j}{|\vec{p}|} \, , \\
P_L^{\mu\nu} &=& -g^{\mu\nu}+\frac{p^\mu p^\nu}{p^2} - P_T^{\mu\nu} \, .
\end{eqnarray}
The spectral functions follow from the usual definitions,
$\rho_{V,A}^{\perp,\parallel} = -{\rm Im}\Pi_{V,A}^{\perp,\parallel}/\pi$.
As in vacuum, the gauge symmetry renders the spectral functions of
a form suggestive of (A)VMD, thus Eq.~(\ref{eq:vmdcor}) holds
for both 3D polarizations.

\subsection{Finite-$T$ Vector Spectral Function}

\begin{figure}[t!]
  \centering
  \subfigure[]{\label{fig:pia}
	\includegraphics[width=.13\textwidth]{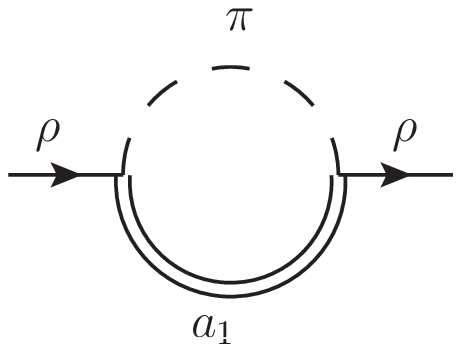}}
  \subfigure[]{\label{fig:sigmaTada}
	\includegraphics[width=.08\textwidth]{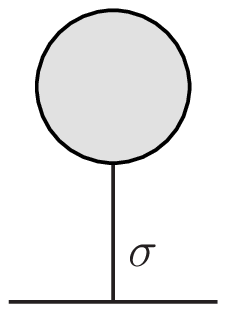}}
  \subfigure[]{\label{fig:sigmaTadb}
    \includegraphics[width=.25\textwidth]{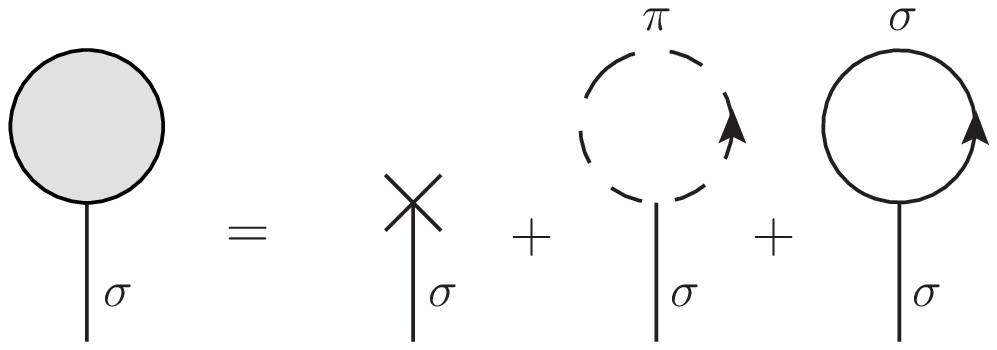}}
\caption{Diagrams added for in-medium self-energy calculations.
(a) $\pi a_1$ loop which contributes to the
finite-temperature $\rho$ self-energy.
(b) $\sigma$ lollipop diagrams which contribute to
the finite-temperature hadronic self-energies.
(c) Detail of the contributions of the shaded loop in (b).
The cross in the first diagram refers to the tree level contributions.}
\label{fig:sigmaTad}
\end{figure}

\begin{figure}[t!]
  \centering
  \subfigure[Unitarity cut (decay)]{
	\includegraphics[width=.15\textwidth]{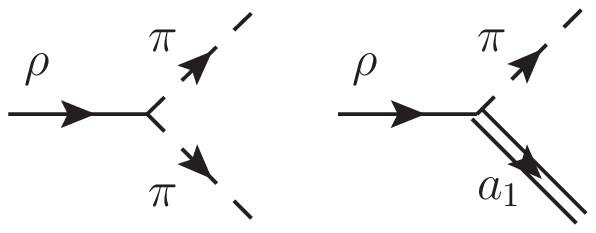}}
  \subfigure[Landau cut (scattering)]{\label{fig:rhocutsB}
    \includegraphics[width=.3\textwidth]{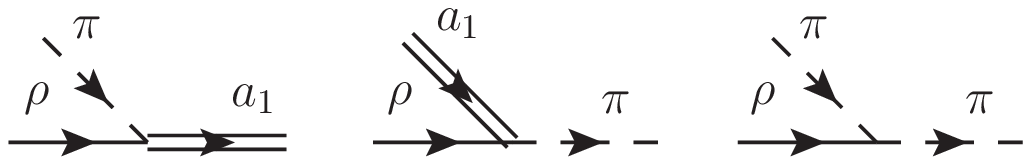}}
\caption{Diagrams of the processes contributing
to broadening the $\rho$ spectral function in medium. The last two diagrams
for the Landau contribution are only present when the $\rho$ is space-like.}
\label{fig:rhocuts}
\end{figure}

To calculate the in-medium spectral function, the $\rho$ self-energies
at finite temperature need to be determined. The same loop diagrams that
were considered in vacuum (\cf~upper panel in Fig.~\ref{fig:dia1}) are
evaluated using standard thermal field theory techniques. In addition,
the $\pi a_1$ loop needs to be accounted for to maintain the chiral
properties at finite $T$, \cf~Fig.~\ref{fig:pia}.
The explicit forms of the loop integrals of each contribution (after
the Matsubara sums are evaluated) are collected in Appendix~\ref{sec:loop}.
The pion tadpole diagram contributes only to the real part while the
other loop diagrams include the effects from both the unitarity and Landau
cuts, representing the $\rho$ decaying into thermal hadrons and $\rho$
scattering off a thermal $\pi$ or $a_1$, respectively, as depicted in
Fig.~\ref{fig:rhocuts}.
The linear realization additionally requires the inclusion of the
lollipop diagrams of Fig.~\ref{fig:sigmaTad} where the shaded loop
indicates the sum of a contact interaction and in-medium $\pi$ and
$\sigma$ loops. These sets of diagrams do not contribute in vacuum as
they are set to zero to properly regularize the pion mass to its physical
value.  The thermal and vacuum widths of the pion and $a_1$ states
within the loops are not taken into account.

\begin{figure}[!t]
  \centering
        \includegraphics[width=.45\textwidth]{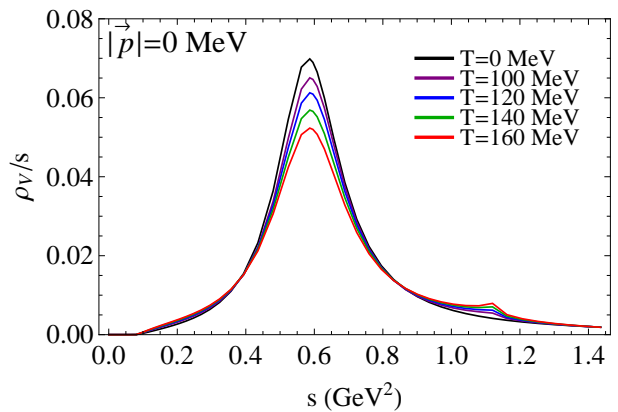}
\caption{Vector spectral function in the non-linear realization
at zero momentum for several temperatures.}
\label{fig:FTvecSFpv0}
\end{figure}

The resulting vector spectral functions for the non-linear realization are
presented in Fig.~\ref{fig:FTvecSFpv0} at different temperatures with
total momentum $|\vec{p}| = 0$ ($\rho^\perp$ and $\rho^\parallel$ are
degenerate under this condition).
We have analytically verified that there is no shift in $M_\rho^2$ at
order $T^2$ for low $T$. This is confirmed in the full numerical calculation
by the lack of an appreciable shift in the spectral peak location. Therefore,
as temperature increases, the $\rho$ spectral function essentially broadens,
although not by much. The broadening is caused by the Bose-enhanced $\rho$
decay into $\pi\pi$ (unitarity cut of $\pi\pi$ loop) and by its
scattering off a thermal $\pi$ into an $a_1$ (Landau cut of the
$\pi a_1$ loop). The width increase of up to $\sim$45\,MeV at $T=160$\,MeV
at the spectral peak is somewhat smaller than in similar calculations
in Refs.~\cite{Rapp:1999qu,Urban:2001uv}, due to a zero-width $a_1$
resonance and a larger $a_1$ mass, respectively.
The small peak at $M=M_{a_1}-m_\pi=1.06$\,GeV is a threshold effect due
the sharp $a_1$ and pion masses: beyond this mass thermal pions cannot
excite an $a_1$ anymore. The inclusion of the hadrons' widths in future
calculations will smoothen this feature out.

\begin{figure}[!t]
  \centering
	\includegraphics[width=.45\textwidth]{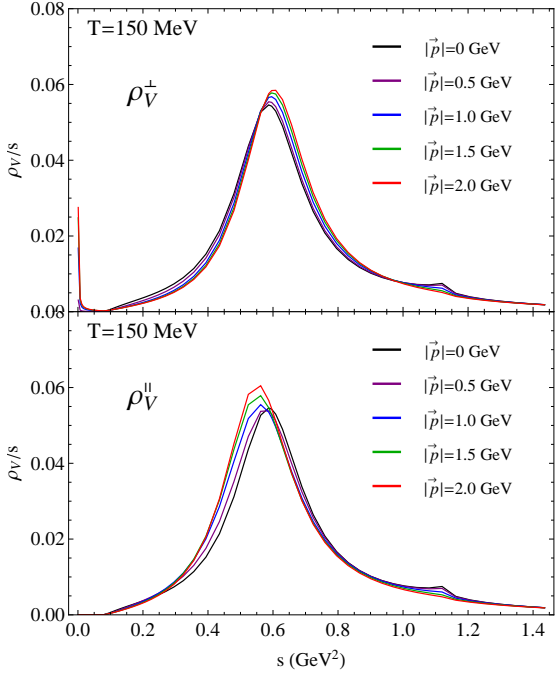}
\caption{Vector spectral function in non-linear realization
at $T$=150 MeV for various 3-momenta. Upper panel:
3D-transverse channel. Lower panel: 3D-longitudinal channel.}
\label{fig:FTvecSFpvf}
\end{figure}

Spectral functions at different momenta and a fixed temperature of
$T=150$\,MeV are shown in Fig.~\ref{fig:FTvecSFpvf}. The degeneracy
between the 3D-transverse and longitudinal components for $|\vec{p}|=0$
is now lifted; the resonance peak in the 3D-transverse channel shifts to
slightly higher energies while it moves downward in the 3D-longitudinal
one. The amount of breaking between the channels increases with momentum.
The divergence in $\rho^\perp/s$ at small invariant masses arises since
$\Sigma_\rho^\perp$ is finite at the photon point. On the other hand,
the longitudinal spectral functions vanish at the photon point.
The results for the linear realization are very similar and thus are not
presented here.

\subsection{Finite-$T$ Axial-Vector Spectral Function}
\label{sec:a1FT}
In the present work, we focus on the case of $|\vec{p}|=0$, therefore
$\rho_A^\perp$ and $\rho_A^\parallel$ are degenerate and equal to
$\rho_A^T$. The in-medium $a_1$ self-energy is calculated using the
same diagrams as in the vacuum, but keeping the vacuum dressing of the
$\rho$ propagator. The vertex corrections are also evaluated in their
vacuum form to be consistent with the level of approximation
that intermediate particles develop no thermal widths in the current
calculations. The explicit forms of the loop integrals for each contribution
are collected in Appendix~\ref{sec:loop}.
Both the unitarity and Landau cut processes, associated with each loop,
${\it e.g.}$, $a_1 \rightarrow \rho\pi$ decay and $a_1 \pi \rightarrow \rho$
scattering, respectively, are accounted for.
As in the vector channel, the medium contribution from the
lollipop diagram of Fig.~\ref{fig:sigmaTada} with appropriate external
legs is additionally needed for the finite-temperature $a_1$ self-energy
in the linear realization.

\begin{figure}[!t]
  \centering
	\includegraphics[width=.45\textwidth]{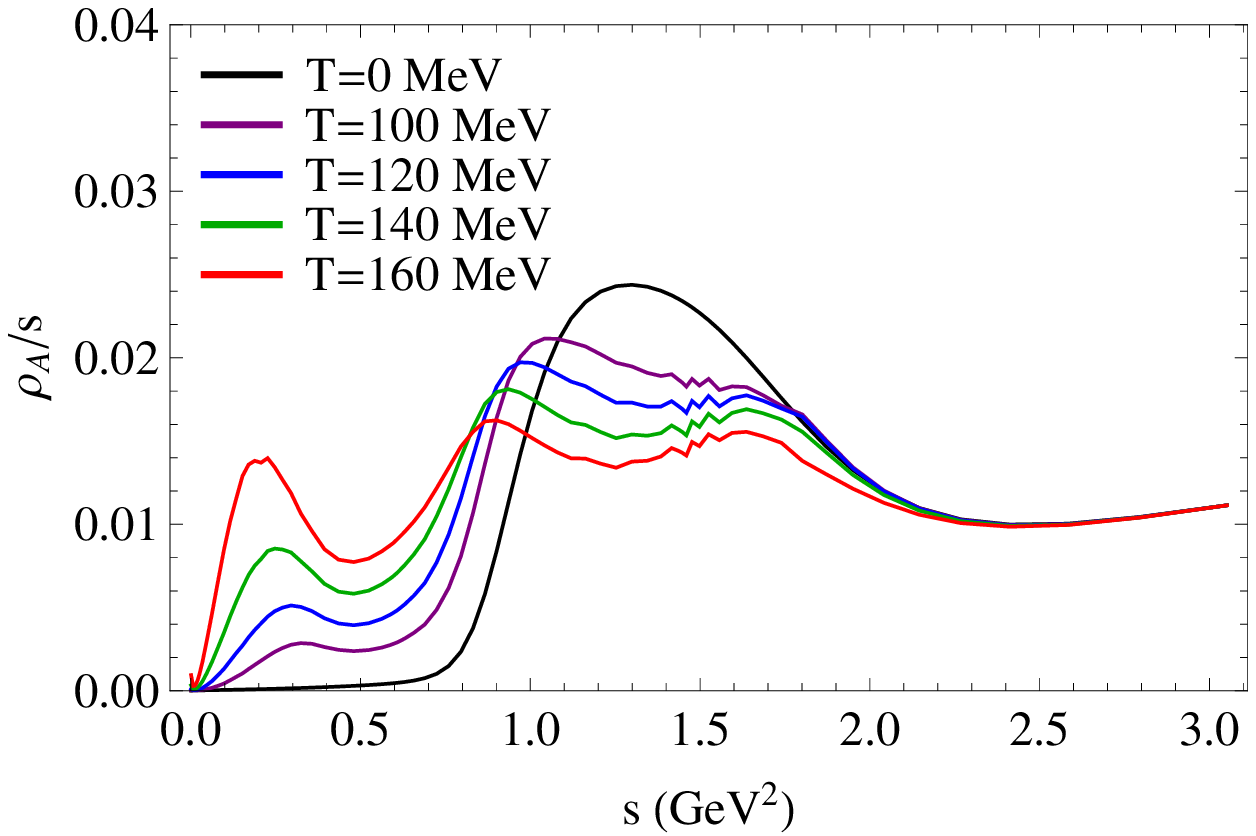}
    \includegraphics[width=.45\textwidth]{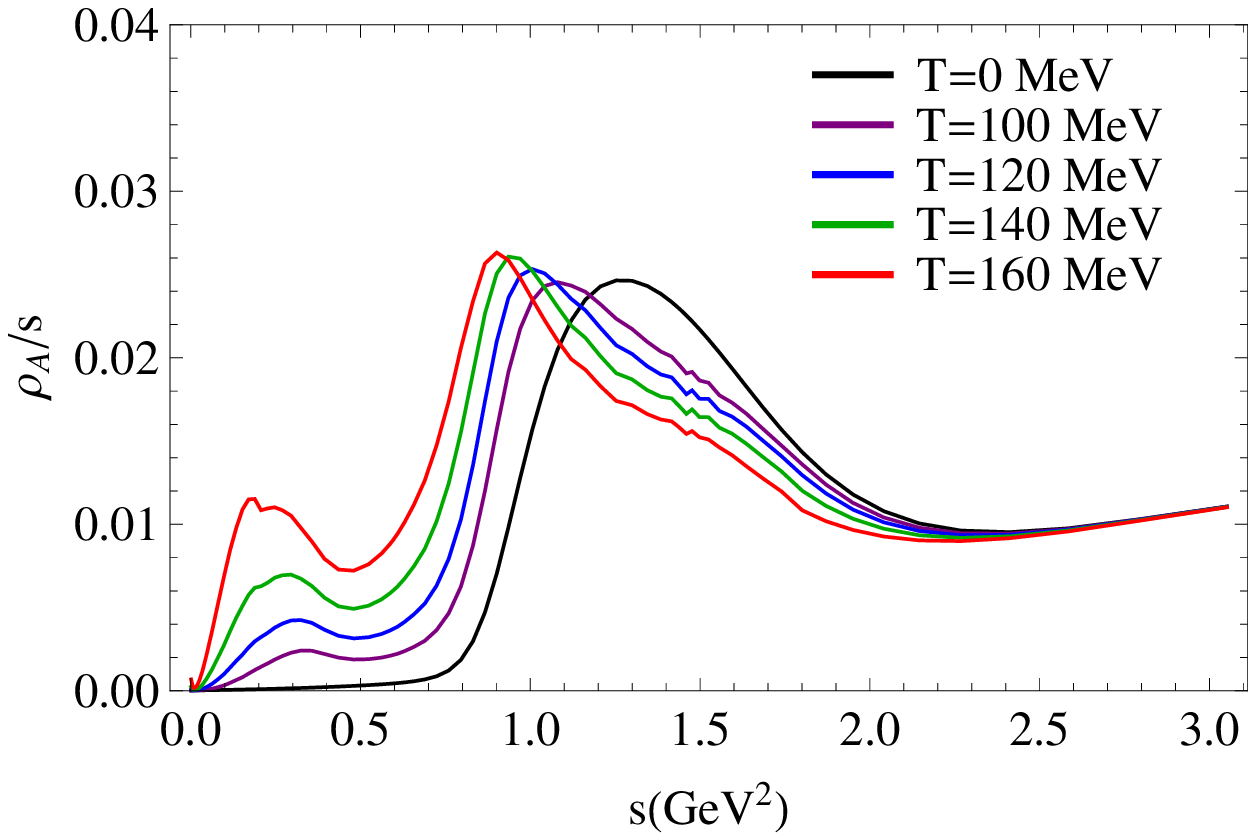}
\caption{Finite-$T$ axial-vector spectral functions at $|\vec{p}|=0$
in non-linear (upper panel) and linear realizations (lower panel).}
\label{fig:AVecSFFT}
\end{figure}
The resulting $AV$ spectral functions are presented at several temperatures
in Fig.~\ref{fig:AVecSFFT} for both realizations of chiral symmetry.
Several common features emerge which may be considered as robust.
First, the $a_1$ resonance peak noticeably shifts to lower masses with
increasing temperature. This is caused by both the attractive tadpole
contributions and the in-medium broadening of the $\pi\rho$ unitarity cut.
This is different from the $\rho$ case where the in-medium tadpoles
generate repulsion. Both realizations also develop a secondary peak or
shoulder at somewhat higher masses, around $s\simeq1.6$-1.7\,GeV$^2$.
In the non-linear version the double-peak structure is rather broad
while in the linear version more strength is in the lower
and somewhat narrower peak. These differences can be traced back to the
behavior of the real part of the inverse $AV$ propagator in vacuum
(recall Fig.~\ref{fig:ReSigmaA}), where the non-linear version
exhibits an extended region close to zero while the linear one has a
zero crossing leading to a well-defined peak.
The second robust feature is the development of a marked low-mass peak
around $\sqrt{s}\simeq0.5$\,GeV, caused by the scattering of an off-shell
$a_1$ off thermal pions into the $\rho$ resonance. This process is related
to the well-known ``chiral mixing"~\cite{Dey:1990ba}, but evaluated with
full thermal and finite-pion mass kinematics (see also
Ref.~\cite{Steele:1996su}), which results in a peak at masses significantly
lower than the nominal $\rho$ mass.

\begin{figure}[!t]
  \centering
  \includegraphics[width=.45\textwidth]{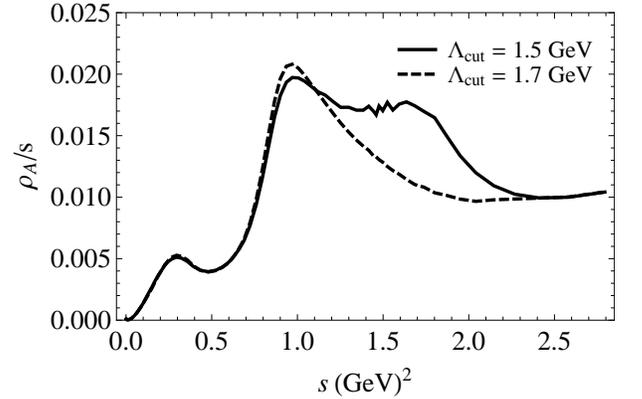}
\caption{Axial-vector spectral function at $T$=120 MeV with
$\Lambda_{\rm cut}$=1.5\,GeV (solid curve) and
$\Lambda_{\rm cut}$=1.7\,GeV (dashed curve).}
\label{fig:cutoff}
\end{figure}
To further scrutinize possible model dependencies, let us return to the
question of the cut-off of the integration over the $\rho$ spectral strength
when computing
the real part of the $a_1$ self-energy from a dispersion integral. In the
context of the vacuum calculations, we argued $\Lambda_{\rm cut}=1.5$\,GeV
to be a reasonable choice, to suppress (unphysical) high-energy contributions
to the imaginary part of the $a_1$ self-energy which rapidly increase due
to  higher-derivative contributions from large invariant $\rho$ masses.
To study the effect of variations in $\Lambda_{\rm cut}$ on the finite-$T$
results, we compare in Fig.~\ref{fig:cutoff} the $a_1$ spectral function in
the non-linear realization at $T=120$\,MeV with two different cut-offs,
1.5\,GeV and 1.7\,GeV. (The counter-term parameters are readjusted for the
larger cut-off to recover a vacuum fit.)
For masses below 1\,GeV, the impact is minimal, but
becomes significant above. Specifically, the larger cut-off produces additional
repulsion in $AV$ self-energy in the region of the vacuum $a_1$ peak and beyond,
leading to a noticeable suppression in spectral strength especially
around $s\simeq1.7$\,GeV, erasing the higher-mass bump. Even though the
$\Lambda_{\rm cut}=1.7$\,GeV bears some resemblance with the finite-$T$
results of the linear realization, the variation must be considered as
a model dependence caused by the ``fragility" of the real part in
the non-linear realization, another symptom for the breakdown of the
theory as higher-order derivative contributions become appreciable.
The high-energy behavior is better controlled with the lower cut-off which
we therefore deem preferable.
%

\section{Quantifying the Approach to Chiral Symmetry Restoration}
\label{sec:chirest}

\begin{figure*}[!t]
  \centering
        \includegraphics[width=.95\textwidth]{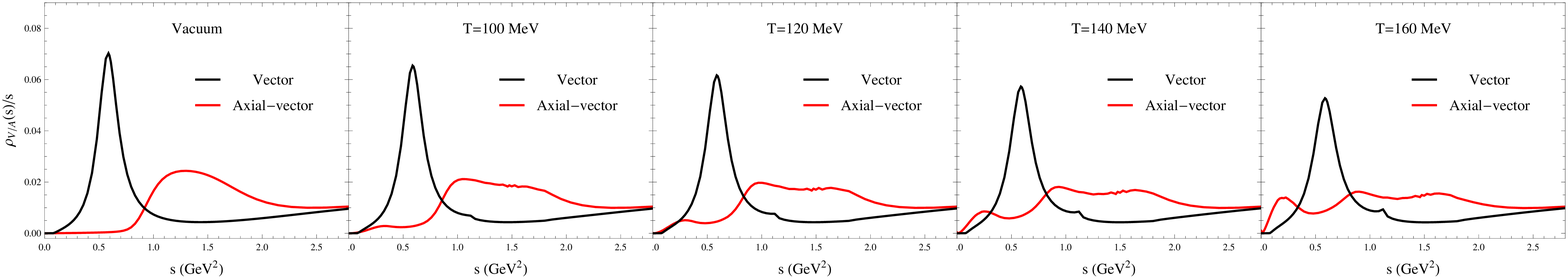}
    \includegraphics[width=.95\textwidth]{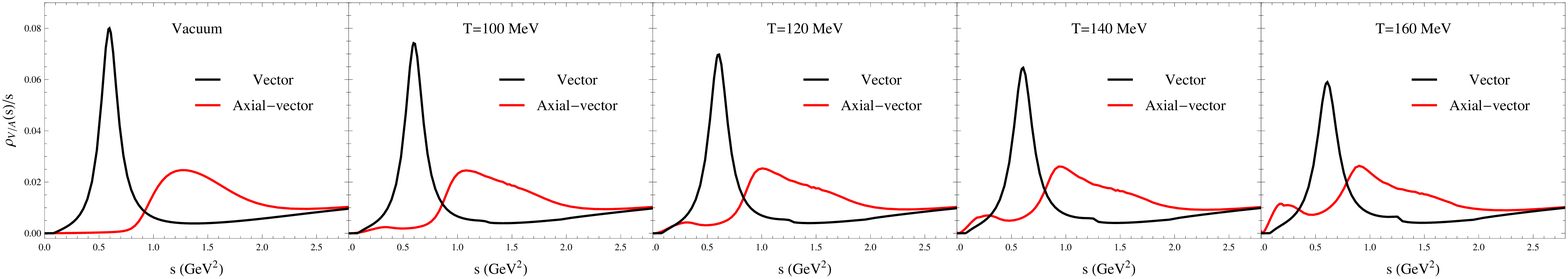}
\caption{Temperature progression of vector and axial-vector spectral functions.
Upper panel: Non-linear realization. Lower panel: Linear realization.}
\label{fig:VAVcomp}
\end{figure*}
Let us start by visually inspecting the temperature progression of the vector
vs.~axialvector spectral functions, \cf~Fig.~\ref{fig:VAVcomp} (for better
clarity of the medium effects we did not include the contributions from
the excited states).
The stability of the $\rho$ mass together with the downward shift of
the $a_1$ indicate an approach to chiral restoration that is characterized
by ``burning off" the chiral mass splitting, rather than a common mass
drop. This mechanism is quite consistent with our independent recent
analysis of Weinberg and chiral sum rules, where an in-medium hadronic
many-body $\rho$ spectral function (that describes dilepton data) and chiral order
parameters from lattice QCD were used as input~\cite{Hohler:2013eba}.
Similar patterns for the $\rho\pi a_1$ system have also been found in
Refs.~\cite{Song:1993af,Urban:2001uv}.

While the suggestive trends discussed above are promising, a key objective of
a microscopic approach, as the one pursued here, is to quantify the amount of
restoration. The natural measure of this are chiral order parameters,
such as the chiral condensate, $\langle \bar{q}q\rangle$, and the pion
decay constant, $F_\pi^2$.
In this section, we elaborate how the $T$ dependence of these two
quantities can be calculated within MYM and discuss the results.

\subsection{Pion Decay Constant}
The most direct calculation of $F_\pi^2(T)$ is following the
same procedure as in vacuum, \ie, from its definition as the effective
coupling between the pion and the axial-vector current.
Alternatively, and more indirectly, we employ the first Weinberg sum
rule by inserting the finite-$T$ vector and axial-vector spectral as
calculated above. We will carry out both methods for both realizations.

In vacuum, the $AV$ longitudinal channel serves as a measure of the
chiral properties of the model through PCAC and has been used as
guideline for identifying vertex corrections necessary to preserve the
symmetry. At finite temperature, this channel can be used to extract
the temperature dependence of $F_\pi^2$ by using Eq.~(\ref{eq:fpiT}).
The key ingredients are the temperature effects on the self-energies
$\Sigma_{a_1}^L$, $\Sigma_{\pi a}$, and $\Sigma_{\pi\pi}$.
These are calculated in the same manner as discussed above for the $a_1$
transverse self-energy with $\rho_A^L$ given by Eq.~(\ref{eq:rhoaL}), and
the loop integrals for each contribution written out in
Appendix~\ref{sec:loop}.

\begin{figure}[!b]
  \centering
        \includegraphics[width=.45\textwidth]{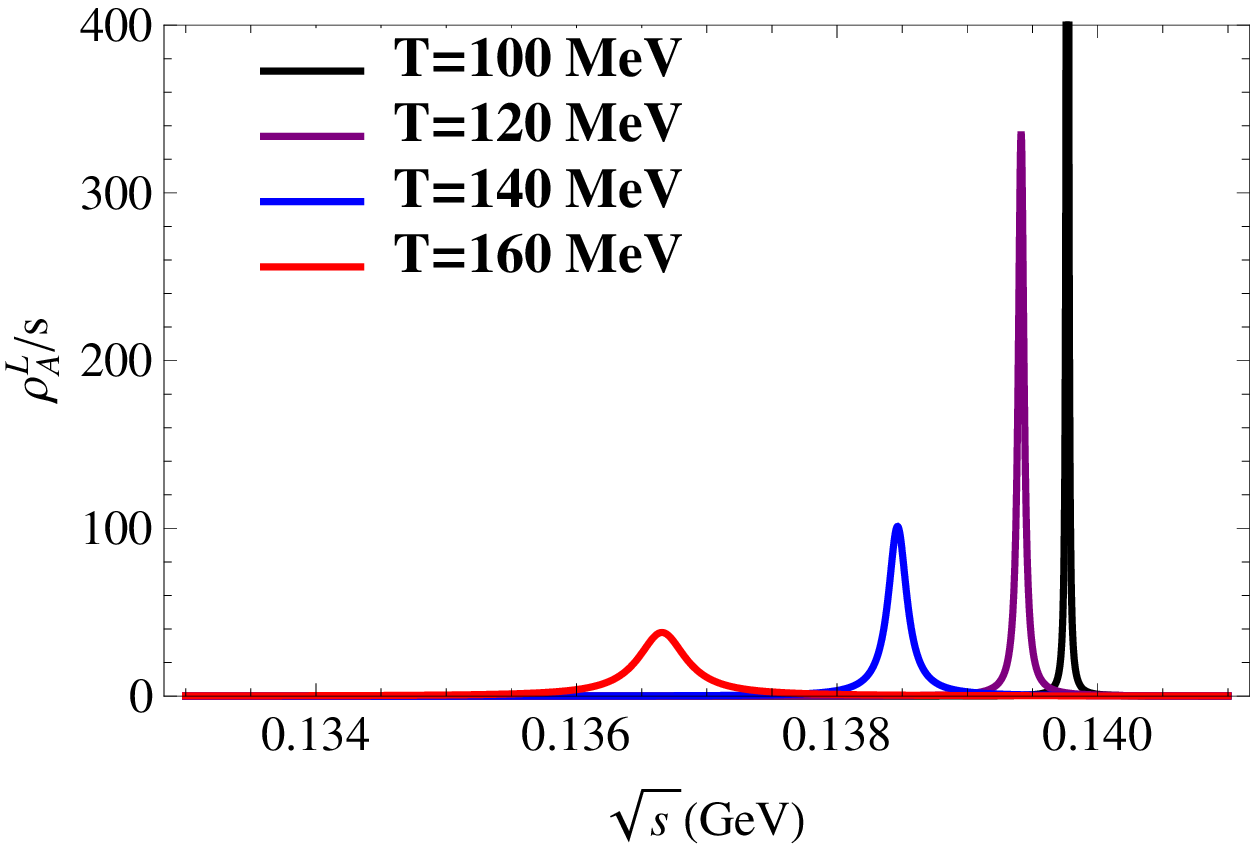}
    \includegraphics[width=.45\textwidth]{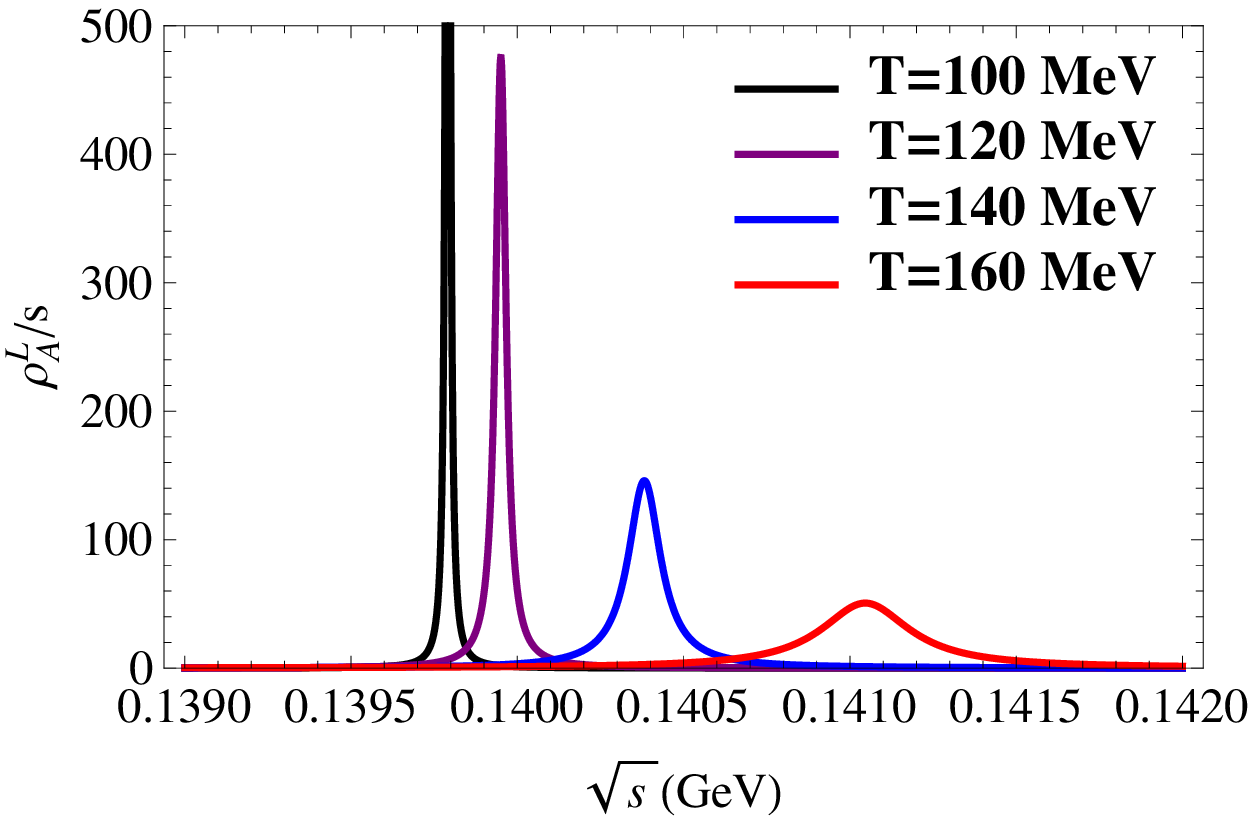}
\caption{Pion spectral function at finite temperature in the
non-linear (upper panel) and linear realization (lower panel).}
\label{fig:pionSF}
\end{figure}
\begin{figure}[!t]
  \centering
    \includegraphics[width=.45\textwidth]{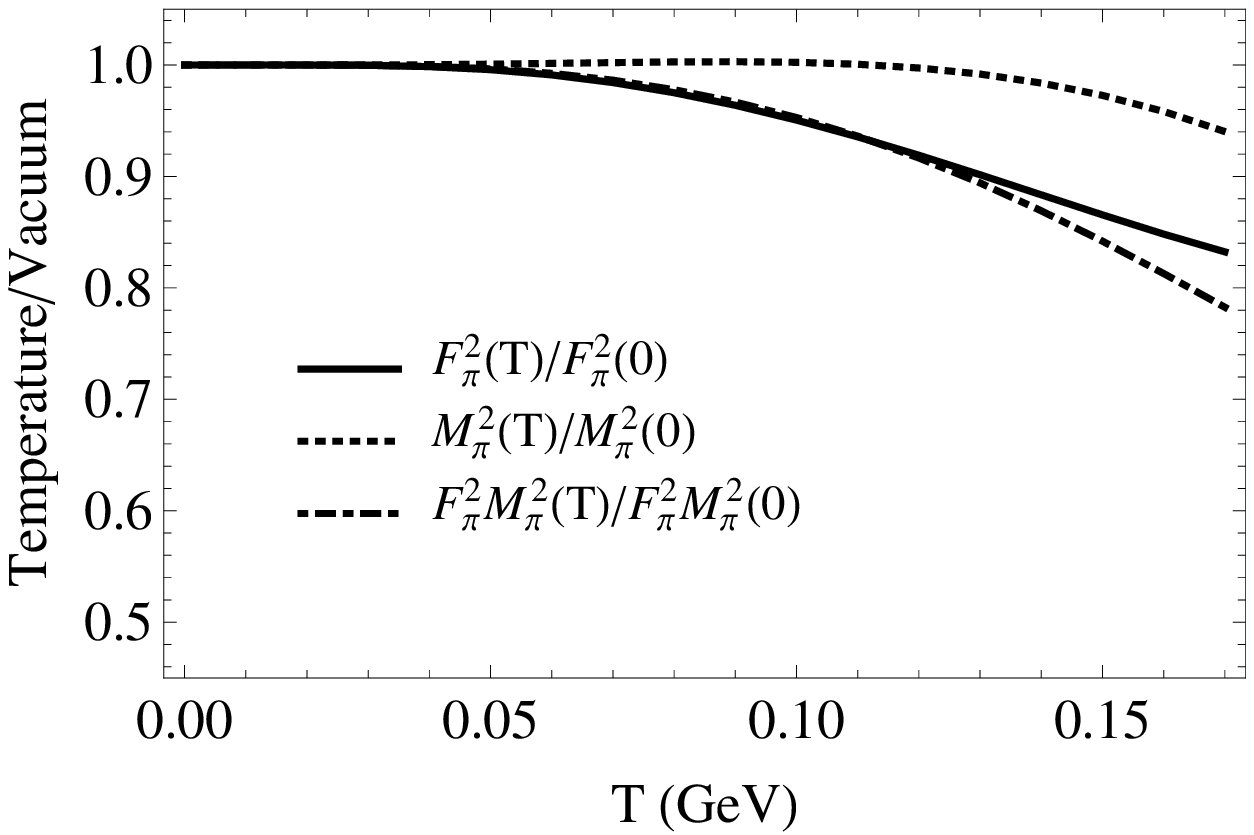}
        \includegraphics[width=.45\textwidth]{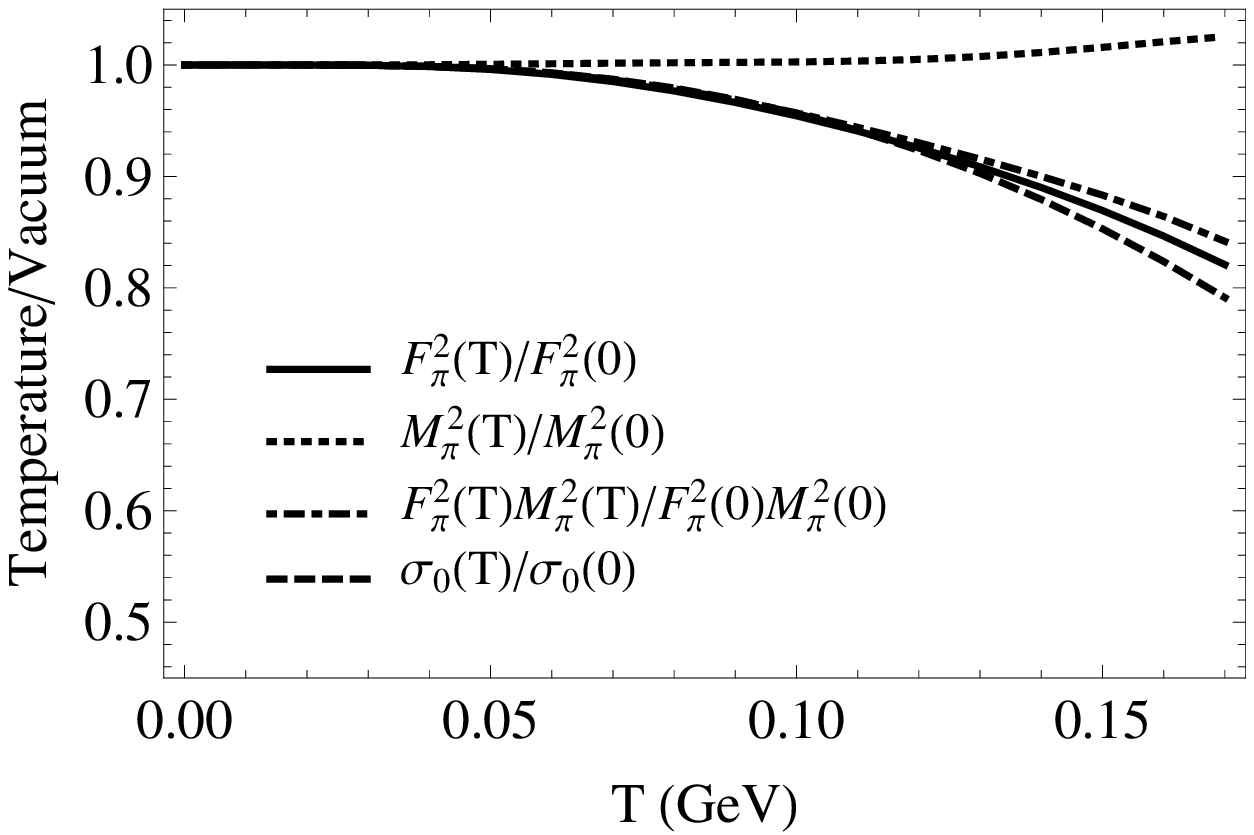}
\caption{Upper panel:
Temperature dependence of $F_\pi^2$, $M_\pi^2$, and their product
for the non-linear realization.
Lower panel: Temperature dependence of $F_\pi^2$, $M_\pi^2$ and
$\sigma_0$ for the linear realization.}
\label{fig:fpi}
\end{figure}

The dominant contribution to the longitudinal spectral function
is the pion spectral peak, displayed in
Fig.~\ref{fig:pionSF} for several temperatures. In vacuum, it is a delta
function (not shown) located at the physical pion mass of $139.6$\,MeV,
while the $T=100$\,MeV peak extents beyond the plot.
The narrowness of the peak implies the imaginary parts of the
self-energies included in $\rho_A^L$ to be small such that $F_\pi^2(T)$
can be well approximated by
\begin{equation}
F_\pi^2 (T) = F_\pi^2(0)\left.{\rm Re}\left[F^2\right]\left(1-\partial {\rm Re}
\Sigma_\pi/\partial p^2\right)^{-1}\right|_{p^2 = M_\pi^2(T)} \, ,
\end{equation}
where $F$ is given by Eq.~(\ref{eq:Feq}).
The resulting temperature dependence of $F_\pi^2$ is shown by the solid
lines in the upper and lower panels of Fig.~\ref{fig:fpi} for non-linear and
linear realizations, respectively.
It turns out to be quite comparable between the two realizations,
being consistent with chiral low-temperature expansions and
reduced by up to 15-17\% at $T=160$\,MeV.

The mechanism which leads to this reduction can be more readily
understood in the non-linear realization. The reduction is controlled
by ${\rm Re}\Sigma_{a\pi}$ due to increasing positive values.
As we saw with the mass shifts in the $\rho$ and
$a_1$ spectral peaks, there are two contributions which can affect this,
one from the loops and one from the tadpoles. The case with $\Sigma_{a\pi}$
is more similar to $\Sigma_\rho$ where the loops and tadpoles provide
opposing contributions, with attractive loops and repulsive tadpoles.
As temperature increases, the two competing sides almost  balance each
but with the repulsive terms winning out.

\begin{figure}[!t]
  \centering
        \includegraphics[width=.45\textwidth]{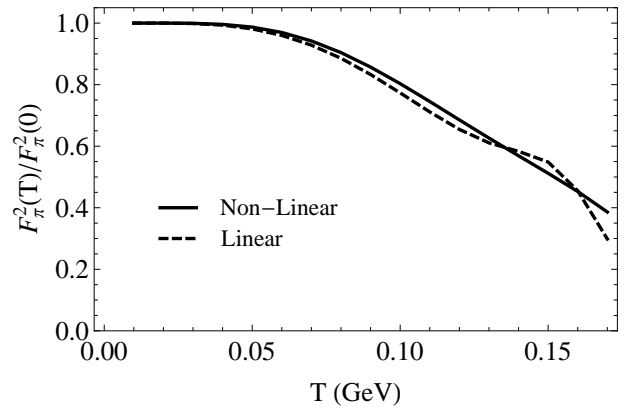}
\caption{Temperature dependence of $F_\pi^2$ relative to its vacuum value
calculated from the first WSR using the non-linear (solid
curve) and linear realization (dashed curve).}
\label{fig:wsr}
\end{figure}

Complementary to the field theory approach is the calculation of
$F_\pi$ from the first Weinberg sum rule (WSR)~\cite{Weinberg:1967kj},
which relates the first negative moment of the $V$ minus $AV$ spectral
function to $F_\pi^2$. For vanishing 3-momentum ($|\vec{p}|=0$) one has
\begin{equation}\label{eq:WSR1}
\int \left( \rho_V^T - \rho_A^T \right) s^{-1} ds = \frac{F_\pi^2}{2} \ .
\end{equation}
Ideally, one would evaluate this equation with just the in-medium $V$ and
$AV$ spectral functions as calculated above. However, already in vacuum
the WSRs are not satisfied when only the $\rho$ and $a_1$
contributions are accounted for. Excited states, $\rho'$ and $a_1'$,
and/or different continuum contributions, are essential to
quantitatively satisfy the WSRs, see, \eg, the
recent discussion in Ref.~\cite{Hohler:2012xd}.
Since the calculation of medium modifications of the excited states is
currently beyond the scope of MYM, we will
approximate their contribution by a $T$ independent constant added to the
left-hand side of the Eq.~(\ref{eq:WSR1}). This constant is chosen so that
the vacuum sum rule is satisfied. Assuming the constant to be temperature
independent will presumably overestimate $F_\pi^2$ in the medium as an
expected chiral restoration of the excited states would lead to a further
reduction (this effect will become stronger for the higher WSRs
which involve higher moments in $s$). Put differently, we only evaluate
the effects of chiral restoration due to in-medium effects on the $\rho$
and $a_1$ states.
The temperature dependence of $F_\pi^2$ calculated in this manner
is depicted in Fig.~\ref{fig:wsr}. While it is rather consistent
between the two realizations, its reduction with temperature is
more pronounced compared to the evaluation from the longitudinal
$AV$ spectral function.
The stronger reduction can have two principal sources: a
lack of vector spectral strength and/or too much axial-vector spectral
strength; low-mass strength, in particular, affects the negative moment
of the first WSR. An obvious origin of missing strength
in the vector channel is the in-medium dressing of the pions, which
is known to generate a low-mass enhancement in the $\rho$ spectral
function especially through the unitarity cut, but also through
the Landau cut at very low mass (although suppressed by small pion widths
at small 3-momenta). In addition, in the current calculation, the
$\pi a_1$ Landau cut only includes a sharp ({\it i.e.},
zero-width) $a_1$. A more complete calculation of these contributions,
which requires further vertex corrections to preserve gauge
symmetry, would induce extra broadening in the $\rho$ spectral peak.
At the same time, pion broadening is expected to affect less the
low-mass strength of the $a_1$ spectral function. On the other hand,
increased imaginary parts at finite $T$ in the $a_1$ self-energy could, in
fact, lead to a suppression of the $AV$ spectral function in the energy regime
around $s>1$\,GeV$^2$ (as illustrated in Fig.~\ref{fig:cutoff} in the context
of (spurious) vacuum contributions).
Thus, it is conceivable that future calculations, involving additional
finite-width effects on all particles in the loops, will
increase (decrease) the $V$  ($AV$) contribution to the WSRs and
``stabilize" them.

\subsection{Chiral Condensate}
In the linear realization, $\langle \bar{q}q\rangle(T)$ can be directly related
to the $T$ dependence of the scalar-field condensate, $\sigma_0(T)$. This is
determined without relying on any of the specific spectral functions of the
field theory, by minimizing the scalar potential with respect to $\sigma_0$.
Diagrammatically, this is equivalent to setting the lollipop diagrams of
Fig.~\ref{fig:sigmaTad} to zero and solving for $\sigma_0$ at each temperature.
However, this procedure is not consistent with the 1-loop treatment that we
implement throughout the present work. The appropriate temperature correction
to the vacuum $\sigma_0$ at the 1-loop level is rather obtained from the lollipop
diagrams alone, \ie, not solving for $\sigma_0$. One has
\begin{equation}
\frac{\sigma_0(T)}{\sigma_0(0)} = 1 + \frac{1}{\sigma_0 M_\sigma^2}
\left(\Sigma_{\rm lolli}^{\pi} + \Sigma_{\rm lolli}^{\sigma}\right) \ ,
\end{equation}
with the $\Sigma_{\rm lolli}$'s quoted in Appendix~\ref{sec:loop}.
The contributions of these diagrams are related to the scalar densities
of the pion and sigma, respectively. Therefore, this procedure is similar
to the standard density expansion where the reduction of the condensate
is driven by the increase of the density of the medium particles
(``vacuum cleaner"). The resulting temperature dependence of $\sigma_0$ is
shown as the dashed line in Fig.~\ref{fig:fpi}. It exhibits a smooth
decrease of up to $\sim$18\% at $T=160$\,MeV.

In the non-linear realization, the chiral condensate cannot be directly
calculated. However, we can take recourse to calculations of $F_\pi(T)$
and $M_\pi(T)$  to find the $T$ dependence of the chiral
condensate through the Gellmann-Oakes-Renner (GOR) relation,
\begin{equation}
\langle \bar{q}q\rangle(T)/\langle \bar{q}q\rangle(0) =
M_\pi^2(T) F_\pi^2(T) / M_\pi^2(0) F_\pi^2 (0) \, .
\end{equation}
The temperature dependence of $M_\pi$ is determined from the zero of
the real part of the inverse pion propagator in Eq.~(\ref{eq:piprop}).
We find a very slightly increasing trend at low temperatures in
both realizations (in line with low-$T$ chiral expansions~\cite{Gasser:1986vb}),
which, however, turns around into a $\sim$5\%
attraction in the nonlinear case at $T=160$\,MeV, while the linear
case reaches a 2\% repulsion at this temperature (see dotted lines in
Fig.~\ref{fig:fpi}).
When evaluating the quark condensate with GOR, the two realizations
slightly differ because of the added repulsion in pion self-energy for the
linear realization. Furthermore, in the linear realization, the scalar
condensate calculated via either GOR or directly gives slightly different
results particularly at higher temperatures. This stems from the
differences between performing a loop
expansion rather than a formal $T^2$ expansion.

Nevertheless, our overall findings for the temperature dependence of the chiral
order parameters give a fairly consistent picture between the two
realizations, indicating that their reduction in temperature is a
robust feature accompanying the approach toward degeneracy in the
spectral functions. Quantitatively, when directly evaluated from the
low-energy properties of the longitudinal $AV$ spectral function, the
reduction reaches up to $\sim$15-20\% at $T=160$\,MeV. It is about
twice as large when evaluated with the first WSR, which we tentatively
attribute to neglecting higher-order width effects expected to
impact the $V$ channel more strongly than the $AV$ one,
thus taming the reduction.

\section{Hadronic Mechanisms of Chiral Symmetry Restoration}
\label{sec:disc}
In this section we put our work into a broader context and
identify future studies to improve our understanding of hadronic
mechanisms leading to chiral restoration.

Hadronic degeneracy is well established in mean-field approaches of
chiral effective hadronic theories, where the masses of, \eg,
the $\pi$-$\sigma$ and $\rho$-$a_1$, degenerate in the chirally restored
phase~\cite{Pisarski:1995xu,Bochkarev:1995gi,Bilic:1997sh,Petropoulos:1998gt,Struber:2007bm}.
It remains, however, rather challenging to conduct these calculations
beyond the mean-field level under the inclusion of finite-width effects
which are essential to obtain realistic spectral functions. As we discussed
above, this is a non-trivial task for the axial-vector spectral function
already in vacuum, at least for local-gauge approaches. On the other hand,
our finite-temperature calculations presented above are carried out in
a loop expansion which does not implement in-medium changes of the underlying
hadronic coupling constants. The next step should thus aim at a self-consistent
treatment.

Let us focus on the linear realization of MYM, which is more readily amenable
for treating the transition region.
At the mean-field level, the masses and couplings are explicit
functions of the condensate $\sigma_0$. We may thus identify for each mass
and coupling a contribution which is finite for $\sigma_0=0$ and one
which vanishes and interpret the latter as due to spontaneous symmetry
breaking.  As $\sigma_0$ is driven to lower values, the symmetry breaking
contributions ``burn" off. The seed for this mechanism is already inherent
in our calculations, as confirmed by the results presented above in
terms of the $a_1$ mass approaching the $\rho$ mass and the accompanying
reduction in the scalar condensate.
Extending this all the way to the restored phase, the bare masses of the
chiral partners, $(M_\rho,M_{a_1})$ and $(M_\pi,M_\sigma)$, will become
identical as can be gleaned from Eq.~(\ref{eq:linpara}). For the self-energies,
at the one loop level, this implies that the $\rho\to\pi\pi$ and pion tadpole
of Fig.~\ref{fig:dia1} degenerate with the $a_1\to\sigma\pi$ loop (third
diagram in Fig.~\ref{fig:axialLin}) and the pion tadpole for the $a_1$
(second diagram in Fig.~\ref{fig:dia1}), while all other diagrams vanish,
in particular, the $\pi\rho$ loop in the $a_1$ self-energy and the $\pi a_1$
loop in the $\rho$ self-energy (since the $\rho\pi a_1$ coupling
vanishes for $\sigma_0=0$). A similar analysis holds for the $\pi$ and $\sigma$
self-energies. With $\pi$-$\sigma$ degeneracy, this implies the $V$ and $AV$
spectral functions to degenerate as well. This analysis suggests that the mechanism
of chiral mixing, which is the leading effect in the low-$T$ limit induced by the
$\pi\rho a_1$ coupling, vanishes at the restoration point, while a central role
is taken by the $\pi$-$\sigma$ degeneracy and the ``burning" of the chiral
mass splitting. In this context, it is interesting to note that a recent
lattice computation of the correlation functions of the nucleon and its chiral
partner, the $N^*(1535)$, also suggests that the main effect of chiral restoration
at finite temperature is a vanishing of the mass splitting while the nucleon mass
itself is little affected~\cite{Aarts:2015mma}.

In practice, a self-consistent field-theoretic implementation with realistic
(broad) spectral distributions remains challenging.
The evaluation of $\sigma_0(T)$ from solving the gap equation
needs to be accompanied by higher thermal-loop calculations.
This will require the dressing of all internal particles with
their thermal widths.
The dressed propagators will necessarily precipitate compensating vertex
corrections to maintain the symmetries, some of which will be similar to
the types considered here. To be able to use the full temperature dependence
of $\sigma_0$ in the vertices may also require a resummation of the vertices,
or at least casting the vertex corrections in terms of the relevant self-energies.

To make contact with experimental measurements of dilepton emission
through the vector channel, the effects of baryons need to be accounted
for. Their quantitative importance in the experimental context is
likely to alter the predicted pseudo-critical temperature, but they
might not introduce a qualitatively new mechanism
of chiral restoration, relative to what our pion gas findings in the
present analysis suggest. In fact, our aforementioned analysis of QCD
and WSRs, using realistic in-medium spectral functions,
indicated that chiral-mass burning is compatible with dilepton
experiments~\cite{Hohler:2013eba}.

\section{Conclusion}
\label{sec:concl}
Based on a Massive Yang-Mills implementation of vector and axial-vector mesons
into the chiral pion Lagrangians which yields vacuum spectral functions compatible
with experiment, we have calculated their modifications in a hot pion gas.
A key feature of such approaches is
that they enable the evaluation of chiral order parameters within the same framework,
and thus provide connections between chiral restoration and spectral modifications
(which can be measured in dilepton experiments).
Toward this end, we have carried out a temperature loop expansion encompassing tadpole and
2-particle loop diagrams as well as vertex corrections associated with the vector and chiral Ward
identities. As a check of the robustness of our predictions, we have evaluated both
spectral functions and chiral order parameters in both the linear and non-linear
realization of chiral symmetry in the pion Lagrangian. A fair consistency of the two
methods for both order parameters and spectral functions is found.
While the $\rho$ meson spectral function undergoes a modest finite-temperature broadening
with a stable pole mass (in agreement with previous calculations of a similar kind), the
in-medium $a_1$ spectral function exhibits a significant mass shift toward the
$\rho$ meson. These medium effects are accompanied by a 15-20\% reduction in the
scalar condensate and pion decay constant at $T$=160\,MeV, suggesting that the reduction
in the $a_1$ mass signals a mechanism of chiral symmetry restoration whereby chiral
mass splittings are burned off. A similar mechanism also emerged from
QCD and Weinberg sum rule analyses with full hadronic many-body $\rho$ spectral
functions, and from recent thermal lattice-QCD computations of the correlation
functions of the nucleon and its chiral partner, $N^*(1535)$. We have conjectured
that such a mechanism is quite different from the so-called chiral mixing and
rather involves the degeneracy in the scalar-pseudoscalar channel as a critical
ingredient.
The full embodiment of this mechanism, ultimately leading to spectral degeneracy in
the restored phase, requires the self-consistent implementation of additional medium
effects beyond the loop expansion, associated with dressed in-medium propagators
(including baryonic effects), accompanying vertex corrections, and a simultaneous
solution of the chiral gap equation (which is more naturally done in the linear
realization of MYM).
Work in these directions is in progress.
\\

\acknowledgments
This work has been supported by the US-NSF under grant No.~PHY-1306359, and
by the A.-v.-Humboldt Foundation (Germany).
\\

\begin{appendix}
\section{Lagrangian couplings for relevant MYM vertices}
\label{sec:appVert}
\begin{figure*}[t!]
  \centering
	\includegraphics[width=.8\textwidth]{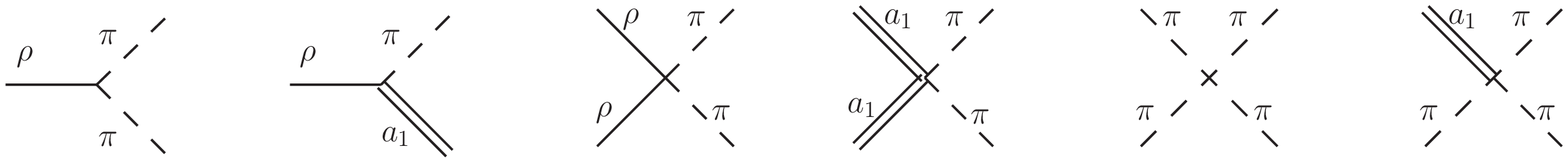}
\caption{Tree level vertices for both the non-linear and linear realizations
 used in calculations in this paper.}
\label{fig:appvertex}
\end{figure*}

In this appendix, we detail the vertex couplings which
were utilized in the calculations of this paper. They follow from the MYM
Lagrangian, Eq.~(\ref{eq:lag}), upon expansion in the physical fields,
as described in Sec.~\ref{sec:MYM}.

We begin with the relevant vertices of the non-linear realization,
diagrammatically depicted in Fig.~\ref{fig:appvertex}.
The resulting Lagrangian terms are as follows:
\begin{widetext}
\begin{eqnarray}
\mathcal{L}_{\rho\pi\pi}^{\rm nLin} &=& g_{\rho\pi\pi} \,\rho_\mu \cdot \pi \times \partial^\mu
\pi+g_{\rho\pi\pi}^{(3)} \, \rho_{\mu\nu} \cdot \partial^\mu \pi \times \partial^\nu \pi \\
\mathcal{L}_{\rho\pi a_1}^{\rm nLin} &=& C\, \rho_\mu \cdot \pi \times a^\mu + D \,
\rho_{\mu\nu} \cdot \pi \times a^{\mu\nu} + E \, \rho_\mu \cdot \partial_\nu
\pi \times a^{\mu\nu} + F \, \rho_{\mu\nu} \cdot \partial^\mu \pi \times a^\nu \\
\mathcal{L}_{\rho\rho\pi\pi}^{\rm nLin} &=& \alpha_1 \left( \pi\cdot\rho_\mu \,
\pi\cdot\rho^\mu-\pi\cdot\pi \, \rho_\mu\cdot \rho^\mu \right)
+\alpha_2 \left(\pi\cdot\rho_{\mu\nu} \,\pi\cdot\rho^{\mu\nu}-\pi\cdot\pi \,
\rho_{\mu\nu}\cdot\rho^{\mu\nu}\right)\\
&&+ \alpha_3 \left(\partial_\mu \pi\cdot\rho^{\mu\nu} \,\pi\cdot\rho_\nu-
\partial_\nu\pi \cdot\rho^{\mu\nu} \,\pi\cdot\rho_\mu\right) +\alpha_4
\left(\partial_\mu \pi\cdot\rho_\nu \,\pi\cdot\rho^{\mu\nu}-\partial_\nu\pi\cdot
\rho_\mu \, \pi\cdot\rho^{\mu\nu}\right)\nonumber\\
&&+\alpha_5\, \partial_\mu\pi\cdot\rho_\nu \,\partial^\nu \pi\cdot\rho^\mu
+\alpha_6\, \partial_\mu\pi\cdot\rho^\mu \, \partial_\nu \pi\cdot\rho^\nu
+\alpha_7 \left(\rho_\mu \cdot \rho^{\mu\nu}\, \pi\cdot\partial_\nu\pi-
\rho_\nu\cdot\rho^{\mu\nu}\, \pi\cdot\partial_\mu\pi\right) \nonumber \\
&&+\alpha_8 \left(\partial_\mu \pi\cdot\partial^\mu  \pi \,\rho_\nu \cdot
\rho^\nu-\partial_\mu\pi\cdot \partial_\nu \pi \,\rho^\mu\cdot\rho^\nu-
\partial_\mu \pi\cdot\rho_\nu \, \partial^\mu \pi \cdot \rho^\nu\right)\nonumber\\
\mathcal{L}_{a_1a_1\pi\pi}^{\rm nLin} &=& \beta_1 \left( \pi\cdot a_\mu \, \pi\cdot
 a^\mu-\pi\cdot\pi\, a_\mu\cdot a^\mu \right)
+\beta_2 \left(\pi\cdot a_{\mu\nu}\, \pi\cdot a^{\mu\nu}-\pi\cdot\pi \,
a_{\mu\nu}\cdot a^{\mu\nu}\right)\\
&&+ \beta_3 \left(\partial_\mu \pi\cdot a^{\mu\nu}\, \pi\cdot a_\nu-
\partial_\nu\pi \cdot a^{\mu\nu}\, \pi\cdot a_\mu\right)+\beta_4\,
\partial_\mu\pi\cdot a_\nu \,\partial^\nu \pi\cdot a^\mu \nonumber \\
&&+\beta_5\, \partial_\mu\pi\cdot a^\mu \,\partial_\nu \pi\cdot a^\nu
+\beta_6 \left(a_\mu \cdot a^{\mu\nu}\, \pi\cdot\partial_\nu\pi-a_\nu
\cdot a^{\mu\nu}\, \pi\cdot\partial_\mu\pi\right) \nonumber \\
&&+\beta_7 \left(\partial_\mu \pi\cdot\partial^\mu \pi \,a_\nu \cdot a^\nu-
\partial_\mu\pi\cdot \partial_\nu \pi \,a^\mu\cdot a^\nu-\partial_\mu \pi
\cdot a_\nu \,\partial^\mu \pi \cdot a^\nu\right)\nonumber \\
\mathcal{L}_{\pi\pi\pi\pi}^{\rm nLin} &=& \gamma_1 \left( \pi\cdot\partial_\mu
\pi \,\pi\cdot\partial^\mu\pi - \pi\cdot\pi \,\partial_\mu\pi\cdot\partial^\mu\pi\right)
+ \gamma_2 \left(\partial_\mu\pi\cdot\partial_\nu\pi\, \partial^\mu\pi
\cdot\partial^\nu\pi - \partial_\mu\pi\cdot\partial^\mu\pi\,
\partial_\nu\pi\cdot\partial^\nu\pi\right)\\
&&+ \gamma_3 \,\pi\cdot\pi \,\pi\cdot\pi \\
\mathcal{L}_{a_1 \pi\pi\pi}^{\rm nLin}&=& \delta_1 \left(a_\mu \cdot\pi \,\partial^\mu\pi
\cdot\pi-a_\mu \cdot\partial^\mu\pi \,\pi\cdot\pi\right)
+\delta_2 \left(a_{\mu\nu} \cdot \partial^\mu\pi \,\partial^\nu\pi\cdot
\pi-a_{\mu\nu} \cdot \partial^\nu \pi \,\partial^\mu \pi\cdot\pi\right) \\
&&+\delta_3 \left(a_\mu\cdot\partial^\mu\pi \,\partial_\nu\pi\cdot\partial^\nu
\pi-a_\mu\cdot\partial_\nu\pi \,\partial^\mu\pi\cdot\partial^\nu \pi\right) \nonumber
\end{eqnarray}
\end{widetext}
In these expressions, two indices on a vector or axial-vector
field represent the field strength,
$\rho_{\mu\nu} = \partial_\mu \rho_\nu - \partial_\nu \rho_\mu$,
and the dot and cross products refer to isospin space.
The couplings can be expressed in terms of the
bare Lagrangian parameters as
\begin{eqnarray}
C &=& -g_{\rho\pi\pi}^2 \frac{F_\pi}{\sqrt{2}}\frac{Z_\pi M_{a_1}}{M_\rho}\\
D &=& \frac{M_\rho Z_\pi}{\sqrt{2} F_\pi M_{a_1}}
\left(1-\frac{M_{a_1}^2}{M_\rho^2 Z_\pi^2}\right)\\
E &=& -g_{\rho\pi\pi}^2 \frac{F_\pi}{\sqrt{2}} \frac{Z_\pi}{M_{a_1} M_\rho}\\
F &=& -\sqrt{2} g_{\rho\pi\pi} F_\pi Z_\pi \frac{M_{a_1}}{M_\rho}
\left(g_{\rho\pi\pi}^{(3)}+\frac{g_{\rho\pi\pi}}{2 M_\rho^2}\right)
\end{eqnarray}

\begin{eqnarray}
\alpha_1 &=& -\frac{1}{2} g_{\rho\pi\pi}^2 Z_\pi^2\\
\alpha_2 &=& -\frac{1}{2 F_\pi^2}\left(1-\frac{Z_\pi^2 M_\rho^2}{M_{a_1}^2}\right)\\
\alpha_3 &=& \frac{g_{\rho\pi\pi}^2 F_\pi^2}{M_\rho^2} \, \alpha_2\\
\alpha_4 &=& \alpha_5 \, Z_\pi^2 + \alpha_8 \, \frac{M_\rho^2}{M_{a_1}^2}\\
\alpha_5 &=& - g_{\rho\pi\pi} g_{\rho\pi\pi}^{(3)}\\
\alpha_6 &=& \alpha_8 - \alpha_5\\
\alpha_7 &=& \alpha_3 + \alpha_4 \\
\alpha_8 &=& - \frac{1}{2} E^2
\end{eqnarray}

\begin{eqnarray}
\beta_1 &=& -\alpha_1 \frac{M_{a_1}^2}{M_\rho^2 Z_\pi^2 }\\
\beta_2 &=& -\alpha_2 \frac{M_{a_1}^2}{M_\rho^2 Z_\pi^2}\\
\beta_3 &=& -\left(\alpha_3+\alpha_4\right) \frac{M_{a_1}^2}{M_\rho^2 Z_\pi^2}\\
\beta_4 &=& \left(\alpha_5 -\alpha_4\frac{g_{\rho\pi\pi}^4 F_\pi^4
Z_\pi^2}{4 M_\rho^4}\right)
\frac{M_{a_1}^2}{M_\rho^2 Z_\pi^2}\\
\beta_5 &=& \beta_7 - \beta_4\\
\beta_6 &=& \beta_3\\
\beta_7 &=& \left(\alpha_8 \frac{M_{a_1}^2}{M_\rho^2 Z_\pi^2}
+\alpha_4 \frac{g_{\rho\pi\pi}^2 F_\pi^2}{M_\rho^2}\right)
\frac{M_{a_1}^2}{M_\rho^2 Z_\pi^2}\\
\end{eqnarray}

\begin{eqnarray}
\gamma_1 &=& -\frac{g_{\rho\pi\pi}^2}{2 M_\rho^2}+\frac{1}{3 F_\pi^2}\\
\gamma_2 &=& -\frac{g_{\rho\pi\pi}^3 g_{\rho\pi\pi}^{(3)} F_\pi^2}
{2 M_\rho^4}-\frac{g_{\rho\pi\pi}^6 F_\pi^4}{16 M_\rho^8}\\
\gamma_3 &=& \frac{M_\pi^2}{12 F_\pi^2}
\end{eqnarray}

\begin{eqnarray}
\delta_1 &=& -\frac{\sqrt{2} g_{\rho\pi\pi} M_{a_1}}{3 F_\pi Z_\pi M_\rho}\left(3-Z_\pi^2\right)\\
\delta_2 &=& \frac{1}{g_{\rho\pi\pi} F_\pi^2 Z_\pi^2}
\left(F - \frac{E M_{a_1}^2}{M_\rho^2 Z_\pi^2}\right)\\
&&-\frac{g_{\rho\pi\pi}^3 F_\pi^2}{2 M_\rho^4} D \nonumber\\
\delta_3 &=& \frac{g_{\rho\pi\pi} C}{M_\rho^4 Z_\pi^4}
\left(1-2 M_\rho^2 Z_\pi^2 \left(2-Z_\pi^2\right)\frac{F}{C}\right)
\end{eqnarray}

Note that in vacuum, when the $\pi a_1$ loop is not included in the $\rho$
self-energy calculation, the $\alpha_i$ couplings need to be adjusted
to preserve gauge symmetry. The result is
$\alpha_1 = -g_{\rho\pi\pi}^2/2$ and $\alpha_i =0$ for $i \neq 1$.

\begin{figure*}[t!]
  \centering
	\includegraphics[width=.8\textwidth]{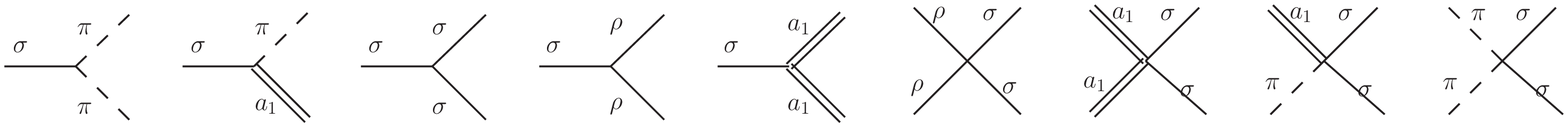}
\caption{Tree level vertices unique to the linear realization
 used in calculations in this paper.}
\label{fig:appvertexlin}
\end{figure*}

For the linear realization, the same vertices
are pertinent, though their Lagrangian structure and
couplings are slightly modified from their non-linear
counter-parts. The Lagrangian terms, followed by the
couplings, are given by
\begin{widetext}
\begin{eqnarray}
\mathcal{L}_{\rho\pi\pi}^{\rm Lin} &=& g_{\rho\pi\pi} \,\rho_\mu \cdot \pi \times \partial^\mu
\pi+g_{\rho\pi\pi}^{(3)} \, \rho_{\mu\nu} \cdot \partial^\mu \pi \times \partial^\nu \pi \\
\mathcal{L}_{\rho\pi a_1}^{\rm Lin} &=& \bar{C}\, \rho_\mu \cdot \pi \times a^\mu + \bar{D} \,
\rho_{\mu\nu} \cdot \pi \times a^{\mu\nu} + \bar{E} \, \rho_\mu \cdot \partial_\nu
\pi \times a^{\mu\nu} + \bar{F} \, \rho_{\mu\nu} \cdot \partial^\mu \pi \times a^\nu \\
\mathcal{L}_{\rho\rho\pi\pi}^{\rm Lin} &=& \bar{\alpha}_1 \left( \pi\cdot\rho_\mu \,
\pi\cdot\rho^\mu-\pi\cdot\pi \, \rho_\mu\cdot \rho^\mu \right)
+\bar{\alpha}_2 \left(2\pi\cdot\rho_{\mu\nu} \,\pi\cdot\rho^{\mu\nu}-\pi\cdot\pi \,
\rho_{\mu\nu}\cdot\rho^{\mu\nu}\right)\\
&&+ \bar{\alpha}_3 \left(\partial_\mu \pi\cdot\rho^{\mu\nu} \,\pi\cdot\rho_\nu\right)
 +\bar{\alpha}_4
\left(\partial_\mu \pi\cdot\rho_\nu \,\pi\cdot\rho^{\mu\nu}\right)\nonumber\\
&&+\bar{\alpha}_5\, \partial_\mu\pi\cdot\rho_\nu \,\partial^\nu \pi\cdot\rho^\mu
+\bar{\alpha}_6\, \partial_\mu\pi\cdot\rho^\mu \, \partial_\nu \pi\cdot\rho^\nu
+\bar{\alpha}_7 \left(\rho_\mu \cdot \rho^{\mu\nu}\, \pi\cdot\partial_\nu\pi\right) \nonumber \\
&&+\bar{\alpha}_8 \left(\partial_\mu \pi\cdot\partial^\mu  \pi \,\rho_\nu \cdot
\rho^\nu-\partial_\mu\pi\cdot \partial_\nu \pi \,\rho^\mu\cdot\rho^\nu-
\partial_\mu \pi\cdot\rho_\nu \, \partial^\mu \pi \cdot \rho^\nu\right)\nonumber\\
\mathcal{L}_{a_1a_1\pi\pi}^{\rm Lin} &=& \bar{\beta}_1 \left( \pi\cdot a_\mu \, \pi\cdot
 a^\mu \right)
+\bar{\beta}_2 \left(2\pi\cdot a_{\mu\nu}\, \pi\cdot a^{\mu\nu}-\pi\cdot\pi \,
a_{\mu\nu}\cdot a^{\mu\nu}\right)\\
&&+ \bar{\beta}_3 \left(\partial_\mu \pi\cdot a^{\mu\nu}\, \pi\cdot a_\nu
\right)+\bar{\beta}_4\,
\partial_\mu\pi\cdot a_\nu \,\partial^\nu \pi\cdot a^\mu \nonumber \\
&&+\bar{\beta}_5\, \partial_\mu\pi\cdot a^\mu \,\partial_\nu \pi\cdot a^\nu
+\bar{\beta}_6 \left(a_\mu \cdot a^{\mu\nu}\, \pi\cdot\partial_\nu\pi
\right) \nonumber \\
&&+\bar{\beta}_7 \left(\partial_\mu \pi\cdot\partial^\mu \pi \,a_\nu \cdot a^\nu-
\partial_\mu\pi\cdot \partial_\nu \pi \,a^\mu\cdot a^\nu-\partial_\mu \pi
\cdot a_\nu \,\partial^\mu \pi \cdot a^\nu\right)\nonumber \\
\mathcal{L}_{\pi\pi\pi\pi}^{\rm Lin} &=& \bar{\gamma}_1 \left( \pi\cdot\partial_\mu
\pi \,\pi\cdot\partial^\mu\pi\right)
+ \bar{\gamma}_2 \left(\partial_\mu\pi\cdot\partial_\nu\pi\, \partial^\mu\pi
\cdot\partial^\nu\pi - \partial_\mu\pi\cdot\partial^\mu\pi\,
\partial_\nu\pi\cdot\partial^\nu\pi\right)\\
&&+ \bar{\gamma}_3 \,\pi\cdot\pi \,\pi\cdot\pi \\
\mathcal{L}_{a_1 \pi\pi\pi}^{\rm Lin}&=& \bar{\delta}_1 \left(a_\mu \cdot\pi \,\partial^\mu\pi
\cdot\pi\right)
+\bar{\delta}_2 \left(a_{\mu\nu} \cdot \partial^\mu\pi \,\partial^\nu\pi\cdot
\pi\right) \\
&&+\bar{\delta}_3 \left(a_\mu\cdot\partial^\mu\pi \,\partial_\nu\pi\cdot\partial^\nu
\pi-a_\mu\cdot\partial_\nu\pi \,\partial^\mu\pi\cdot\partial^\nu \pi\right) \nonumber
\end{eqnarray}
\end{widetext}

\begin{eqnarray}
\bar{C} &=& - g_{\rho\pi\pi}^2 \sigma_0 \frac{M_{a_1}}{M_\rho}\\
\bar{D} &=& \frac{\sigma_0 M_\rho}{F_\pi^2 M_{a_1}} \lambda_1 \\
\bar{E} &=& -\frac{g_{\rho\pi\pi}^2 \sigma_0}{M_\rho M_{a_1}} \\
\bar{F} &=& -2 g_{\rho\pi\pi} \sigma_0 \frac{M_{a_1}}{M_\rho}
\left(g_{\rho\pi\pi}^{(3)} + \frac{g_{\rho\pi\pi}}{2 M_\rho^2}\right)
\end{eqnarray}

\begin{eqnarray}
\bar{\alpha}_1 &=& -\frac{1}{2} g_{\rho\pi\pi}^2 Z_\pi^2\\
\bar{\alpha}_2 &=& -\frac{1}{4 F_\pi^2}
\left(1-\frac{M_\rho^2 Z_\pi^2}{M_{a_1}^2}\right)\\
\bar{\alpha}_3 &=& \frac{8 g_{\rho\pi\pi}^2 \sigma_0^2}
{M_\rho^2 Z_\pi^2} \, \bar{\alpha}_2 \\
\bar{\alpha}_4 &=& 2 Z_\pi^2 \bar{\alpha}_5 + 2\frac{M_{a_1}^2}
{M_\rho^2} \bar{\alpha}_8 \\
\end{eqnarray}

\begin{eqnarray}
\bar{\alpha}_5 &=& - g_{\rho\pi\pi} g_{\rho\pi\pi}^{(3)} \\
\bar{\alpha}_6 &=& \bar{\alpha}_8 - \bar{\alpha}_5 \\
\bar{\alpha}_7 &=& \bar{\alpha}_3 + \bar{\alpha}_4 \\
\bar{\alpha}_8 &=& -\frac{1}{2} \bar{E}^2
\end{eqnarray}

\begin{eqnarray}
\bar{\beta}_1 &=& -\bar{\alpha_1} \frac{M_{a_1}^2}
{M_\rho^2 Z_\pi^2} \\
\bar{\beta}_2 &=& -\bar{\alpha_2} \frac{M_{a_1}^2}
{M_\rho^2 Z_\pi^2} \\
\bar{\beta}_3 &=& -\left(\bar{\alpha}_3 + \bar{\alpha}_4\right)
\frac{M_{a_1}^2}{M_\rho^2 Z_\pi^2} \\
\bar{\beta}_4 &=& \left(\bar{\alpha_5} +\bar{\alpha}_4
\frac{g_{\rho\pi\pi}^4 \sigma_0^4}{2 M_\rho^4 Z_\pi^2}
\right) \frac{M_{a_1}^2}{M_\rho^2 Z_\pi^2} \\
\bar{\beta}_5 &=& \bar{\beta}_7 - \bar{\beta}_4\\
\bar{\beta}_6 &=& - \left(\bar{\alpha}_3 - \bar{\alpha}_4
g_{\rho\pi\pi}^2 \right)
\frac{M_{a_1}^2}{M_\rho^2 Z_\pi^2} \\
\bar{\beta}_7 &=& \left(\bar{\alpha}_8 \frac{M_{a_1}^2}
{M_\rho^2 Z_\pi^2} + \bar{\alpha}_4 \frac{g_{\rho\pi\pi}^2 \sigma_0^2}
{M_\rho^2 Z_\pi^2}\right) \frac{M_{a_1}^2}{M_\rho^2 Z_\pi^2} \\
\end{eqnarray}

\begin{eqnarray}
\bar{\gamma}_1 &=& \frac{g_{\rho\pi\pi}^2}{2 M_\rho^2}
\left(Z_\pi^2 -1 \right)\\
\bar{\gamma}_2 &=& -\frac{g_{\rho\pi\pi}^3 g_{\rho\pi\pi}^{(3)}
\sigma_0^2}{M_\rho^4 Z_\pi^2} -\frac{g_{\rho\pi\pi}^6
\sigma_0^4}{4 M_\rho^2 Z_\pi^4}\left(1+Z_\pi^2\right)\\
\bar{\gamma}_3 &=& - \frac{1}{4} \lambda Z_\pi^4
\end{eqnarray}
\begin{eqnarray}
\bar{\delta}_1 &=& \frac{g_{\rho\pi\pi}^3 \sigma_0 M_{a_1}}{M_\rho^3} \\
\bar{\delta}_2 &=& -\frac{g_{\rho\pi\pi}}{M_\rho^2}
\left(\bar{F}-\bar{E} \frac{M_{a_1}^2}{M_\rho^2 Z_\pi^2}\right)\\
&&-\frac{2\sqrt{2} g_{\rho\pi\pi}^3 \sigma_0^3}
{F_\pi M_\rho^4 Z_\pi^3} \bar{D} \nonumber\\
\bar{\delta}_3 &=& \frac{g_{\rho\pi\pi} \bar{C}}
{M_\rho^4 Z_\pi^4}
\left(1-2 M_\rho^2 Z_\pi^2 \left(2-Z_\pi^2\right)
\frac{\bar{F}}{\bar{C}}\right)
\end{eqnarray}

Note that the 3-point vertices are identical between the
two realizations but are written out here to show their
explicit dependence on $\sigma_0$.  In addition, vertices incorporating
the $\sigma$ field are needed which are depicted in Fig.~\ref{fig:appvertexlin}.
The Langrangian terms for these new vertices and their couplings
are given by:

\begin{widetext}
\begin{eqnarray}
\mathcal{L}_{\sigma\pi\pi} &=& a_1\, \sigma \,\pi \cdot \pi + a_2\, \sigma \,\partial_\mu \pi
\cdot \partial^\mu \pi + a_3\, \partial_\mu \sigma \,\pi \cdot \partial^\mu \pi\\
\mathcal{L}_{\sigma\pi a_1} &=& b_1 \, \partial_\mu \sigma \, \pi \cdot a^\mu
+ b_2 \, \sigma \, \partial_\mu \pi \cdot a^\mu + b_3 \partial_\mu \sigma \,
\partial_\nu \pi \cdot a^{\mu\nu}\\
\mathcal{L}_{\sigma\sigma\sigma} &=& -\lambda \sigma_0 \,\sigma^3\\
\mathcal{L}_{\sigma\rho\rho} &=& c_1 \, \sigma \, \rho_{\mu\nu} \cdot \rho^{\mu\nu}\\
\mathcal{L}_{\sigma a_1 a_1} &=& d_1\,  \sigma \, a_\mu \cdot a^\mu +d_2 \, \sigma \, a_{\mu\nu}
\cdot a^{\mu\nu} + d_3 \, \partial_\mu \sigma \, a_\nu \cdot a^{\mu\nu}\\
\mathcal{L}_{\rho\rho\sigma\sigma} &=& e_1 \, \sigma^2 \, \rho_{\mu\nu} \cdot \rho^{\mu\nu} \\
\mathcal{L}_{a_1 a_1 \sigma\sigma} &=& f_1 \, \sigma^2 \, a_\mu \cdot a^\mu + f_2 \,
\sigma^2 \, a_{\mu\nu} \cdot a^{\mu\nu} + f_3 \, \sigma \partial_\mu \sigma \, a_\nu \cdot
a^{\mu\nu}\\\
\mathcal{L}_{a_1 \pi\sigma\sigma} &=& g_1\, \sigma^2\, \partial_\mu \pi \cdot a^\mu
+ g_2 \, \sigma \partial_\mu \sigma \, \partial_\nu \pi \cdot a^{\mu\nu}\\
\mathcal{L}_{\pi\pi\sigma\sigma} &=& h_1\, \sigma^2 \pi \cdot \pi + h_2\, \sigma^2
\partial_\mu \pi \cdot \partial^\mu \pi\\
\end{eqnarray}
\end{widetext}

\begin{eqnarray}
a_1 &=& -Z_\pi^2 \lambda \sigma_0 \\
a_2 &=& -\frac{g_{\rho\pi\pi}^2 \sigma_0}{Z_\pi^2 M_\rho^2}\\
a_3 &=& -\frac{g_{\rho\pi\pi}^2 \sigma_0}{M_\rho^2}
\end{eqnarray}
\begin{eqnarray}
b_1 &=& g_{\rho\pi\pi} \frac{M_{a_1}}{M_\rho}\\
b_2 &=& g_{\rho\pi\pi} \frac{M_{a_1}}{M_\rho} \left(1-\frac{2}{Z_\pi^2}\right)\\
b_3 &=& -\frac{2 M_{a_1}}{M_\rho} \lambda_2
\end{eqnarray}
\begin{eqnarray}
c_1 &=& \frac{\sigma_0 M_\rho^2}{2 F_\pi^2  M_{a_1}^2}
\lambda_1\\
d_1 &=& \frac{g_{\rho\pi\pi}^2 \sigma_0 M_{a_1}^2}{Z_\pi^2 M_\rho^2}\\
d_2 &=& -\frac{\sigma_0}{2 F_\pi^2 Z_\pi^2} \lambda_1\\
d_3 &=& \frac{2 g_{\rho\pi\pi} \sigma_0 M_{a_1}^2}{M_\rho^2}
\lambda_2\\
\end{eqnarray}
\begin{eqnarray}
e_1 &=& \frac{M_\rho^2}{4 F_\pi^2 M_{a_1}^2 Z_\pi^4} \lambda_1\\
f_1 &=& \frac{g_{\rho\pi\pi}^2 M_{a_1}^2}{2 Z_\pi^2 M_\rho^2} \\
f_2 &=& -\frac{1}{4 F_\pi^2 Z_\pi^2} \lambda_1\\
f_3 &=& \frac{2 g_{\rho\pi\pi} M_{a_1}^2}{M_\rho^2} \lambda_2\\
\end{eqnarray}
\begin{eqnarray}
g_1 &=& \frac{g_{\rho\pi\pi}^3 \sigma_0 M_{a_1}}{M_\rho^3 Z_\pi^2}\\
g_2 &=& \frac{2 g_{\rho\pi\pi}^2 \sigma_0 M_{a_1}}{M_\rho^3}
\lambda_2\\
h_1 &=& -\frac{1}{2} \lambda Z_\pi^2\\
h_2 &=& \frac{g_{\rho\pi\pi}^4 \sigma_0^2}{2 M_\rho^4 Z_\pi^2}
\end{eqnarray}

with
\begin{eqnarray}
\lambda_1 &=& 1-\frac{M_{a_1}^2}{M_\rho^2 Z_\pi^2} ,\\
\lambda_2 &=& g_{\rho\pi\pi}^{(3)} + \frac{g_{\rho\pi\pi}^3
\sigma_0^2}{2M_\rho^4 Z_\pi^2} \ .
\end{eqnarray}

\section{Counter-term contributions to each self-energy}
\label{sec:counter}

In Sec.~\ref{sec:calc}, we presented the MYM Lagrangian
for the counter-terms. In this appendix, we quote the counter-term
contribution to each self-energy having expanded the Lagrangian
in terms of the physical fields. The expressions are written for the
linear realization to indicate their dependence on $\sigma_0$, but
they are the same for the non-linear realization upon replacing
$\sigma_0 = F_\pi Z_\pi/\sqrt{2}$.

\begin{widetext}
\begin{equation}
\begin{split}
\Sigma_\rho^{\rm (CT)}(p^2) &= \kappa_V^2 \left( -p^2 \left(\delta Z_A^{(2)} -
\delta \gamma^{(2)} \frac{2\sigma_0^2}{F_\pi^2 Z_\pi^2} \right)
+ p^4 \left(\delta Z_A^{(4)} + \frac{1}{2} \delta \gamma^{(4)}
\frac{2\sigma_0^2}{F_\pi^2 Z_\pi^2}\right)
+p^6 \left(\delta Z_A^{(6)} + \frac{1}{2} \delta \gamma^{(6)}
 \frac{2\sigma_0^2}{F_\pi^2 Z_\pi^2}\right)\right) ,\\
\Sigma_{a_1}^{\rm (CT)}(p^2) & = - p^2 \left(\kappa_A^2\left(\delta Z_A^{(2)} +
\delta \gamma^{(2)} \frac{2\sigma_0^2}{F_\pi^2 Z_\pi^2}\right)
- \frac{g_{\rho\pi\pi}^2 \sigma_0^2 M_{a_1}^2}{M_\rho^2 Z_\pi^2}
\delta Z_\pi^{(2)}\right) \\&\quad \quad \quad \quad
+ p^4 \left(\kappa_A^2\left(\delta Z_A^{(4)}
 - \frac{1}{2}
\delta \gamma^{(4)}\frac{2\sigma_0^2}{F_\pi^2 Z_\pi^2}\right)
+ \frac{g_{\rho\pi\pi}^2 \sigma_0^2 M_{a_1}^2}{2 M_\rho^2 Z_\pi^2}
\delta Z_\pi^{(4)}\right)
  + p^6 \kappa_A^2 \left(\delta Z_A^{(6)} - \frac{1}{2}
\delta \gamma^{(6)}\frac{2\sigma_0^2}{F_\pi^2 Z_\pi^2}\right) ,\\
\Sigma_{a_1}^{L {\rm (CT)}}(p^2) &= -\frac{g_{\rho\pi\pi}^2 \sigma_0^2 M_{a_1}^2}
{M_\rho^2 Z_\pi^2} \delta Z_\pi^{(2)} - \frac{g_{\rho\pi\pi}^2\sigma_0^2 M_{a_1}^2}
{M_\rho^2 Z_\pi^2} p^2 \delta Z_\pi^{(4)} ,\\
\Sigma_{a \pi}^{\rm (CT)}(p^2) &= -\frac{g_{\rho\pi\pi}\sigma_0 M_{a_1}}
{M_\rho Z_\pi^2} \delta Z_\pi^{(2)} - \frac{g_{\rho\pi\pi} \sigma_0 M_{a_1}}
{M_\rho Z_\pi^2} p^2 \delta Z_\pi^{(4)} ,\\
\Sigma_{\pi\pi}^{\rm (CT)}(p^2) &= - \frac{Z_\pi^2}{2} \delta m_\pi^2
-\frac{p^2}{Z_\pi^2} \delta Z_\pi^{(2)}
-\frac{p^4}{Z_\pi^2} \delta Z_\pi^{(4)}  .
\end{split}
\end{equation}
\end{widetext}

\section{Vertex correction diagram discussion}
\label{sec:VCdisc}
In order to preserve the chiral symmetry in the $a_1$ self-energy
in the presence of a broad $\rho$, the vertex correction (VC)
diagrams of Fig.~\ref{fig:vertexdia} along with couplings
modified from their Lagrangian values are needed.
This is because the broadening of the $\rho$ amounts
to a partial resummation of a subset of diagrams. In this appendix,
the details on how these couplings are determined
for both the longitudinal and transverse channels are spelled out.

\subsection{Longitudinal channel}
\begin{figure}[t!]
  \centering
	\includegraphics[width=.45\textwidth]{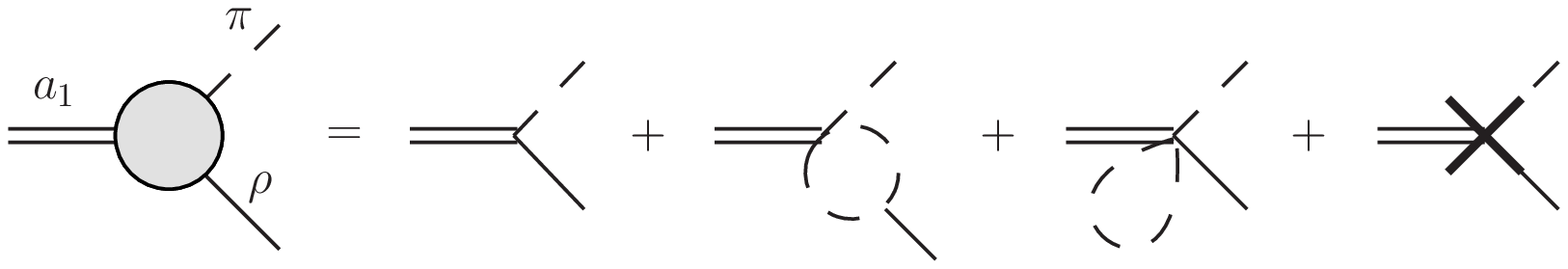}
    \includegraphics[width=.45\textwidth]{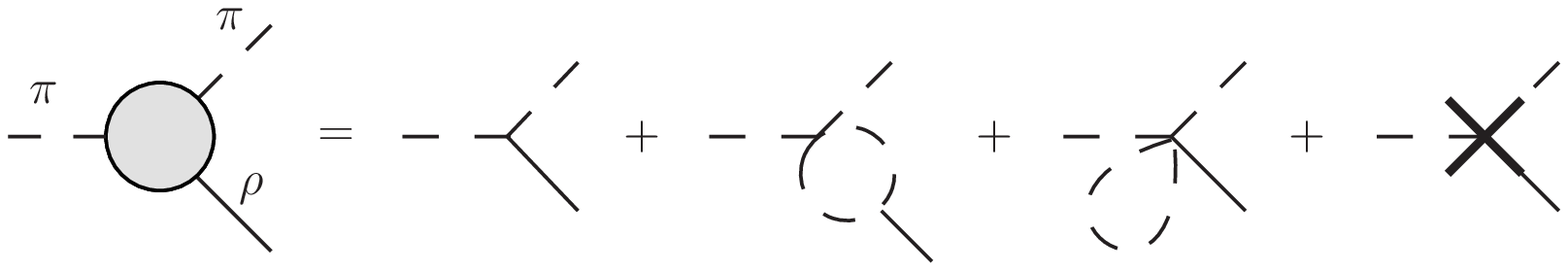}
\caption{Diagrams depicting the $a\pi\rho$ and $\pi\pi\rho$ vertices
expanded to 1-loop order;
the cross represents counter-term contributions.}
\label{fig:3ptvertexdia}
\end{figure}

\begin{figure}[t]
  \centering
	\includegraphics[width=.45\textwidth]{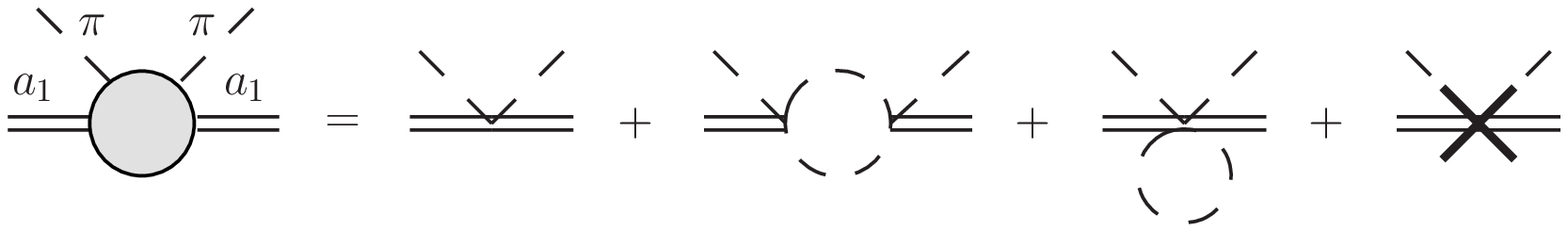}
    \includegraphics[width=.45\textwidth]{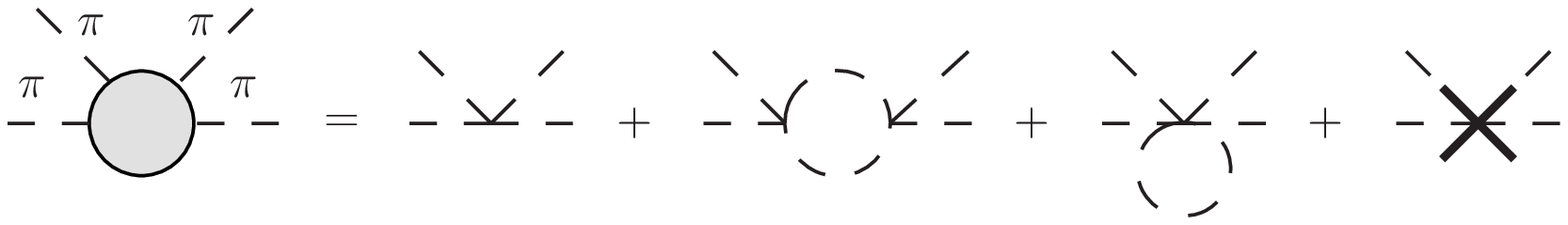}
\caption{Diagrams depicting the $aa\pi\pi$ and $\pi\pi\pi\pi$ vertices
expanded to 1-loop order;
the cross represents counter-term contributions.}
\label{fig:4ptvertexdia}
\end{figure}

The critical features in deciding whether chiral symmetry
is maintained are the chiral limit relations between
$\Sigma_a^L$, $\Sigma_{a\pi}$,
and $\Sigma_{\pi\pi}$ as given by Eq.~(\ref{eq:pcacrel}).
For the discussion here,
we will examine the $\pi\rho$ loop in each self-energy.
With the inclusion of the broad $\rho$, these relationships are
not maintained. Therefore we must expand the $a\pi\rho$ and
the $\pi\rho\pi$ vertices in the basic $\pi\rho$ loop
(first diagram of bottom row in Fig.~\ref{fig:dia1})
to 1-loop order. This is depicted diagrammatically
for the non-linear realization in Fig.~\ref{fig:3ptvertexdia}
where the first diagram is the tree level (point-like) interaction
and the latter three together form the VC.
Using these modified vertices in the $\pi\rho$ loop leads to new
diagrams indicated in Fig.~\ref{fig:vertexdia}.
A similar expansion into a tree level and a loop contribution can be
made for the four-point vertices of $aa\pi\pi$ and $\pi\pi\pi\pi$
which is shown for the non-linear realization
in Fig.~\ref{fig:4ptvertexdia},  leading to the 3$\pi$ loop
of Fig.~\ref{fig:vertexdia}. Note that the VCs have
the same structure as the $\rho$ self-energy,
\ie, a $\pi\pi$ loop, a $\pi$ tadpole, and a counter-term.

With the inclusion of the VCs the $\pi\rho$ loop in the self-energies
$\Sigma_a^L$, $\Sigma_{a\pi}$, and $\Sigma_{\pi\pi}$ can be expressed
as
\begin{widetext}
\begin{equation}
\begin{split}
\label{eq:longa1}
\Sigma_a^{L (\pi\rho)} (p^2) & = 2i
\int\limits_0^{\Lambda_{\rm cut}^2} \frac{-1}{\pi}\, dM^2 \int \frac{d^4k}{(2\pi)^4} \,
{\rm Im}\!\left[D_\rho(M^2) V_a^2(M^2)+\frac{1}{2} V_{aa}^{3\pi}(M^2)\right]
\frac{1}{k^2 - M_\pi^2} \frac{1}{q^2-M^2}
\left(-1+\frac{\left(p\cdot q\right)^2}{p^2 M^2}\right) \ ,
\\
\Sigma_{a\pi}^{(\pi\rho)} (p^2) & = -2i \int\limits_0^{\Lambda_{\rm cut}^2} \frac{-1}{\pi}\,
 dM^2 \int \frac{d^4k}{(2\pi)^4} \,
{\rm Im}\!\left[D_\rho(M^2) V_a(M^2)V_p(M^2)+\frac{1}{2} V_{ap}^{3\pi}(M^2)\right]
\frac{1}{k^2 - M_\pi^2} \frac{1}{q^2-M^2}
\left(-1+\frac{\left(p\cdot q\right)^2}{p^2 M^2}\right) ,
\\
\Sigma_{\pi\pi}^{(\pi\rho)} (p^2) & = 2i \int\limits_0^{\Lambda_{\rm cut}^2} \frac{-1}{\pi}\,
 dM^2 \int \frac{d^4k}{(2\pi)^4} \,
{\rm Im}\!\left[D_\rho(M^2) V_p^2(M^2)+\frac{1}{2} V_{pp}^{3\pi}(M^2)\right]
\frac{1}{k^2 - M_\pi^2} \frac{1}{q^2-M^2}
\left(-p^2+\frac{\left(p\cdot q\right)^2}{M^2}\right) ,
\end{split}
\end{equation}
\end{widetext}
where $q^\mu = p^\mu-k^\mu$, and a dispersion relation for the $\rho$ propagator
has been used including the hard cut-off referred to in the main text.
The 3-point vertices are represented by
$V_a$ and $V_p$ and are decomposed into tree level and VC contributions
as
\begin{equation}
\begin{split}
V_a &= V_a^{\rm tree} + V_a^{VC} ,\\
V_p &= V_p^{\rm tree} + V_p^{VC} ,
\end{split}
\end{equation}
where
\begin{equation}
\begin{split}
V_a^{\rm tree} (q^2) &= \left(C + F q^2\right) ,\\
V_p^{\rm tree} (q^2) &= 2\left(g_{\rho\pi\pi} + g_{\rho\pi\pi}^{(3)} q^2\right) ,
\end{split}
\end{equation}
for the non-linear realization, while the results for
the linear version are found by replacing $C$ and $F$ with $\bar{C}$
and $\bar{F}$.
These couplings are given by the equations in Appendix~\ref{sec:appVert}.
The 4-point vertices, $V^{3\pi}$, only have the VC contribution and are
labeled by the external mesons.

In order to ensure that these expressions satisfy the relations
of Eq.~(\ref{eq:pcacrel}), the goal is to express the longitudinal
self-energies as a coefficient times an integral which
is the same between the different channels.
Note that in the absence of the $\rho$ self-energy and the VCs
this is obtainable and the PCAC relations are realized.

To compensate for the $\rho$ self-energy in the propagator,
the VCs of the 3- and 4-point vertices need to be
proportional to $\Sigma_\rho$\footnote{This restriction
can be relaxed if the deviation still respects the relations
for chiral symmetry. Nevertheless, in this work we will follow the more
stringent requirement.}.
One can show that the $\pi\pi$
loop contribution to the VCs is related to the $\pi\pi$
loop of the $\rho$ self-energies, $\Sigma_\rho^{T (\pi\pi)}$, as
\begin{equation}
\label{eq:vcpipi}
\begin{split}
V_a^{VC (\pi\pi)}(q^2) & = -\frac{3}{4} \frac{2 \delta_1 - \delta_3 q^2}
{g_{\rho\pi\pi} + g_{\rho\pi\pi}^{(3)} q^2} \Sigma_\rho^{T (\pi\pi)} (q^2) ,\\
V_p^{VC (\pi\pi)}(q^2) & = -3 \frac{2 \gamma_1 + \gamma_2 q^2}
{g_{\rho\pi\pi} + g_{\rho\pi\pi}^{(3)} q^2} \Sigma_\rho^{T (\pi\pi)}(q^2) ,\\
V_{aa}^{3\pi (\pi\pi)} (q^2)& = \frac{9}{8} \left(\frac{2 \delta_1 - \delta_3 q^2}
{g_{\rho\pi\pi} + g_{\rho\pi\pi}^{(3)} q^2}\right)^2 \Sigma_\rho^{T (\pi\pi)}(q^2) , \\
V_{ap}^{3\pi (\pi\pi)} (q^2)& = \frac{9}{2}
\frac{(2 \delta_1 - \delta_3 q^2)(2 \gamma_1 + \gamma_2 q^2)}
{(g_{\rho\pi\pi} + g_{\rho\pi\pi}^{(3)} q^2)^2}\Sigma_\rho^{T (\pi\pi)}(q^2) , \\
V_{pp}^{3\pi (\pi\pi)}(q^2) & = 18 \left(\frac{2 \gamma_1 + \gamma_2 q^2}
{g_{\rho\pi\pi} + g_{\rho\pi\pi}^{(3)} q^2}\right)^2 \Sigma_\rho^{T (\pi\pi)}(q^2),
\end{split}
\end{equation}
for the non-linear realization and
\begin{equation}
\label{eq:vcpipilin}
\begin{split}
V_{a {\rm Lin}}^{VC (\pi\pi)} (q^2)& = -\frac{1}{4} \frac{2 \bar{\delta}_1 - 3\bar{\delta}_3 q^2}
{g_{\rho\pi\pi} + g_{\rho\pi\pi}^{(3)} q^2} \Sigma_\rho^{T (\pi\pi)} (q^2) ,\\
V_{p {\rm Lin}}^{VC (\pi\pi)}(q^2) & = - \frac{2 \bar{\gamma}_1 + 3\bar{\gamma}_2 q^2}
{g_{\rho\pi\pi} + g_{\rho\pi\pi}^{(3)} q^2} \Sigma_\rho^{T (\pi\pi)}(q^2) ,\\
V_{aa {\rm Lin}}^{3\pi (\pi\pi)} (q^2)& = \frac{1}{8} \left(\frac{2 \bar{\delta}_1 - 3\bar{\delta}_3 q^2}
{g_{\rho\pi\pi} + g_{\rho\pi\pi}^{(3)} q^2}\right)^2 \Sigma_\rho^{T (\pi\pi)}(q^2) ,\\
V_{ap {\rm Lin}}^{3\pi (\pi\pi)} (q^2)& = \frac{1}{2}
\frac{(2 \bar{\delta}_1 - 3\bar{\delta}_3 q^2)(2 \bar{\gamma}_1 + 3\bar{\gamma}_2 q^2)}
{(g_{\rho\pi\pi} + g_{\rho\pi\pi}^{(3)} q^2)^2}\Sigma_\rho^{T (\pi\pi)} (q^2) ,\\
V_{pp {\rm Lin}}^{3\pi (\pi\pi)}(q^2) & = 2 \left(\frac{2 \bar{\gamma}_1 + 3\bar{\gamma}_2 q^2}
{g_{\rho\pi\pi} + g_{\rho\pi\pi}^{(3)} q^2}\right)^2 \Sigma_\rho^{T (\pi\pi)}(q^2),
\end{split}
\end{equation}
for the linear realization.
Again the couplings are as defined above.
These results are then generalized to combine all contributions
to the VCs through a $\rho$ self-energy which encompasses a $\pi\pi$
loop, a $\pi$ tadpole, and counter-term contributions, \eg,
\begin{equation}
V_a^{VC} (q^2) = -\frac{3}{4} \frac{2 \delta_1 - \delta_3 q^2}
{g_{\rho\pi\pi} + g_{\rho\pi\pi}^{(3)} q^2} \Sigma_\rho^{T} (q^2),
\end{equation}
for the first expression in Eq.~(\ref{eq:vcpipi}).
This is obtained by choosing the appropriate couplings
associated with the $\pi$ tadpole and counter-terms. The
apparent energy dependence of the coupling will be justified
{\it a posteriori}.

Using these relations between the VCs and $\Sigma_\rho^T$ allows
for the expressions in Eq.~(\ref{eq:longa1}) to be simplified.
Furthermore, by choosing the coupling parameters to be
\begin{equation}
\label{eq:vcpara}
\begin{split}
\delta_1 & = -\frac{2 g_{\rho\pi\pi}C}{3 M_\rho^2} , \quad
\delta_3  = \frac{4 g_{\rho\pi\pi}^{(3)} C}{3 M_\rho^2} \ ,
\\
\gamma_1 & = -\frac{g_{\rho\pi\pi}^2}{3 M_\rho^2} , \quad
\gamma_2  = -\frac{2 g_{\rho\pi\pi}g_{\rho\pi\pi}^{(3)}}{3 M_\rho^2} \ ,
\\
\bar{\delta}_1 &= -\frac{2 g_{\rho\pi\pi} \bar{C}}{M_\rho^2}, \quad
\bar{\delta}_3 = \frac{4 g_{\rho\pi\pi}^{(3)} \bar{C}}{3M_\rho^2} \ ,
\\
\bar{\gamma}_1 & = -\frac{g_{\rho\pi\pi}^2}{M_\rho^2} \ , \quad
\bar{\gamma}_2  = -\frac{2 g_{\rho\pi\pi}g_{\rho\pi\pi}^{(3)}}{3M_\rho^2} \ ,
\end{split}
\end{equation}
which are different from their Lagrangian values,
these expressions have the desired form of a coefficient times
a generic integral. This integral is different from the case
of a zero-width $\rho$, but it is still the same for the different channels.
With these choices of parameters, we can also verify that Eq.~(\ref{eq:pcacrel})
is satisfied, thus chiral symmetry is preserved. Returning to
the proportionality factor between the VCs and $\Sigma_\rho^T$, we see
that for the given choice of parameters in Eq.~(\ref{eq:vcpara}),
this factor becomes energy
independent which justifies the generalization that we postulated.

\subsection{Transverse channel}
For the transverse channel,
$\Sigma_a^T$ has a more complicated form,  given by
\begin{widetext}
\begin{equation}
\begin{split}
\Sigma_a^{T (\pi\rho)} (p^2) &= 2 i \int -\frac{1}{\pi} dM^2
\int \frac{d^4k}{(2\pi)^4} \left( {\rm Im}\left[ D_\rho(M^2)
(V_a^{T 1}(M^2))^2+\frac{1}{2} V_1^{3\pi}(M^2)\right] \frac{1}{k^2-M_\pi^2}
\frac{1}{q^2- M^2}\right.\\
& + {\rm Im} \left[2D_\rho(M^2) V_a^{T 1}(M^2) V_a^{T 2}(M^2)
+\frac{1}{2} V_2^{3\pi}(M^2)\right] \frac{p\cdot q}{k^2-M_\pi^2}
\frac{1}{q^2-M^2} \\
&+ {\rm Im} \left[ D_\rho(M^2) (V_a^{T 2}(M^2))^2
+\frac{1}{2} V_3^{3\pi}(M^2)\right]
\frac{(p\cdot q)^2}{k^2 -M_\pi^2}\frac{1}{q^2 - M^2}\\
&+ {\rm Im} \left[D_\rho(M^2)\left((V_a^{T 1}(M^2))^2
-(V_a^{T 2}(M^2))^2 p^2 M^2\right) + \frac{1}{2} V_4^{3\pi}(M^2)\right]
\frac{1}{k^2-M_\pi^2}\frac{1}{q^2 -M^2} \frac{1}{3 M^2}
\left(-q^2+\frac{(p\cdot q)^2}{p^2}\right)\Bigg).
\end{split}
\end{equation}
\end{widetext}
where $q^\mu = p^\mu - k^\mu$.
Note that when evaluating this integral using dimensional regulation,
the 1/3 in the last term becomes a 1/($\delta$-1) where
$\delta$ is the number of dimensions.
The vertex functions, $V_a^{T 1}$ and $V_a^{T 2}$, can again
be decomposed into a tree level part and a VC part as
\begin{equation}
\begin{split}
V_a^{T 1} &= V_1^{\rm tree} + V_1^{VC} ,\\
V_a^{T 2} &= V_2^{\rm tree} + V_2^{VC} .
\end{split}
\end{equation}
The tree level terms are expressed as
\begin{equation}
\begin{split}
V_1^{\rm tree} (q^2) &= C-E p^2 + F q^2 ,\\
V_2^{\rm tree} (q^2) &= 2 D + E - F ,
\end{split}
\end{equation}
for the non-linear realization, with the expressions
for the linear realization found by replacing $C$, $D$,
$E$, and $F$ with $\bar{C}$, $\bar{D}$, $\bar{E}$,
and $\bar{F}$. The functions $V^{3\pi}$ are associated
with the $3\pi$ diagram's contribution and are entirely
a VC contribution.

To determine the functional form of the VCs, we utilize
the same procedure as described above for the longitudinal
channel of relating the VC loops to $\Sigma_\rho^T$
with the proportionality factor guided by the result from the $\pi\pi$
loop. To make the connection with chiral symmetry, we use the same values
for the couplings, $\delta_1$, $\delta_3$, $\gamma_1$, $\gamma_2$ and the
barred versions, as was determined in the longitudinal mode. Additionally,
the transverse self-energy depends on the parameter $\delta_2$, which is not
fixed by the previous considerations, but is chosen to be 0.
This leads to the following expressions for the vertex functions
in the non-linear realization,
\begin{equation}
\label{eq:transvc}
\begin{split}
V_1^{VC} (q^2) &= -\frac{3}{4} \frac{2 \delta_1 - \delta_3 q^2}
{g_{\rho\pi\pi} + g_{\rho\pi\pi}^{(3)} q^2} \Sigma_\rho^T (q^2) , \\
V_2^{VC} (q^2) &= -\frac{3}{4} \frac{\delta_3}
{g_{\rho\pi\pi} + g_{\rho\pi\pi}^{(3)} q^2} \Sigma_\rho^T (q^2) ,\\
V_1^{3\pi} (q^2) &= \frac{9}{8} \left(\frac{2 \delta_1 - \delta_3 q^2}
{g_{\rho\pi\pi} + g_{\rho\pi\pi}^{(3)} q^2}\right)^2 \Sigma_\rho^T(q^2) ,\\
V_2^{3\pi} (q^2) &= \frac{9}{4} \frac{\left(2 \delta_1 - \delta_3 q^2\right)\delta_3}
{\left(g_{\rho\pi\pi} + g_{\rho\pi\pi}^{(3)} q^2\right)^2} \Sigma_\rho^T(q^2) ,\\
V_3^{3\pi} (q^2) &= \frac{9}{8} \frac{\delta_3^2}
{\left(g_{\rho\pi\pi} + g_{\rho\pi\pi}^{(3)} q^2\right)^2} \Sigma_\rho^T(q^2) ,\\
V_4^{3\pi} (q^2) &= \frac{9}{8} \frac{\left(2 \delta_1 - \delta_3 q^2\right)^2
-\delta_3^2 q^2 p^2}
{\left(g_{\rho\pi\pi} + g_{\rho\pi\pi}^{(3)} q^2\right)^2} \Sigma_\rho^T(q^2), \\
\end{split}
\end{equation}
and in the linear realization,
\begin{equation}
\begin{split}
V_{1 {\rm Lin}}^{VC} (q^2) &= -\frac{1}{4} \frac{2 \bar{\delta}_1 - 3\bar{\delta}_3 q^2}
{g_{\rho\pi\pi} + g_{\rho\pi\pi}^{(3)} q^2} \Sigma_\rho^T (q^2) ,\\
V_{2 {\rm Lin}}^{VC} (q^2) &= -\frac{3}{4} \frac{\bar{\delta}_3}
{g_{\rho\pi\pi} + g_{\rho\pi\pi}^{(3)} q^2} \Sigma_\rho^T (q^2) ,\\
V_{1 {\rm Lin}}^{3\pi} (q^2) &= \frac{1}{8} \left(\frac{2 \bar{\delta}_1 - 3\bar{\delta}_3 q^2}
{g_{\rho\pi\pi} + g_{\rho\pi\pi}^{(3)} q^2}\right)^2 \Sigma_\rho^T(q^2) , \\
V_{2 {\rm Lin}}^{3\pi} (q^2) &= \frac{3}{4} \frac{\left(2 \bar{\delta}_1 - 3\bar{\delta}_3 q^2\right)\bar{\delta}_3}
{\left(g_{\rho\pi\pi} + g_{\rho\pi\pi}^{(3)} q^2\right)^2} \Sigma_\rho^T(q^2) ,\\
V_{3 {\rm Lin}}^{3\pi} (q^2) &= \frac{9}{8} \frac{\bar{\delta}_3^2}
{\left(g_{\rho\pi\pi} + g_{\rho\pi\pi}^{(3)} q^2\right)^2} \Sigma_\rho^T(q^2) ,\\
V_{4 {\rm Lin}}^{3\pi} (q^2) &= \frac{1}{8} \frac{\left(2 \bar{\delta}_1 - 3\bar{\delta}_3 q^2\right)^2
-9\bar{\delta}_3^2 q^2 p^2}
{\left(g_{\rho\pi\pi} + g_{\rho\pi\pi}^{(3)} q^2\right)^2} \Sigma_\rho^T(q^2). \\
\end{split}
\end{equation}

Lastly, we note that the broad $\rho$ and VCs modify the
$\rho\pi a_1$ coupling such that $\Sigma_a^T$ develops a
mass shift at low temperatures of $\mathcal{O}(T^2)$. To address this
issue, we exploit the freedom associated with the partial
resummation and the counter-terms, in particular those associated
with $V_2^{VC}$. Choosing this term (rather than $V_1^{VC}$
which figures into the longitudinal channel, see first line
of Eq.~(\ref{eq:vcpipi}) or (\ref{eq:vcpipilin})) ensures that the following
procedure does not affect the longitudinal mode which would disrupt
the chiral relations in that channel. Because the divergence
in the $\rho$ self-energy grows like $q^6$, the counter-terms
have the general form of
\begin{equation}
V_2^{VC ({\rm counter})} (q^2) = A_1 q^2 + A_2 q^4
+ A_3 q^6,
\end{equation}
where the coefficients $A_i$ contain both the finite and infinite
contributions of the counter-terms. Just as the nominal couplings are
different from their Lagrangian values, so can the
counter-terms. Therefore, we exploit this freedom and
redefine $V_2^{VC}$ as
\begin{equation}
\begin{split}
V_2^{VC} (q^2) &= -\frac{3}{4} \frac{\delta_3}
{g_{\rho\pi\pi} + g_{\rho\pi\pi}^{(3)} q^2} \Sigma_\rho^T (q^2)
+\eta_1 q^2 + \eta_2 q^4 + \eta_3 q^6,\\
V_{2 {\rm Lin}}^{VC} (q^2) &= -\frac{3}{4} \frac{\bar{\delta}_3}
{g_{\rho\pi\pi} + g_{\rho\pi\pi}^{(3)} q^2} \Sigma_\rho^T (q^2)
+\bar{\eta}_1 q^2 + \bar{\eta}_2 q^4 + \bar{\eta}_3 q^6,
\end{split}
\end{equation}
where the finite contributions of the $A_i$'s are partitioned
between a part which contributes to $\Sigma_\rho^T$ 
(making use of a derivative expansion for the fraction)
and the $\eta_i$'s ($\bar{\eta}_i$'s).
We determine $\eta_1$ ($\bar{\eta}_1$) by requiring that the low temperature
real part of $\Sigma_a^T$ at an energy of $M_{a_1}$ is identical
to the case without VCs and $\rho$ self-energies to recover the correct
chiral behavior of no mass shift at ${\cal O}(T^2)$.
In principle, $\eta_2$ ($\bar{\eta}_2$) and $\eta_3$ ($\bar{\eta}_3$)
are two additional parameters. However, at this point we only make use
of adjusting $\eta_2$ ($\bar{\eta}_2$) and set $\eta_3$ ($\bar{\eta}_3$)
to zero for simplicity. The parameter values are $\eta_1 = 1.51 {\rm GeV}^{-1}$,
$\eta_2 = -1.805 {\rm GeV}^{-3}$, $\eta_3 =0$, $\bar{\eta}_1 =1.108
{\rm GeV}^{-1}$, $\bar{\eta}_2 = -2.042 {\rm GeV}^{-3}$,
and $\bar{\eta}_3 =0$.

\section{Loop integrals for finite temperature self-energy calculation}
\label{sec:loop}

In this section, we present the loop integrals to be evaluated for the
finite temperature calculations
of the self-energies. In each case, we have already performed
the sum over Matsubara frequencies. Terms which are not proportional
to a thermal Bose factor are associated with vacuum contributions.

The self-energy of a $h\to h_1h_2\to h$ loop diagram has the generic form
\begin{widetext}
\begin{equation}
\label{eq:ftloop}
\begin{split}
&\Sigma (p_0,|\vec{p}|) = \int
\frac{d^3 k}{(2\pi)^3} \frac{1}{4 E_{h_1}(\vec{k}) E_{h_2}(\vec{p}-\vec{k})} \\
& \times \left(\frac{1}{p_0 - E_{h_1}(\vec{k}) - E_{h_2}(\vec{p}-\vec{k}) + i \epsilon}
\left[\left(\frac{1}{2}+f^B(E_{h_1}(\vec{k}))\right) {\cal V}(E_{h_1}(\vec{k}))
+ \left(\frac{1}{2} + f^B(E_{h_2}(\vec{p}-\vec{k}))\right)
{\cal V}(p_0-E_{h_2}(\vec{p}-\vec{k}))\right]\right.\\
& + \frac{1}{p_0 + E_{h_1}(\vec{k}) - E_{h_2}(\vec{p}-\vec{k}) + i \epsilon}
\left[\left(\frac{1}{2}+f^B(E_{h_1}(\vec{k}))\right){\cal V}(-E_{h_1}(\vec{k}))
-\left(\frac{1}{2} + f^B(E_{h_2}(\vec{p}-\vec{k}))\right){\cal V}(p_0 - E_{h_2}(\vec{p}-\vec{k}))
\right]
\\
& - \frac{1}{p_0 - E_{h_1}(\vec{k}) + E_{h_2}(\vec{p}-\vec{k}) + i \epsilon}
\left[\left(\frac{1}{2}+f^B(E_{h_1}(\vec{k})\right) {\cal V}(E_{h_1}(\vec{k}))
-\left(\frac{1}{2} + f^B(E_{h_2}(\vec{p}-\vec{k}))\right) {\cal V}(p_0 + E_{h_2}(\vec{p}-\vec{k}))
\right]
\\
& \left.- \frac{1}{p_0+ E_{h_1}(\vec{k})+ E_{h_2}(\vec{p}-\vec{k}) + i \epsilon}
\left[\left(\frac{1}{2}+f^B(E_{h_1}(\vec{k}))\right){\cal V}(-E_{h_1}(\vec{k}))
+\left(\frac{1}{2}+f^B(E_{h_2}(\vec{p}-\vec{k}))\right) {\cal V}(p_0 + E_{h_2}(\vec{p}-\vec{k}))
\right] \right).
\end{split}
\end{equation}
\end{widetext}
Throughout, we use the notation $E_h(\vec{k}) = (|\vec{k}|^2 + M_h^2)^{1/2}$
for the on-shell energies of hadron $h$. In the integral, $h_1$ and $h_2$ are
determined by the diagram in question, and will be specified for each case below.
The Bose distributions are given by $f^B(E) = (e^{E/T}-1)^{-1}$.
The function ${\cal V}(E)$ is related to the vertices and is unique for each
diagram; it generically depends on combinations of $p_0$, $\vec{p}$, and $\vec{k}$,
while its dependence on an energy argument is specified in the expression above
and in the following ones.
In the self-energy, Eq.~(\ref{eq:ftloop}),
the first term is associated with the decay process (or unitarity cut) including
final-state Bose enhancement.
The second and third terms are associated with
scattering processes (or Landau cuts) where the thermal mesons are
$h_1$ and $h_2$, respectively. The fourth term is the unitarity cut for negative
energies, ensuring the retarded property of the self-energy.

The tadpole diagrams are completely real and have a simpler structure and will
be quoted in due course.

\subsection{Vector channel}
We begin with the $\rho$ self-energy.
At finite temperature, we consider the case of finite 3-momentum, inducing
different 3D-transverse, $\Sigma_\rho^\perp$, and -longitudinal,
$\Sigma_\rho^\parallel$, self-energies.
For the non-linear
realization, each of these self-energies is comprised of three diagrams:
the $\pi\pi$ loop (first
diagram of the top row in Fig.~\ref{fig:dia1}), the $\pi$-tadpole
(second diagram of the top row in Fig.~\ref{fig:dia1}), and
the $\pi a_1$ loop (diagram in in Fig.~\ref{fig:pia}).
The linear realization allows for three new
diagrams because of the explicit $\sigma$ field:
a $\sigma$ tadpole (first diagram in Fig.~\ref{fig:axialLin}),
a $\sigma \rho$ loop (second diagram in Fig.~\ref{fig:axialLin}),
and a lollipop diagram (first diagram in Fig.~\ref{fig:sigmaTada}).

For the $\pi\pi$ loop, the self-energy is in the form of Eq.~(\ref{eq:ftloop})
with $M_{h_1}=M_{h_2} = M_\pi$. For $\Sigma_\rho^\perp$, the vertex function
reads
\begin{equation}
{\cal V}_\perp^{(\pi\pi)}(k_0) = 2
\left(g_{\rho\pi\pi} + g_{\rho\pi\pi}^{(3)} p^2\right)^2
|\vec{k}|^2 (1-x^2)
\end{equation}
while for $\Sigma_\rho^\parallel$ it is given by
\begin{equation}
\begin{split}
{\cal V}_\parallel^{(\pi\pi)}(k_0) &= 4
\left(g_{\rho\pi\pi} + g_{\rho\pi\pi}^{(3)} p^2\right)^2
\\
&\times \left(-k_0^2+|\vec{k}|^2 x^2+ \frac{(p \cdot k)^2}{p^2}\right) \ .
\end{split}
\end{equation}
In these expressions, $p^2 = p_0^2 - |\vec{p}|^2$,
$k^2 = k_0^2 -|\vec{k}|^2$, $p\cdot k = p_0 k_0 - \vec{p}\cdot\vec{k}$,
and $x = \cos(\theta)$ with
$\theta$ being the angle between $\vec{p}$ and $\vec{k}$.
We will use this notation throughout this section.
The above two contributions are identicl for the two realizations.

For the $\pi$ tadpole the two self-energies
in the non-linear realization are
\begin{widetext}
\begin{equation}
\begin{split}
\Sigma_\rho^{\perp (\pi {\rm Tad})}  (p_0, |\vec{p}|)&= -2\int \frac{d^3 k}{(2\pi)^3}
\frac{1+2 f^B(E_\pi(\vec{k}))}{E_\pi(\vec{k})}
\left[\alpha_1+2 \alpha_2 p^2-\alpha_8 \left(M_\pi^2+\frac{1}{2}
\left(|\vec{k}|^2-\frac{(\vec{p}\cdot\vec{k})^2}{|\vec{p}|^2}\right)\right)\right] ,\\
\Sigma_\rho^{\parallel (\pi {\rm Tad})} (p_0, |\vec{p}|) &=
-2\int \frac{d^3 k}{(2\pi)^3} \frac{1+2 f^B(E_\pi(\vec{k}))}{E_\pi(\vec{k})}
\left[\alpha_1+2 \alpha_2 p^2 - \alpha_8 \left(\frac{p_0^2 E_\pi(\vec{k})^2}{p^2}
+\left(\vec{p}\cdot\vec{k}\right)^2\left(\frac{1}{p^2}+\frac{1}{|\vec{p}|^2}\right)
-|\vec{k}|^2\right)\right] ,
\end{split}
\end{equation}
and for the linear realization
\begin{equation}
\begin{split}
\Sigma_{\rho {\rm Lin}}^{\perp (\pi {\rm Tad})}  (p_0, |\vec{p}|)&= -2\int \frac{d^3 k}{(2\pi)^3}
\frac{1+2 f^B(E_\pi(\vec{k}))}{E_\pi(\vec{k})}
\left[\bar{\alpha}_1+ \bar{\alpha}_2 p^2-\bar{\alpha}_8 \left(M_\pi^2+\frac{1}{2}
\left(|\vec{k}|^2-\frac{(\vec{p}\cdot\vec{k})^2}{|\vec{p}|^2}\right)\right)\right] ,\\
\Sigma_{\rho {\rm Lin}}^{\parallel (\pi {\rm Tad})} (p_0, |\vec{p}|) &=
-2\int \frac{d^3 k}{(2\pi)^3} \frac{1+2 f^B(E_\pi(\vec{k}))}{E_\pi(\vec{k})}
\left[\bar{\alpha}_1+ \bar{\alpha}_2 p^2 - \bar{\alpha}_8 \left(\frac{p_0^2 E_\pi(\vec{k})^2}{p^2}
+\left(\vec{p}\cdot\vec{k}\right)^2\left(\frac{1}{p^2}+\frac{1}{|\vec{p}|^2}\right)
-|\vec{k}|^2\right)\right].
\end{split}
\end{equation}

For the $\pi a_1$ loop, the self-energy is of the form
in Eq.~(\ref{eq:ftloop}) with $M_{h_1}= M_\pi$ and $M_{h_2} = M_{a_1}$. For
$\Sigma_\rho^\perp$, the 3D-transverse vertex function is given by
\begin{equation}
\begin{split}
{\cal V}_\perp^{(\pi a_1)}(k_0) &= 2\left(C-(2 D-E-F)
 p\cdot k+2 D p^2 - E k^2\right)^2
+(1-x^2)\frac{|\vec{k}|^2}{M_{a_1}^2}\\
& \times \left(
p^2(2 C F-M_{a_1}^2(4 D(D+E-F)+F^2)+F^2 p^2)
+E^2 M_{a_1}^2(k^2-2 p\cdot k-M_{a_1}^2)
\right),
\end{split}
\end{equation}
and for the 3D-longitudinal part
\begin{equation}
\begin{split}
{\cal V}_\parallel^{(\pi a_1)}(k_0) &=
2\left(C-(2 D-E-F)
 p\cdot k+2 D p^2 - E k^2\right)^2
+\frac{1}{M_{a_1}^2}
\left(-k_0^2 +|\vec{k}|^2 x^2 +\frac{(p \cdot k)^2}{p^2}
\right)\\
& \times \left(
p^2(2 C F-M_{a_1}^2(4 D(D+E-F)+F^2)+F^2 p^2)
+E^2 M_{a_1}^2(k^2-2 p\cdot k-M_{a_1}^2)
\right)
\end{split}
\end{equation}
\end{widetext}
For the linear realization ${\cal V}_\perp$ and ${\cal V}_\parallel$
are identical but with the couplings $C$, $D$, $E$, $F$ replaced
by $\bar{C}$, $\bar{D}$, $\bar{E}$, $\bar{F}$.

The $\sigma$ tadpole provides the same contribution to both
the 3D-transverse and -longitudinal self-energies and is
given by
\begin{equation}
\Sigma_{\rho {\rm Lin}}^{\perp/\parallel (\sigma {\rm Tad})}  (p_0, |\vec{p}|)=
e_1 p^2 \int \frac{d^3 k}{(2\pi)^3}
\frac{1+2 f^B(E_\sigma(\vec{k}))}{E_\sigma(\vec{k})} .
\end{equation}

The $\sigma \rho$ loop contribution has the form
of Eq.~(\ref{eq:ftloop}) with $M_{h_1}=M_\sigma$ and
$M_{h_2} = M_{\rho}$. The functions ${\cal V}$ for the two
polarizations are:
\begin{widetext}
\begin{equation}
\begin{split}
{\cal V}_{\perp {\rm Lin}}^{(\sigma \rho)}(k_0)&= (4 c_1)^2 \left[\left(
p^2 - p\cdot k\right)^2 -\frac{1}{2} p^2 |k|^2(1-x^2)\right] ,\\
{\cal V}_{\parallel {\rm Lin}}^{(\sigma \rho)}(k_0)&= (4 c_1)^2 \left[\left(
p^2 - p \cdot k\right)^2 -p^2\left(-k_0^2+\frac{(p \cdot k)^2}
{p^2} + \frac{(\vec{p}\cdot \vec{k})^2}{|\vec{p}|^2}\right)\right] .
\end{split}
\end{equation}
\end{widetext}

Finally, for the $\rho$ self-energy, the contribution from the
lollipop diagrams is identical for the two polarizations,
\begin{equation}
\Sigma_{\rho {\rm Lin}}^{\perp/\parallel ({\rm lolli})}  (p_0, |\vec{p}|)=
\frac{4 c_1 p^2}
{M_\sigma^2} \left(\Sigma_{\rm lolli}^\pi + \Sigma_{\rm lolli}^\sigma\right) \ .
\end{equation}
The terms $\Sigma_{\rm lolli}^\pi$ and $\Sigma_{\rm lolli}^\sigma$
are associated with the pion and sigma loops in the diagram of
Fig.~\ref{fig:sigmaTadb} and read
\begin{equation}
\begin{split}
\Sigma_{\rm lolli}^\pi &= -Z_\pi \sigma_0 \left(\lambda
+\frac{g_{\rho\pi\pi}^2 M_\pi^2}{M_\rho^2 Z_\pi^2}\right)\\
&\times\int \frac{d^3k}{(2\pi)^3} \frac{1+2 f^B(E_\pi(\vec{k}))}
{2 E_\pi(\vec{k})} ,\\
\Sigma_{\rm lolli}^\sigma &= -3 \lambda \sigma_0
\int \frac{d^3k}{(2\pi)^3} \frac{1+2 f^B(E_\sigma(\vec{k}))}
{2 E_\sigma(\vec{k})}.
\end{split}
\end{equation}

\subsection{Axial-vector channel}

In the $AV$ and pion channels,
there are four self-energies, the $a_1$ self-energy for the
4D-transverse and -longitudinal channels ($\Sigma_{a_1}^T$
and $\Sigma_{a_1}^L$) the pion self-energy ($\Sigma_{\pi\pi}$),
and the $a_1-\pi$ transition self-energy ($\Sigma_{a\pi}$).
Each self-energy has two contributions in both realizations,
a $\pi \rho$ loop and a $\pi$-tadpole, as shown in the
bottom row of Fig.~\ref{fig:dia1}, and an additional
four contributions which are unique to the linear realization,
a $\pi \sigma$ loop, a $\sigma$ tadpole, a $\sigma a_1$
loop, and a lollipop diagram (\cf~the last three
diagrams in Fig.~\ref{fig:axialLin} and Fig.~\ref{fig:sigmaTada},
all with appropriate external legs).
In this paper, only the zero-momentum case is analyzed for
these self-energies. Thus expressions will be presented only for
this case.

The $\pi\rho$ loop within the $AV$ channel is more involved
than the expressions already presented because of the
resummed $\rho$ propagator and associated VCs,
which involve an additional integral. Because we are concerned only
with the vacuum broadening of the $\rho$ meson and the vacuum
VCs, this additional integral is only over the
invariant mass of the $\rho$ meson.

The transverse $a_1$ self-energy can be written in the form
\begin{widetext}
\begin{equation}
\begin{split}
\Sigma_{a_1}^{T (\pi \rho)} (p_0) &=
\int\limits_{0}^{\Lambda_{\rm cut}^2} \frac{-1}{\pi} d M^2
\left( {\rm Im}\left[ D_\rho(M^2)\right] \tilde{\Sigma}_1(p_0) \right.
\\
&\left.+ {\rm Im} \left[ D_\rho(M^2)
\left(\left(V_a^{T 1}(M^2)\right)^2-(V_a^{T 2}(M^2))^2 p_0^2 M^2\right)
+ \frac{1}{2} V_4^{3\pi}(M^2)\right] \frac{1}{3} \tilde{\Sigma}_2(p_0)\right),
\end{split}
\label{eq:SigTa1pirho}
\end{equation}
\end{widetext}
where the integration is over the invariant mass of the resummed $\rho$ meson.
The vertex functions, $V_a^{T 1}$, $V_a^{T 2}$, and the $V_i^{3\pi}$s,
are evaluated in the vacuum (specific to the realization) and have
been quoted in the previous section.
The temperature dependence of the integral in Eq.~(\ref{eq:SigTa1pirho})
resides in the functions $\tilde{\Sigma}_1$ and $\tilde{\Sigma}_2$.
They are of the form of Eq.~(\ref{eq:ftloop}) with $M_{h_1} = M_\pi$
and $M_{h_2} =M$ (the $\rho$'s invariant mass) and vertex functions
\begin{widetext}
\begin{equation}
\begin{split}
{\cal V}_1(k_0) &= 2\left[ \left(V_a^{T 1}(q^2) +p_0(p_0 - k_0) V_a^{T 2}(q^2)\right)^2
+ \frac{1}{2}D_\rho^{-1}(q^2)\left(V_1^{3\pi}(q^2)
+p_0(p_0 -k_0) V_2^{3\pi}
  + p_0^2 (p_0-k_0)^2 V_3^{3\pi}\right)\right]
\end{split}
\end{equation}
\end{widetext}
for the former (with $q^2 = (p_0 - k_0)^2 - |\vec{k}|^2$) and
${\cal V}_2(k_0)=2|\vec{k}|^2/M^2$ for the latter. Additionally, to preserve
the retarded properties of the self-energy and because we have integrate
over $M^2$, the calculation of $\tilde{\Sigma}_1$ requires using the complex
conjugate, ${\cal V}_1^*(p_0 + E_{h_2})$.  The VC loop integrals are again
the same as presented in the previous section.

For $\Sigma_{a_1}^L$, $\Sigma_{a\pi}$ and $\Sigma_{\pi\pi}$,
the $\pi\rho$ loop contribution (in both realizations) has the form
\begin{widetext}
\begin{equation}
\begin{split}
\label{eq:longa1-2}
\Sigma_a^{L (\pi\rho)} (p_0) & =  \int\limits_{0}^{\Lambda_{\rm cut}^2} \frac{-1}{\pi}\, dM^2
{\rm Im}\!\left[D_\rho(M^2) V_a^2(M^2)+\frac{1}{2} V_{aa}^{3\pi}(M^2)\right]
\tilde{\Sigma}_2(p_0) ,\\
\Sigma_{a\pi}^{(\pi\rho)} (p_0) & = - \int\limits_{0}^{\Lambda_{\rm cut}^2} \frac{-1}{\pi}\,
 dM^2
{\rm Im}\!\left[D_\rho(M^2) V_a(M^2)V_p(M^2)+\frac{1}{2} V_{ap}^{3\pi}(M^2)\right]
\tilde{\Sigma}_2(p_0) ,\\
\Sigma_{\pi\pi}^{(\pi\rho)} (p_0) & = \int\limits_{0}^{\Lambda_{\rm cut}^2} \frac{-1}{\pi}\,
 dM^2
{\rm Im}\!\left[D_\rho(M^2) V_p^2(M^2)+\frac{1}{2} V_{pp}^{3\pi}(M^2)\right]
 \tilde{\Sigma}_2( p_0),
\end{split}
\end{equation}
\end{widetext}
where the integration is over the invariant mass of the
dressed $\rho$ meson.
This is similar to the vacuum structure of Eq.~(\ref{eq:longa1}).
The function $\tilde{\Sigma}_2$ is the same as described above,
encoding the temperature dependence.
The pertinent vertex functions for each realization are given
by their vacuum forms as presented in the previous section.

For the longitudinal self-energies, the $\pi\rho$ loop also
has an additional term which is completely real and thereby cannot
be expressed in a form similar to the above expressions.
These extra terms are
\begin{widetext}
\begin{equation}
\begin{split}
\Sigma_{a_1}^{L ({\rm extra})} (p_0) &=  \frac{C^2}{M_\rho^2}
\int \frac{d^3 k}{(2\pi)^3}
\frac{1+2 f^B(E_\pi(\vec{k}))}{E_\pi(\vec{k})} ,
\\
\Sigma_{a\pi}^{({\rm extra})} (p_0) &=  -\frac{g_{\rho\pi\pi} C}{M_\rho^2}
\int \frac{d^3 k}{(2\pi)^3}
\frac{1+2 f^B(E_\pi(\vec{k}))}{E_\pi(\vec{k})} ,
\\
\Sigma_{\pi\pi}^{({\rm extra})} (p_0) &= \frac{g_{\rho\pi\pi}^2}
{M_\rho^2}\left(p_0^2+M_\pi^2\right)
\int \frac{d^3 k}{(2\pi)^3}
\frac{1+2 f^B(E_\pi(\vec{k}))}{E_\pi(\vec{k})} \ .
\end{split}
\end{equation}
They are the same in the linear realization except for replacing $C$
by $\bar{C}$.

The pion tadpole contributions in the non-linear realization
are as follows:
\begin{equation}
\begin{split}
\Sigma_{a_1}^{T (\pi {\rm Tad})} (p_0) &= -2 \int \frac{d^3 k}{(2\pi)^3}
\frac{1+2 f^B(E_\pi(\vec{k}))}{E_\pi(\vec{k})}
\left(\beta_1 + 2 \beta_2 p_0^2 - \beta_7 \left(M_\pi^2+\frac{1}{3} |\vec{k}|^2\right)
\right) ,\\
\Sigma_{a_1}^{L (\pi {\rm Tad})} (p_0) &= 2 \int \frac{d^3 k}{(2\pi)^3}
\frac{1+2 f^B(E_\pi(\vec{k}))}{E_\pi(\vec{k})}
\left(\beta_1  + \beta_7 |\vec{k}|^2
\right) ,\\
\Sigma_{a \pi}^{(\pi {\rm Tad})} (p_0) &= - \int \frac{d^3 k}{(2\pi)^3}
\frac{1+2 f^B(E_\pi(\vec{k}))}{E_\pi(\vec{k})}
\left(\delta_1  + \delta_3 |\vec{k}|^2
\right) ,\\
\Sigma_{\pi\pi}^{(\pi {\rm Tad})} (p_0) &= 2 \int \frac{d^3 k}{(2\pi)^3}
\frac{1+2 f^B(E_\pi(\vec{k}))}{E_\pi(\vec{k})}
\left(\gamma_1(p_0^2 +M_\pi^2) -2 \gamma_2 p_0^2 |\vec{k}|^2
 - 5 \gamma_3 \right) \ ,
\end{split}
\end{equation}
and in the linear realization
\begin{equation}
\begin{split}
\Sigma_{a_1 {\rm Lin}}^{T (\pi {\rm Tad})} (p_0) &=
\int \frac{d^3 k}{(2\pi)^3}
\frac{1+2 f^B(E_\pi(\vec{k}))}{E_\pi(\vec{k})}
\left(\bar{\beta}_1 - 2 \bar{\beta_2} p_0^2 + 2\bar{\beta}_7
\left(M_\pi^2+\frac{1}{3} |\vec{k}|^2\right)
\right) ,\\
\Sigma_{a_1 {\rm Lin}}^{L (\pi {\rm Tad})} (p_0) &=
- \int \frac{d^3 k}{(2\pi)^3}
\frac{1+2 f^B(E_\pi(\vec{k}))}{E_\pi(\vec{k})}
\left(\bar{\beta}_1  - 2 \bar{\beta}_7 |\vec{k}|^2
\right) ,\\
\Sigma_{a \pi {\rm Lin}}^{(\pi {\rm Tad})} (p_0) &=
\frac{1}{2} \int \frac{d^3 k}{(2\pi)^3}
\frac{1+2 f^B(E_\pi(\vec{k}))}{E_\pi(\vec{k})}
\left(\bar{\delta}_1  - 2 \bar{\delta}_3 |\vec{k}|^2
\right) ,\\
\Sigma_{\pi\pi {\rm Lin}}^{(\pi {\rm Tad})} (p_0) &=
- \int \frac{d^3 k}{(2\pi)^3}
\frac{1+2 f^B(E_\pi(\vec{k}))}{E_\pi(\vec{k})}
\left(\bar{\gamma}_1(p_0^2+M_\pi^2) +4 \bar{\gamma}_2 p_0^2 |\vec{k}|^2
+10\bar{\gamma}_3 \right) \ .
\end{split}
\end{equation}

For the $\sigma$ tadpole, the self-energies are given by
\begin{equation}
\begin{split}
\Sigma_{a_1 {\rm Lin}}^{T (\sigma {\rm Tad})}(p_0) &=
\left(f_1 +2 f_2 p_0^2\right) \int \frac{d^3 k}{(2\pi)^3}
\frac{1+2 f^B(E_\sigma(\vec{k}))}{E_\sigma} , \\
\Sigma_{a_1 {\rm Lin}}^{L (\sigma {\rm Tad})}(p_0) &=
-f_1  \int \frac{d^3 k}{(2\pi)^3}
\frac{1+2 f^B(E_\sigma(\vec{k}))}{E_\sigma} , \\
\Sigma_{a\pi {\rm Lin}}^{(\sigma {\rm Tad})}(p_0) &=
\frac{g_1}{2}  \int \frac{d^3 k}{(2\pi)^3}
\frac{1+2 f^B(E_\sigma(\vec{k}))}{E_\sigma} ,\\
\Sigma_{\pi\pi {\rm Lin}}^{(\sigma {\rm Tad})}(p_0) &=
-(h_1+h_2 p_0^2)  \int \frac{d^3 k}{(2\pi)^3}
\frac{1+2 f^B(E_\sigma(\vec{k}))}{E_\sigma} \ .
\end{split}
\end{equation}

The $\pi \sigma$ loop has the structure of Eq.~(\ref{eq:ftloop})
with $M_{h_1}=M_\pi$ and $M_{h_2} = M_\sigma$ and vertex functions
\begin{equation}
\begin{split}
{\cal V}_{a_1 {\rm Lin}}^{T (\pi \sigma)} (k_0) &=
\frac{1}{3} |\vec{k}|^2 \left(b_1 -b_2 -b_3 p_0^2\right)^2 ,\\
{\cal V}_{a_1 {\rm Lin}}^{L (\pi \sigma)} (k_0) &=
\left(b_1 p_0 +k_0 (b_2 - b_1)\right)^2 ,\\
{\cal V}_{a \pi {\rm Lin}}^{(\pi \sigma)} (k_0) &=
-\frac{1}{p_0^2} \left(b_1 p_0^2 -\left(b_1-b_2\right) p_0 k_0\right)
\left(2 a_1+2 \left(a_2-a_3\right) k_0 p_0 +a_3\left(k^2+p_0^2\right)\right) ,\\
{\cal V}_{\pi\pi {\rm Lin}}^{(\pi \sigma)} (k_0) &=
\left(2 a_1 +2 a_2 p_0 k_0 +a_3 (p_0^2-2 p_0 k_0+k^2)\right)^2 .
\end{split}
\end{equation}

The $\sigma a_1$ loop has the structure given
in Eq.~(\ref{eq:ftloop}) with $M_{h_1} = M_\sigma$,
$M_{h_2} = M_{a_1}$ and vertex functions for each
channel defined as:
\begin{equation}
\begin{split}
{\cal V}_{a_1 {\rm Lin}}^{T (\sigma a_1)} (k_0) &=
\left(2 d_1 +d_3 k^2 - 4 d_2 k_0 p_0 + 4 d_2 p_0^2\right)^2\\
&+\frac{1}{3} \frac{|\vec{k}|^2}{M_{a_1}^2} \left( 4 d_1^2
+ 4 d_1 d_3 (M_{a_1}^2 + p_0^2)
+ d_3^2(k^2 M_{a_1}^2 -2 M_{a_1}^2 k_0 p_0 - M_{a_1}^2 p_0^2
+p_0^4)+16 d_2 p_0^2 M_{a_1}^2(d_3-d_2)\right) ,\\
{\cal V}_{a_1 {\rm Lin}}^{L (\sigma a_1)} (k_0) &=
-\left(2 d_1 +d_3 (p_0^2 - 2 p_0 k_0 +k^2)\right)^2\\
&- \frac{p_0-k_0}{p_0 M_{a_1}^2} \left(4 d_1^2 + 4 d_1 d_3 M_{a_1}^2
+d_3^2 M_{a_1}^2(p_0^2-2 p_0 k_0+k^2)\right) ,\\
{\cal V}_{a \pi {\rm Lin}}^{(\sigma a_1)} (k_0) &=
-\frac{1}{M_{a_1}^2 p_0^2} \left(2 d_1\left( b_1
\left(k^2 p_0^2+k_0^2 p_0^2-k_0 p_0(k^2-M_{a_1}^2 +p_0^2)\right)
+ b_2(k_0^2 p_0^2-2 k_0 p_0^3-M_{a_1}^2 p_0^2+p_0^4)
+b_3 M_{a_1}^2 p_0^2 |\vec{k}|^2\right)\right.\\
&\left.+ d_3 M_{a_1}^2 p_0^2 |\vec{k}|^2 \left(b_2-b_1
+b_3(k^2 -2 k_0 p_0 +p_0^2)\right)\right) ,\\
{\cal V}_{\pi\pi {\rm Lin}}^{(\sigma a_1)} (k_0) &=
(b_1-b_2-b_3(p_0^2-2 p_0 k_0+k^2))^2\left(-p_0^2+\frac{(p_0^2-p_0 k_0)^2}
{M_{a_1}^2}\right)\\
&+\frac{1}{M_{a_1}^2}\left(p_0^2-2 p_0 k_0+k^2-M_{a_1}^2\right)
\Big(2(b_1-b_2-b_3(p_0^2-2 p_0 k_0+k^2))(b_3(p_0^2-p_0 k_0)-b1)
(p_0^2-p_0 k_0)\\
&\left.+\left(b_3(p_0^2-p_0 k_0)-b_1\right)^2
(p_0^2-2 p_0 k_0 +k^2)\right) .
\end{split}
\end{equation}
\end{widetext}

The last contributions are the lollipop diagrams which
read
\begin{equation}
\begin{split}
\Sigma_{a_1 {\rm Lin}}^{T ({\rm Lolli})} (p_0) &=
\left(2 d_1 + 4 d_2 p_0^2\right) \frac{1}{M_\sigma^2}
\left(\Sigma_{\rm lolli}^\pi + \Sigma_{\rm lolli}^\sigma\right) ,\\
\Sigma_{a_1 {\rm Lin}}^{L ({\rm Lolli})} (p_0) &= \frac{2 d_1}{M_\sigma^2}
\left(\Sigma_{\rm lolli}^\pi + \Sigma_{\rm lolli}^\sigma\right) ,\\
\Sigma_{a \pi {\rm Lin}}^{({\rm Lolli})} (p_0) &=
\frac{b_2}{M_\sigma^2}
\left(\Sigma_{\rm lolli}^\pi + \Sigma_{\rm lolli}^\sigma\right) ,\\
\Sigma_{\pi\pi {\rm Lin}}^{({\rm Lolli})} (p_0) &=
\frac{-2(a_1+a_2 p_0^2)}{M_\sigma^2}
\left(\Sigma_{\rm lolli}^\pi + \Sigma_{\rm lolli}^\sigma\right).
\end{split}
\end{equation}

\end{appendix}


\begin{thebibliography}{99}


\bibitem{Borsanyi:2010bp}
  S.~Borsanyi {\it et al.}  [Wuppertal-Budapest Collaboration],
  JHEP {\bf 1009}, 073 (2010).

\bibitem{Bazavov:2011nk}
 A.~Bazavov {\it et al.},
 Phys.\ Rev.\ D {\bf 85} 054503 (2012).

\bibitem{Porter:1997rc}
  R.~J.~Porter {\it et al.} [DLS Collaboration],
  Phys.\ Rev.\ Lett.\  {\bf 79}, 1229 (1997).



\bibitem{Adamova:2006nu}
  D.~Adamova {\it et al.} [CERES/NA45 Collaboration]
  Phys.\ Lett.\ B {\bf 666}, 425 (2008).

\bibitem{Arnaldi:2008fw}
  R.~Arnaldi {\it et al.}  [NA60 Collaboration],
  Eur.\ Phys.\ J.\ C {\bf 61}, 711 (2009).

\bibitem{Agakishiev:2011vf}
  G.~Agakishiev {\it et al.} [HADES Collaboration],
  Phys.\ Rev.\ C {\bf 84}, 014902 (2011).

\bibitem{Geurts:2012rv}
  F.~Geurts [STAR Collaboration],
  Nucl.\ Phys.\ A {\bf 904-905}, 217c (2013).


\bibitem{Rapp:2009yu}
  R.~Rapp, J.~Wambach and H.~van Hees,
  in {\it Relativistic Heavy-Ion Physics}, edited by R.~Stock
  and Landolt B\"{o}rnstein (Springer, Berlin, 2010), New Series
  {\bf I/23A}, p. 4-1;
  [arXiv:0901.3289 [hep-ph]].



\bibitem{Rapp:2013nxa}
  R.~Rapp,
  Adv.\ High Energy Phys.\  {\bf 2013}, 148253 (2013).


\bibitem{Hohler:2013eba}
  P.~M.~Hohler and R.~Rapp,
  Phys.\ Lett.\ B {\bf 731}, 103 (2014).

\bibitem{Ayala:2014rka}
  A.~Ayala, C.~A.~Dominguez, M.~Loewe and Y.~Zhang,
  Phys.\ Rev.\ D {\bf 90}, 034012 (2014).

\bibitem{Yang:1954ek}
  C.-N.~Yang and R.L.~Mills,
  Phys.\ Rev.\  {\bf 96}, 191 (1954).

\bibitem{Gasser:1983yg}
  J.~Gasser and H.~Leutwyler,
  Annals Phys.\  {\bf 158}, 142 (1984).

\bibitem{Gomm:1984at}
  H.~Gomm, O.~Kaymakcalan and J.~Schechter,
  Phys.\ Rev.\ D {\bf 30}, 2345 (1984).


\bibitem{Bando:1987br}
  M.~Bando, T.~Kugo and K.~Yamawaki,
  Phys.\ Rept.\  {\bf 164}, 217 (1988).



\bibitem{Harada:2003jx}
  M.~Harada and K.~Yamawaki,
  Phys.\ Rept.\  {\bf 381}, 1 (2003).

\bibitem{Song:1993ae}
  C.~Song,
  Phys.\ Rev.\ C {\bf 47}, 2861 (1993).

\bibitem{Ko:1994en}
  P.~Ko and S.~Rudaz,
  Phys.\ Rev.\ D {\bf 50}, 6877 (1994).

\bibitem{Song:1993af}
  C.~Song,
  Phys.\ Rev.\ D {\bf 48}, 1375 (1993).

\bibitem{Song:1993ipa}
  C.~Song,
  Phys.\ Rev.\ D {\bf 49}, 1556 (1994).

\bibitem{Pisarski:1995xu}
  R.~D.~Pisarski,
  Phys.\ Rev.\ D {\bf 52}, 3773 (1995);
  Nucl.\ Phys.\ A {\bf 590}, 553C (1995);
  hep-ph/9503330.

\bibitem{Barate:1998uf}
  R.~Barate {\it et al.}  [ALEPH Collaboration],
  Eur.\ Phys.\ J.\ C {\bf 4}, 409 (1998).

\bibitem{Ackerstaff:1998yj}
  K.~Ackerstaff {\it et al.}  [OPAL Collaboration],
  Eur.\ Phys.\ J.\ C {\bf 7}, 571 (1999).

\bibitem{Harada:2008hj}
  M.~Harada, C.~Sasaki and W.~Weise,
  Phys.\ Rev.\ D {\bf 78}, 114003 (2008).

\bibitem{Urban:2001ru}
  M.~Urban, M.~Buballa and J.~Wambach,
  Nucl.\ Phys.\ A {\bf 697}, 338 (2002).

\bibitem{Wagner:2008gz}
  M.~Wagner and S.~Leupold,
  Phys.\ Rev.\ D {\bf 78}, 053001 (2008).

\bibitem{Parganlija:2010fz}
  D.~Parganlija, F.~Giacosa and D.H.~Rischke,
  Phys.\ Rev.\ D {\bf 82}, 054024 (2010).



\bibitem{Urban:2001uv}
  M.~Urban, M.~Buballa and J.~Wambach,
  Phys.\ Rev.\ Lett.\  {\bf 88}, 042002 (2002).

\bibitem{Struber:2007bm}
  S.~Str\"uber and D.~H.~Rischke,
  Phys.\ Rev.\ D {\bf 77}, 085004 (2008).


\bibitem{Hohler:2013ena}
  P.~M.~Hohler and R.~Rapp,
  Phys.\ Rev.\ D {\bf 89}, 125013 (2014).

\bibitem{Kroll:1967it}
  N.~M.~Kroll, T.~D.~Lee and B.~Zumino,
  Phys.\ Rev.\  {\bf 157}, 1376 (1967).

\bibitem{Weinberg:1996kr}
  S.~Weinberg,
  {\it The quantum theory of fields. Vol. 2: Modern applications},
  Cambridge Univ. Pr. (1996).





\bibitem{Hohler:2012xd}
  P.M.~Hohler and R.~Rapp,
  Nucl.\ Phys.\ A {\bf 892}, 58 (2012).

\bibitem{Shuryak:1993kg}
  E.V.~Shuryak,
  Rev.\ Mod.\ Phys.\  {\bf 65}, 1 (1993).

\bibitem{Adolph:2015pws}
  C.~Adolph {\it et al.} [COMPASS Collaboration],
  arXiv:1501.05732 [hep-ex].


\bibitem{Zielinski:1984au}
  M.~Zielinski
{\it et al.},
  Phys.\ Rev.\ Lett.\  {\bf 52}, 1195 (1984).

\bibitem{Roca:2006am}
  L.~Roca, A.~Hosaka and E.~Oset,
  Phys.\ Lett.\ B {\bf 658}, 17 (2007).

\bibitem{Agashe:2014kda}
  K.~A.~Olive {\it et al.}  [Particle Data Group Collaboration],
  Chin.\ Phys.\ C {\bf 38}, 090001 (2014).

\bibitem{Froggatt:1977hu}
  C.~D.~Froggatt and J.~L.~Petersen,
  Nucl.\ Phys.\ B {\bf 129}, 89 (1977).

\bibitem{Amendolia:1983di}
  S.~R.~Amendolia, B.~Badelek, G.~Batignani, G.~A.~Beck, E.~H.~Bellamy, E.~Bertolucci, D.~Bettoni and H.~Bilokon {\it et al.},
  Phys.\ Lett.\ B {\bf 138}, 454 (1984).

\bibitem{Amendolia:1984nz}
  S.~R.~Amendolia, B.~Badelek, G.~Batignani, G.~A.~Beck, F.~Bedeschi, E.~H.~Bellamy, E.~Bertolucci and D.~Bettoni {\it et al.},
  Phys.\ Lett.\ B {\bf 146}, 116 (1984).

\bibitem{Barkov:1985ac}
  L.~M.~Barkov, A.~G.~Chilingarov, S.~I.~Eidelman, B.~I.~Khazin, M.~Y.~Lelchuk, V.~S.~Okhapkin, E.~V.~Pakhtusova and S.~I.~Redin {\it et al.},
  Nucl.\ Phys.\ B {\bf 256}, 365 (1985).

\bibitem{Rapp:1999qu}
  R.~Rapp and C.~Gale,
  Phys.\ Rev.\ C {\bf 60}, 024903 (1999).


\bibitem{Dey:1990ba}
  M.~Dey, V.L.~Eletsky and B.L.~Ioffe,
  Phys.\ Lett.\ B {\bf 252} 620 (1990).

\bibitem{Steele:1996su}
  J.V.~Steele, H.~Yamagishi and I.~Zahed,
  Phys.\ Lett.\ B {\bf 384} 255 (1996).

\bibitem{Weinberg:1967kj}
  S.~Weinberg,
  Phys.\ Rev.\ Lett.\  {\bf 18} (1967) 507.

\bibitem{Gasser:1986vb}
  J.~Gasser and H.~Leutwyler,
  Phys.\ Lett.\ B {\bf 184}, 83 (1987).

\bibitem{Bochkarev:1995gi}
  A.~Bochkarev and J.~I.~Kapusta,
  Phys.\ Rev.\ D {\bf 54}, 4066 (1996).

\bibitem{Bilic:1997sh}
  N.~Bilic and H.~Nikolic,
  Eur.\ Phys.\ J.\ C {\bf 6}, 515 (1999).

\bibitem{Petropoulos:1998gt}
  N.~Petropoulos,
  J.\ Phys.\ G {\bf 25}, 2225 (1999).


\bibitem{Aarts:2015mma}
  G.~Aarts, C.~Allton, S.~Hands, B.~J\"{a}ger, C.~Praki and J.~I.~Skullerud,
  Phys.\ Rev.\ D {\bf 92}, 014503 (2015).

\end{thebibliography}
\end{document}